\documentclass[a4paper,11pt]{article}
\pdfoutput=1 

\usepackage{jheppub} 
\usepackage{mathrsfs}
\usepackage{graphicx}
\usepackage{csquotes}
\usepackage{latexsym}
\usepackage{bbm}
\usepackage[mathscr]{euscript}
\usepackage{subcaption}
\usepackage{amsmath,amssymb,bm} 
\usepackage{braket}
\usepackage{slashed}
\usepackage{moresize}

\usepackage{float}

\newcommand{\beq}{\begin{equation}}
\newcommand{\eeq}{\end{equation}}
\newcommand{\ba}{\begin{array}{ccc}}

\def\bea{\begin{eqnarray}}
\def\eea{\end{eqnarray}}
\usepackage[colorlinks=true]{hyperref} 
\hypersetup{
    unicode=false,          
    pdftoolbar=true,        
    pdfmenubar=true,        
    pdffitwindow=false,     
    pdfstartview={FitH},    
    pdfsubject={},   
    pdfcreator={},   
    pdfproducer={}, 
    pdfkeywords={} {} {}, 
    pdfnewwindow=true,      
    colorlinks=true,       
    linkcolor=magenta, 
    citecolor=blue,        
    filecolor=magenta,      
    urlcolor=blue           
}

\renewcommand{\approx}{\simeq}

\renewcommand{\approx}{\simeq}

\renewcommand{\approx}{\simeq}

\usepackage{xspace}
\usepackage[all]{xy}
\usepackage{amsmath}
\usepackage{float}

\title{Quantum models with the Yang-Lee phase transition}
\author[\mathcal{P}]{Erick Arguello Cruz,}
\author[\mathcal{T}]{Grigory Tarnopolsky}
\affiliation[\mathcal{PT}]{Department of Physics, Carnegie Mellon University, Pittsburgh, PA 15213, USA}

\emailAdd{earguell@andrew.cmu.edu}
\emailAdd{gtarnopo@andrew.cmu.edu}

\abstract{
In this article, we present four different $1+1$D quantum models that realize  
the Yang-Lee (YL) phase transition under a deformation that preserves $\mathcal{P}\mathcal{T}$ symmetry. These are the antiferromagnetic Ising spin chain
in transverse and longitudinal magnetic fields, the massive Schwinger model, the Blume-Capel model, and the three-state quantum clock model. Using the state-operator correspondence, we identify the YL critical point, compute the scaling dimensions of the lowest operators in each model, and find perfect agreement with the exact results for the YL criticality in two dimensions.  
Using  bosonization for the Schwinger model and the Polyakov-Hubbard transformation for the other models, we show that in all of these quantum models the YL critical point is described, as expected, by a massless bosonic field with an $i \phi^3$ interaction. 
In the quantum clock model, this critical field interacts with a massive bosonic field, and we identify the massless and massive states in the Hamiltonian spectrum. In addition,  we numerically compute the two-point function of $\phi$ at the Yang-Lee critical point and show that it grows with distance, in agreement with theoretical expectations.

}

\begin{document}
\maketitle
\flushbottom

\newpage
\section{Introduction and Summary}
The Yang-Lee criticality was originally discovered through a numerical analysis of the density of the Yang-Lee zeros for classical Ising ferromagnets \cite{Kortman:1971zz}. In the paramagnetic phase of the Ising model, the zeros of the partition function in the complex magnetic-field plane, known as the Yang-Lee zeros, lie on the imaginary axis away from the real axis \cite{Yang:1952be, Lee:1952ig}. It was later realized by Michael Fisher that the edge criticality of the Yang-Lee zeros is described by the Euclidean $\phi^3$ field theory with an imaginary coupling \cite{Fisher:1978pf}:
\begin{align}
S = \int d^{d}x \left(\frac{1}{2}(\partial_{\mu}\phi)^2 + i (h-h_{\textrm{crit}})\phi + \frac{ig}{3!} \phi^3\right)\,, \label{Phi3act}
\end{align}
where $d$ is the dimension of space and $g, h \in \mathbb{R}$. In order to reach the Yang-Lee (YL) critical point, one has to tune the imaginary magnetic field $ih$ to the critical value $ih_{\textrm{crit}}$. The action (\ref{Phi3act}) is obtained by starting from the Euclidean $\phi^4$ massive field theory, and perturbing it by the imaginary magnetic field term $i h\phi$. This term breaks the Ising model $\mathbb{Z}_{2}$ symmetry $\phi \to -\phi$, but preserves the antiunitary $\mathcal{P}\mathcal{T}$ symmetry, where $\mathcal{P}: \; \phi \to -\phi$ and $\mathcal{T}: \; i \to -i$ \cite{Becker:1991nr, Becker:1991nq}.   From this field-theoretical description it is clear that the Yang-Lee critical point appears back-to-back with the Ising one, and there exists a renormalization group (RG) flow from the Ising conformal field theory (CFT) to the Yang-Lee CFT.  The YL phase transition can be regarded as a transition associated with spontaneous breaking of  the $\mathcal{P}\mathcal{T}$ symmetry \cite{Gehlen.JPA.1991, Gehlen.JMPB.1994}\footnote{We note that the $\mathcal{P}\mathcal{T}$ symmetry of the Yang-Lee model can be spontaneously broken in a finite-size system, whereas $\mathbb{Z}_{2}$ symmetry of the Ising model can be spontaneously broken only in the thermodynamic limit.}. 

The $d = 6 - \epsilon$ expansion of the massless Euclidean $i\phi^3$ theory  provides good estimates for the scaling dimensions of the low-dimensional operators \cite{Fisher:1978pf, Fei:2014yja, Fei:2014xta, Gracey:2015tta, Borinsky:2021jdb, Kompaniets:2021hwg, Schnetz:2025wtu, Gracey:2025rnz, ArguelloCruz:2025zuq}, as well as for the free energy and other observables \cite{Tarnopolsky:2016vvd, Giombi:2024zrt, Diatlyk:2026oxm, Fraser-Taliente:2026iuj} in the Yang-Lee CFT
in all integer dimensions $d=2,3,4,5$. 
Recently, using the novel fuzzy sphere numerical method   \cite{Zhu:2022gjc, Hu.PRL.2023, Hu.PRB.2025, Fardelli.SP.2025, Lauchli:2025fii}, it became possible to uncover the CFT structure of the low-dimensional operators in the Yang-Lee CFT in $d=3$  \cite{ArguelloCruz:2025zuq, Fan:2025bhc, EliasMiro:2025msj}. The results agree well with $d=6-\epsilon$ and high-temperature expansions \cite{Butera:2012tq}.

In $d=2$, the non-unitary minimal model $M(2,5)$ provides an exact description of the Yang-Lee CFT \cite{Cardy:1985yy}. In this case, the YL CFT has only two primary operators:  the identity operator $I$, with  scaling dimension $\Delta_{I}=0$, and $\phi$, with  negative scaling dimension $\Delta_{\phi}= -2/5$ \cite{Belavin:1984vu, Cardy:1985yy}.
All  other descendant and quasi-primary operators are obtained by acting with Virasoro generators on these two primary ones  \cite{Belavin:1984vu}. As expected, in $d=2$, there exists an RG flow from the Ising  CFT, described by the unitary minimal model $M(3,4)$, to the YL CFT, described by $M(2,5)$ \cite{Fonseca:2001dc}.
In $d=2$, multicritical points with $\mathbb{Z}_{2}$ symmetry  are described by massless Euclidean $\phi^{2n}$ field theories with $n=3,4,\dots$  and correspond to  the A-series of the unitary minimal models $M(n+1,n+2)$  \cite{Zamolodchikov:1986db}.  It was recently argued that $i\phi^{2n+1}$ theories correspond to non-unitary minimal models $M(2,2n+3)$  for $n=3,4,\dots$ \cite{Katsevich:2024jgq, Katsevich:2025ojk} (see also \cite{Zambelli:2016cbw, Lencses:2022ira, Lencses:2024wib, Benedetti:2026tpa}). 

Since the Yang-Lee model is non-unitary, the  Monte Carlo methods, which provide highly accurate results for the Ising model, are not applicable to 
the YL case. Hence,  current numerical efforts to study the YL critical point using microscopic lattice descriptions are based on the Hamiltonian approach \cite{Uzelac.JPA.1981, Itzykson.EPL.1986, Gehlen.JPA.1991, Gehlen.JMPB.1994,  ArguelloCruz:2025zuq, Fan:2025bhc, EliasMiro:2025msj}. These methods usually consider the ferromagnetic quantum Ising model in a transverse magnetic field and perturb it by an imaginary longitudinal magnetic field to reach the YL critical point.  Therefore, it is of interest to explore the universality of the YL CFT and verify that it can be realized in other quantum microscopic models with $\mathcal{PT}$ symmetry. With this goal in mind, in this paper we consider four different $1+1$D quantum models in addition to the canonical  ferromagnetic quantum Ising model in a transverse magnetic field.  Namely, we consider the antiferromagnetic Ising model in transverse and longitudinal magnetic fields \cite{NovotyJMM1986, ovchinnikov2003}, the massive Schwinger model \cite{SchwingerPR1962, LowensteinAnn1971, CasherPRD1974}, the spin-$1$ Blume-Capel (BC) model \cite{Gefen.PRB.1981, Alcaraz.PRB.1985, Balbao.JPA.1987, Gehlen.JMPB.1994} and the three-state quantum clock model (CM) \cite{Horn.PRD.1979, Ostlund.PRB.1981, Huse.PRB.1981}\, \footnote{Recently, $\mathcal{PT}$-symmetric  XXZ$_q$ spin chains with quantum-group symmetry were studied in \cite{Gorbenko:2025wzs}.}.  Using the state-operator correspondence and the finite-size scaling analysis, we numerically demonstrate that each of these models has a YL critical point under a particular deformation that preserves $\mathcal{PT}$ symmetry.  Using bosonization for the Schwinger model \cite{Coleman:1974bu, Coleman:1975pw, Coleman:1976uz} and the Polyakov-Hubbard transformation  \cite{Polyakov1969Micr,  Hubbard1972} for the other models, we show that, in all of these models, the YL criticality is described by the massless Euclidean  $i\phi^3$ field theory, whereas the presence of  $\mathcal{PT}$ symmetry alone is not sufficient to guarantee a YL phase transition. Moreover, the Landau-Ginzburg-Wilson (LGW) description for the three-state quantum clock model explains why the YL phase transition appears only for the positive sign of the $\mathcal{PT}$-symmetric deformation.

The paper is organized as follows. In Section \ref{PHtrans}, we review the derivation of the Euclidean $\phi^4$ field theory from the classical Ising model  using the Polyakov-Hubbard transformation.  We obtain the $i\phi^3$ theory by adding an imaginary magnetic field and determine the critical value of the magnetic field  as a function of temperature using the mean-field approach.  Next, in Section \ref{FMISandYL}, we focus on the quantum Ising model in a transverse magnetic field and 
review the Hamiltonian description of Yang-Lee criticality. In Section \ref{2ptYLCFT}, we numerically compute the Yang-Lee CFT structure constant $C_{\phi \phi \phi}$ and two-point correlation functions of $\phi$.  We show that the two-point correlation function of $\phi$ in the identity state grows with distance, in accordance with the CFT prediction, since the operator $\phi$ has negative scaling dimension. In Section \ref{AFMIS}, we explain and verify the Ising criticality in the non-Hermitian antiferromagnetic quantum Ising model in an imaginary longitudinal magnetic field. In Section \ref{AFMYL}, we investigate  YL criticality in the antiferromagnetic quantum Ising model perturbed by an imaginary staggered magnetic field. Finally, in Sections \ref{YLSchwinger}, \ref{secYLinBC} and \ref{YLCMcrit}, we establish YL criticality in the Schwinger, spin-1 Blume-Capel and three-state quantum clock models under particular deformations. In the final section, \ref{IsCMcrit}, we explore Ising criticality in the deformed three-state quantum clock model.

\section{Polyakov-Hubbard transformation for the classical Ising model}
\label{PHtrans}
We start by reviewing the Polyakov-Hubbard (PH) transformation \cite{Polyakov1969Micr,  Hubbard1972} for the classical Ising model on a lattice. Consider a classical Ising model on an arbitrary Bravais lattice $L$
with Hamiltonian 
\begin{align}
\mathcal{H} = -\frac{1}{2}\sum_{\textbf{r},\textbf{r}'\in L }V_{\textbf{r}\textbf{r}'} \sigma_{\textbf{r}}\sigma_{\textbf{r}'}, \quad \sigma_{\textbf{r}} = \pm 1\,,
\end{align}
where $V_{\textbf{r}\textbf{r}'}$ is an arbitrary translationally invariant symmetric exchange interaction,  so $V_{\textbf{r}\textbf{r}'} = V_{\textbf{r}'\textbf{r}}$  and $V_{\textbf{r}\textbf{r}'}= V_{\textbf{r}-\textbf{r}'}$. The partition function is obtained by summing the weight $e^{-\beta \mathcal{H}}$ over all spin configurations:
\begin{align}
Z = \sum_{\{ \sigma_{\textbf{r}} = \pm 1\}} \exp\Big(\frac{1}{2}\sum_{\textbf{r},\textbf{r}' \in L}\beta V_{\textbf{r}\textbf{r}'} \sigma_{\textbf{r}}\sigma_{\textbf{r}'}\Big)\,,
\end{align}
where $\beta =1/T$ is the inverse temperature.
We can rewrite this partition function by introducing a real variable $\phi_{\textbf{r}}$ on each lattice site $\textbf{r}$:
\begin{align}
Z = \sum_{\{ \sigma_{\textbf{r}} = \pm 1\}} \mathcal{N} \int_{-\infty}^{+\infty} \prod_{\textbf{r}}d\phi_{\textbf{r}} \exp\Big(-\frac{1}{2\beta} \sum_{\textbf{r},\textbf{r}' \in L}   \phi_{\textbf{r}}(V^{-1})_{\textbf{r}\textbf{r}'}  \phi_{\textbf{r}'} + \sum_{\textbf{r}\in L} \sigma_{\textbf{r}}  \phi_{\textbf{r}}\Big) \,,
\end{align}
where $\mathcal{N}$ is a normalization constant. 
We can now sum over the Ising spin configurations and obtain the partition function 
\begin{align}
Z =  \mathcal{N} \int_{-\infty}^{+\infty} \prod_{\textbf{r}}d\phi_{\textbf{r}} \exp\Big(-\frac{1}{2\beta} \sum_{\textbf{r},\textbf{r}'}   \phi_{\textbf{r}}(V^{-1})_{\textbf{r}\textbf{r}'}  \phi_{\textbf{r}'} + \sum_{\textbf{r}} \ln (2\cosh(\phi_{\textbf{r}}))\Big) \,.
\end{align}
We define the discrete Fourier-transformed fields as
\begin{align}
\phi_{\textbf{k}} = \frac{1}{\sqrt{N}}\sum_{\textbf{r}\in L} \phi_{\textbf{r}}e^{-i\textbf{k}\textbf{r}}\,, \quad 
\phi_{\textbf{r}} = \frac{1}{\sqrt{N}}\sum_{\textbf{k} \in \textrm{BZ}} \phi_{\textbf{k}}e^{i\textbf{k}\textbf{r}}\,,
\end{align}
where $N$ is the total number of sites in the lattice $L$, and the momentum vector $\textbf{k}$ belongs to the Brillouin zone (BZ). Using these definitions, we obtain
\begin{align}
 \sum_{\textbf{r},\textbf{r}' \in L}   \phi_{\textbf{r}}(V^{-1})_{\textbf{r}\textbf{r}'}  \phi_{\textbf{r}'} &= \sum_{\textbf{k} \in \textrm{BZ}} V^{-1}(\textbf{k}) \phi_{\textbf{k}}\phi_{-\textbf{k}}\,,
\end{align}
where $V^{-1}(\textbf{k}) = 1/V(\textbf{k})$ and $V(\textbf{k}) \equiv  \sum_{\textbf{r}}V_{\textbf{r}} e^{i\textbf{k}\textbf{r}}$. Finally, the partition function reads 
\begin{align}
Z =  \mathcal{N} \int_{-\infty}^{+\infty} \prod_{\textbf{r}}d\phi_{\textbf{r}} \exp(-S[\phi_{\textbf{r}}]) \,,
\end{align}
where the Euclidean Landau-Ginzburg-Wilson (LGW) action $S[\phi_{\textbf{r}}]$ is
\begin{align}
 S[\phi_{\textbf{r}}] = \frac{1}{2}\sum\limits_{\textbf{k} \in \textrm{BZ}} \varepsilon(\textbf{k}) |\phi_{\textbf{k}}|^2 +\sum_{\textbf{r}\in L}\left(\frac{\phi_{\textbf{r}}^2}{2\beta V(0)} -\ln (2\cosh(\phi_{\textbf{r}}))\right)\,, \label{Sphi4micro}
\end{align}
with  $\varepsilon(\textbf{k})\equiv  (V^{-1}(\textbf{k}) - V^{-1}(0))/\beta$ being the kinetic energy. 
We assume that the expansion of $\varepsilon(\textbf{k})$ at small momenta starts with the term $\textbf{k}^2$. The higher order terms in $\textbf{k}$ break orthogonal invariance down to the discrete point-group symmetry of the lattice, but we can neglect them because these terms are irrelevant in IR.
We can also expand the term $\ln (2\cosh(\phi_{\textbf{r}}))$  in a series. Then, using the Wilson RG procedure \cite{Wilson197475}, we integrate out the high-energy modes and neglect higher power terms of $\phi$, since they are less relevant. As a result, one can argue that this microscopic action  is described in the IR by the Euclidean $\phi^4$ theory with  action 
\begin{align}
 S[\phi(\textbf{r})] = \int d^{d}\textbf{r} \left(\frac{1}{2}(\partial_{\mu} \phi)^2 + \frac{1}{2}m^2 \phi^2 + \frac{\lambda}{4!}\phi^4\right)\,.
\end{align}
By changing the temperature, one adjusts the UV mass term in the microscopic action (\ref{Sphi4micro}) and can tune it so that the physical mass is zero and the system is at the Ising critical point.

Let us consider a nearest-neighbor potential on a simple square lattice:
\begin{align}
V_{\textbf{r},\textbf{r}'}= J(\delta_{\textbf{r},\textbf{r}'+a\hat{x}} +
\delta_{\textbf{r},\textbf{r}'-a\hat{x}}+ \delta_{\textbf{r},\textbf{r}'+a\hat{y}} +
\delta_{\textbf{r},\textbf{r}'-a\hat{y}})\,.
\end{align}
In this case, we find  $V(\textbf{k})=2J(\cos(k_{x}a) + \cos(k_{y}a))$, where $k_{x},k_{y} \in [-\pi/a, \pi/a]$. Such a potential is not positive definite, since it has points $\textbf{k}$ where it is negative, and therefore the Gaussian integrals over  $\phi_{\textbf{k}}$ for such $\textbf{k}$ are not well-defined.  One can easily fix this issue by adding to $V_{\textbf{r},\textbf{r}'}$ an identity term, namely  $V_{\textbf{r},\textbf{r}'} \to V_{\textbf{r},\textbf{r}'} + 2cJ \delta_{\textbf{r},\textbf{r}'}$, where $c$ is a constant. This term is just a constant energy shift, $cJ \sum_{\textbf{r},\textbf{r}'}\delta_{\textbf{r},\textbf{r}'}\sigma_{\textbf{r}}\sigma_{\textbf{r}'} = cJ \sum_{\textbf{r}} (\sigma_{\textbf{r}})^2 = cJN$ and does not affect the physics. With this modification, we find  $V(\textbf{k})=2J(c+\cos(k_{x}a) + \cos(k_{y}a))$, which is positive for $c > 2$,  and thus the Gaussian integrals over all the modes $\phi_{\textbf{k}}$ are well defined.

Extremizing the microscopic action in (\ref{Sphi4micro}), $\delta S[\phi]/\delta \phi = 0$, and assuming a uniform field configuration $\phi_{\textbf{r}}=\phi$, we obtain the mean-field equation for the Ising magnetization,
\begin{align}
\frac{\phi }{\beta V(0)} = \tanh(\phi)\,.
\end{align}
This equation indicates a phase transition at $(\beta V(0))_{\textrm{crit}}=1$. 

Repeating the steps above for the classical Ising Hamiltonian in an imaginary magnetic field, 
\begin{align}
\mathcal{H} = -\frac{1}{2}\sum_{\textbf{r},\textbf{r}'\in L }V_{\textbf{r}\textbf{r}'} \sigma_{\textbf{r}}\sigma_{\textbf{r}'} - i h \sum_{\textbf{r} \in L} \sigma_{\textbf{r}}\,, \label{clIs_h}
\end{align}
we obtain the following microscopic LGW action \cite{Cardy:2023lha}
\begin{align}
 S[\phi_{\textbf{r}}] = \frac{1}{2}\sum\limits_{\textbf{k} \in \textrm{BZ}} \varepsilon(\textbf{k}) |\phi_{\textbf{k}}|^2 +\sum_{\textbf{r}\in L}\left(\frac{\phi_{\textbf{r}}^2}{2\beta V(0)} -\ln (2\cosh(\phi_{\textbf{r}} + i\beta h))\right)\,. \label{Sphi3micro}
\end{align}
The uniform imaginary magnetic field breaks the $\mathbbm{Z}_{2}$ symmetry $\phi_{\textbf{r}}\to -\phi_{\textbf{r}}$ of the action in (\ref{Sphi3micro}), but preserves the $\mathcal{PT}$ symmetry: $\phi_{\textbf{r}}\to -\phi_{\textbf{r}}$ and $i\to -i$. 

The microscopic action (\ref{Sphi3micro}) can be brought to the massless $i\phi^3$ theory by shifting the field $\phi$ by a constant complex value $i\phi_{*}$ and tuning the magnetic field to its critical value $h_{\textrm{crit}}$ \cite{Fisher:1978pf}. This is equivalent to imposing  two conditions on the action:
\begin{align}
\left.\frac{\delta S[\phi_{\textbf{r}}]}{\delta \phi_{\textbf{r}}} \right|_{\substack{\phi_{\textbf{r}} = i\phi_{*} \\ h=h_{\textrm{crit}} }}= 0, \quad
\left.\frac{\delta^2 S[\phi_{\textbf{r}}]}{\delta^{2} \phi_{\textbf{r}}} \right|_{\substack{\phi_{\textbf{r}} = i\phi_{*} \\ h=h_{\textrm{crit}} }}= 0\,. \label{YLmfeq}
\end{align}
These equations determine the mean-field $(T, h)$ phase diagram for the Yang-Lee phase transition:
\begin{align}
\frac{ih_{\textrm{crit}}}{J} = \pm \frac{i}{\beta J} \Big(\arccos(\sqrt{\beta V(0)})- \sqrt{\beta V(0)(1-\beta V(0))}\Big)\,.
\end{align}
Although this equation is not quantitatively accurate, it provides the correct qualitative behavior, and 
shows that, for any temperature $T$ above the Ising critical temperature $T_{\textrm{crit}}$, there exists a YL critical point at some value of $ih_{\textrm{crit}}$.

\section{Ferromagnetic quantum Ising model in imaginary magnetic field: a Yang-Lee recipe}
\label{FMISandYL}

In order to study the YL phase transition numerically, we start from the ferromagnetic quantum Ising model in a transverse magnetic field 
\begin{align}
    H_{\textrm{Ising}}= - J \sum_{n=0}^{N-1} Z_{n}Z_{n+1} -h_{x} \sum_{n=0}^{N-1} X_{n} \,,
    \label{eq_H_IM}
\end{align}
where $J>0$ is a ferromagnetic exchange coupling, $h_{x}$ is the transverse magnetic field, and  
\begin{align}
  \qquad X = \left(  \begin{array}{cc}
        0 & 1 \\
        1 & 0
    \end{array} \right), \quad  Z = \left(  \begin{array}{cc}
        1 & 0 \\
        0 & -1
    \end{array} \right),
\end{align}
are the Pauli matrices. In this and the other quantum models below, we always assume periodic boundary conditions (PBC), so that $Z_{N}\equiv Z_{0}$.  This Hamiltonian can be obtained from the classical Ising model in the  highly anisotropic limit \cite{Kogut:RMP51.659}.  In this case, the transfer matrix of the classical  Ising model can be written as $T = e^{-a H_{\textrm{Ising}}}$, where $H_{\textrm{Ising}}$ is the Hamiltonian in (\ref{eq_H_IM}), $a J = \beta J_{x}$, and $\tanh(a h_{x})= e^{-2\beta J_{y}}$, with $a\to 0$.  In this limit, the classical Ising transition temperature is obtained from the equation $\sinh(2\beta J_{x})\sinh(2\beta J_{y})=1$, which corresponds to the quantum phase transition at $h_{x}/J=1$.
For $h_{x}/J >1$, the quantum Ising model is in the paramagnetic phase. The quantum Hamiltonian (\ref{eq_H_IM}) has a $\mathbb{Z}_{2}$ spin-flip symmetry generated by the operator
\begin{align}
    \mathcal{P} = \prod_{n=0}^{N-1} X_{n}\,.
\end{align}

\begin{figure}[h]
    \centering
    \includegraphics[width=7cm]{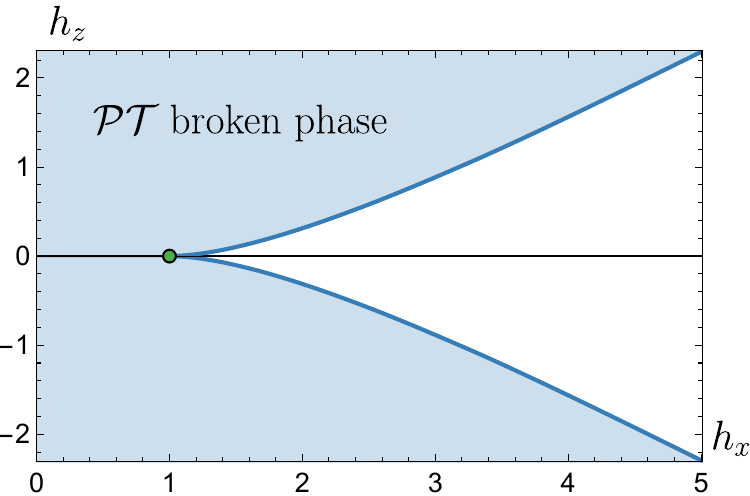}
    \caption{Phase diagram of the Yang-Lee quantum model in \eqref{eq_H_YL}. The green dot marks the Ising critical point at $h_{x,c}=1$. The blue lines correspond to the YL critical points, and the shaded region is the $\mathcal{PT}$ broken phase, for different values of $h_{x}$ and $h_{z}$, with $J=1$.}
    \label{fig:YL_PD}
\end{figure}

The imaginary magnetic filed $ih$ in the classical Ising Hamiltonian (\ref{clIs_h}) corresponds to an imaginary longitudinal magnetic field $i h_{z}$ in the quantum Ising Hamiltonian:  
\begin{align}
    H_{\textrm{YL}} = H_{\textrm{Ising}} - i h_{z}\sum_{n=0}^{N-1} Z_{n} \,,
    \label{eq_H_YL}
\end{align}
where $ah_{z}=\beta h $. The imaginary term imposes the following changes of the Hamiltonian
\begin{enumerate}
    \item[i)] Breaks the $\mathbb{Z}_{2}$ symmetry,
    \item[ii)] breaks Hermicity, and
    \item[iii)] preserves $\mathcal{PT}$ symmetry, where $\mathcal{T}$ corresponds to complex conjugation: $i \to -i$.
    \label{YL_recipe}
\end{enumerate}
It is well known that this quantum Hamiltonian  has a YL phase transition  at  a critical value $h_{z,c}/J$, which depends on $h_{x}/J$, for $h_{x}/J >1$  \cite{Uzelac.JPA.1981, Itzykson.EPL.1986, Gehlen.JMPB.1994, Gehlen.JPA.1991}.
The phase diagram of \eqref{eq_H_YL} obtained numerically, is shown in Figure \ref{fig:YL_PD}.

The energy spectrum of the quantum Hamiltonian (\ref{eq_H_YL}) provides direct access to 
the operators scaling dimensions and the effective central charge of the YL CFT, via the operator-state correspondence \cite{CardyJPA17.1984, BlotePRL56.1986, CardyNPB.1986}. These CFT data, computed using the Hamiltonian (\ref{eq_H_YL}), are in agreement with the exact predictions of the minimal model $M(2,5)$ \cite{Uzelac.JPA.1981, Itzykson.EPL.1986, Gehlen.JPA.1991, Gehlen.JMPB.1994, ArguelloCruz:2025zuq, Fan:2025bhc}. At the YL critical point, the ground-state energy $E_{0}$ corresponds to the state $|\phi\rangle$, and the first excited energy $E_{1}$ corresponds to the identity state $|I\rangle$, since the operator $\phi$ has negative scaling dimension $\Delta_{\phi}=-2/5$.

\section{Numerical Computation of two-point functions and the structure constant in the YL CFT}
\label{2ptYLCFT}

In this section, we present the numerical calculation of the Yang-Lee CFT structure constant $C_{\phi\phi\phi}$ and the two-point functions of the operator $\phi$, using the Hamiltonian \eqref{eq_H_YL}, and show that they agree with the theoretical predictions.
In particular, we show that the correlation function $\langle I | \phi_{0} \phi_{x}|I\rangle$ on a cylinder grows with the distance $x$, since the primary operator $\phi$ has negative scaling dimension $\Delta_{\phi}=-2/5$. We also note that this correlation function is computed in the identity state $|I\rangle$, which is the first excited state above the ground state  $|\phi\rangle$.

In the IR, the lattice operators $X$ and $Z$ are represented as superpositions of operators in the $M(2,5)$ minimal model \cite{ArguelloCruz:2025zuq, Fan:2025bhc}
\begin{equation}
\begin{aligned}
X &=  a_{1}I+ib_{1}\phi +\dots, \qquad Z =  ia_{2}I +b_{2}\phi +\dots, \label{eq_lat_CFT} 
\end{aligned}   
\end{equation}
where $a_{i},b_{i}$ are real constants, and $\dots$ denote operators  with higher scaling dimensions.
We note that $X$ and $Z$ are even  and odd under  $\mathcal{PT}$ symmetry, respectively. 

\begin{figure}[htbp]
    \centering
    \includegraphics[width=10cm]{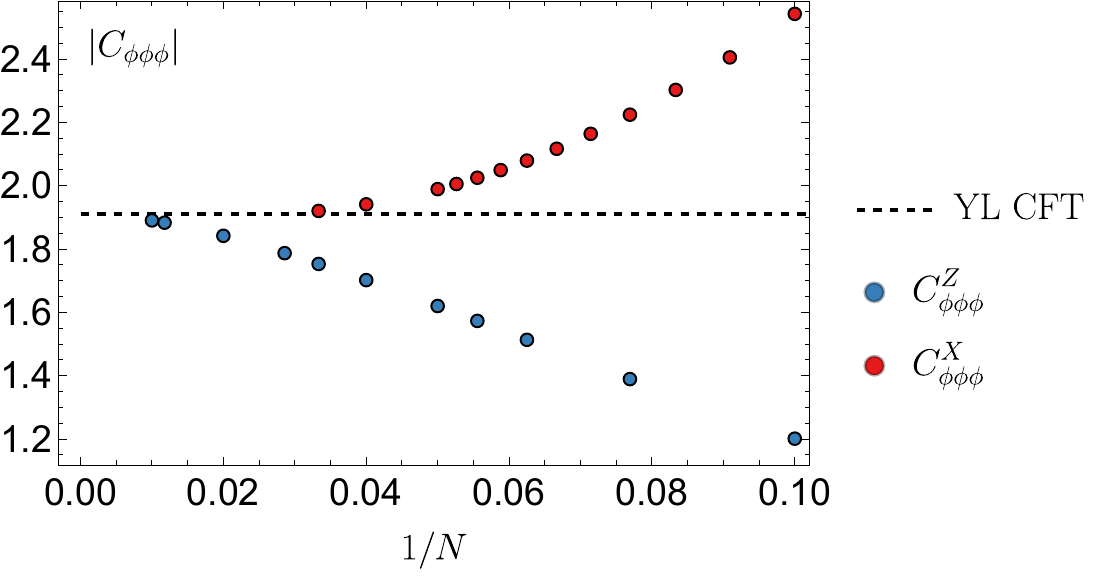}
    \caption{Structure constant as a function of system size obtained from \eqref{eq_CPP}, evaluated at $J=1$, $h_{x}=1.25$ and $h_{z}=h_{z,c}$. The dashed line represents the theoretical value \eqref{CpppTheory}.}
    \label{fig:Cppp}
\end{figure}

First, we use the relations (\ref{eq_lat_CFT}) to numerically extract the YL CFT structure constant $C_{\phi \phi \phi}$ \cite{Wydro_2009, Fan:2025bhc}, which is imaginary and known exactly \cite{Dotsenko:1985hi, Cardy:1985yy} 
\begin{align}
    C_{\phi\phi\phi} =i \left(\frac{\Gamma\left(\frac{6}{5}\right)^2\Gamma\left(\frac{1}{5}\right)\Gamma\left(\frac{2}{5}\right)}{\Gamma\left(\frac{3}{5}\right)\Gamma\left(\frac{4}{5}\right)^3}\right)^{1/2} \approx 1.9113127\ i \,. \label{CpppTheory}
\end{align}
To this end, we use the general relation between  CFT structure constants and matrix elements of an operator $O_{\alpha}(\tau, x)$ on a cylinder of radius  $R=N/(2\pi)$   \cite{CardyNPB.1986}
\begin{align}
C_{\alpha\beta \gamma} = R^{\Delta_{\alpha}} \langle O_{\beta}|O_{\alpha}(0, 0)|O_{\gamma}\rangle\,, \label{StrMatElem}
\end{align}
where the cylinder bra and ket states are defined as 
\begin{align}
\langle O_{\Delta}| = \lim_{\tau \to \infty} R^{\Delta} e^{\frac{2\pi}{N}\tau \Delta} \langle 0 | O_{\Delta}(\tau,x)\,, \quad 
|O_{\Delta}\rangle = \lim_{\tau \to -\infty} R^{\Delta} e^{-\frac{2\pi}{N}\tau \Delta} O_{\Delta}(\tau,x)|0\rangle\,,
\end{align}
and the map from plane to cylinder is $z=e^{\frac{2\pi}{N} (\tau + ix)}$. Using (\ref{eq_lat_CFT}) and (\ref{StrMatElem}) we find 
\begin{equation}
\begin{aligned}
&\langle \phi | X |\phi \rangle = a_{1} + ib_{1}R^{-\Delta_{\phi}}C_{\phi \phi \phi} + \dots, \;\; \langle I | X |\phi \rangle = ib_{1}R^{-\Delta_{\phi}} + \dots, \;\; 
\langle I | X |I \rangle = a_{1}\,,
\end{aligned}   
\end{equation}
where $\dots$ denote higher-order terms in $1/R$ and $\langle I| $ and $\langle \phi |$ are the left eigenstates of the YL Hamiltonian (\ref{eq_H_YL})\footnote{The YL Hamiltonian (\ref{eq_H_YL}) is not Hermitian but symmetric, therefore, the left eigenstates can be obtained from the right eigenstates by transposition. }.  
Thus, the structure constant $C_{\phi\phi\phi}$ can be computed as \cite{ArguelloCruz:2025zuq, Fan:2025bhc} 
\begin{align}
\label{eq_CPP}
    C_{\phi\phi\phi}=\lim_{N\to \infty } \frac{\langle \phi | X | \phi \rangle - \langle I | X | I \rangle}{
    \langle I | X | \phi\rangle
    }  = \lim_{N\to \infty } \frac{\langle \phi | Z | \phi \rangle - \langle I | Z | I \rangle}{
    \langle I | Z | \phi\rangle
    } \,.
\end{align}
Our numerical results are $C_{\phi\phi\phi}^{X}= 1.9092985\ i$ and $C_{\phi\phi\phi}^{Z}=1.9148405\ i$, which agree with the theoretical value in (\ref{CpppTheory})  within an error of order $10^{-3}$. The results at fixed system size, up to $N=100$, were obtained using the Density Matrix Renormalization Group (DMRG) method via the publicly available iTensor library \cite{itensor, itensor-r0.3} and are shown in Figure \ref{fig:Cppp}. The thermodynamic limits were obtained using the Bulirsch–Stoer (BST) algorithm \cite{bulirsch1966numerical, henkel1988finite} with  $w=2$\;\footnote{In this work, all thermodynamic values were obtained using the BST algorithm with $w=2$, unless explicitly stated otherwise.}.  
It is worth noting that the BST algorithm has narrow poles that depend on $w$ and the maximum system size $N_{\textrm{max}}$ \cite{henkel1988finite}; however, we find that $w=2$ gives the most stable results for almost all of our extrapolations.

\begin{figure}[htbp]
    \centering
    \includegraphics[width=15cm]{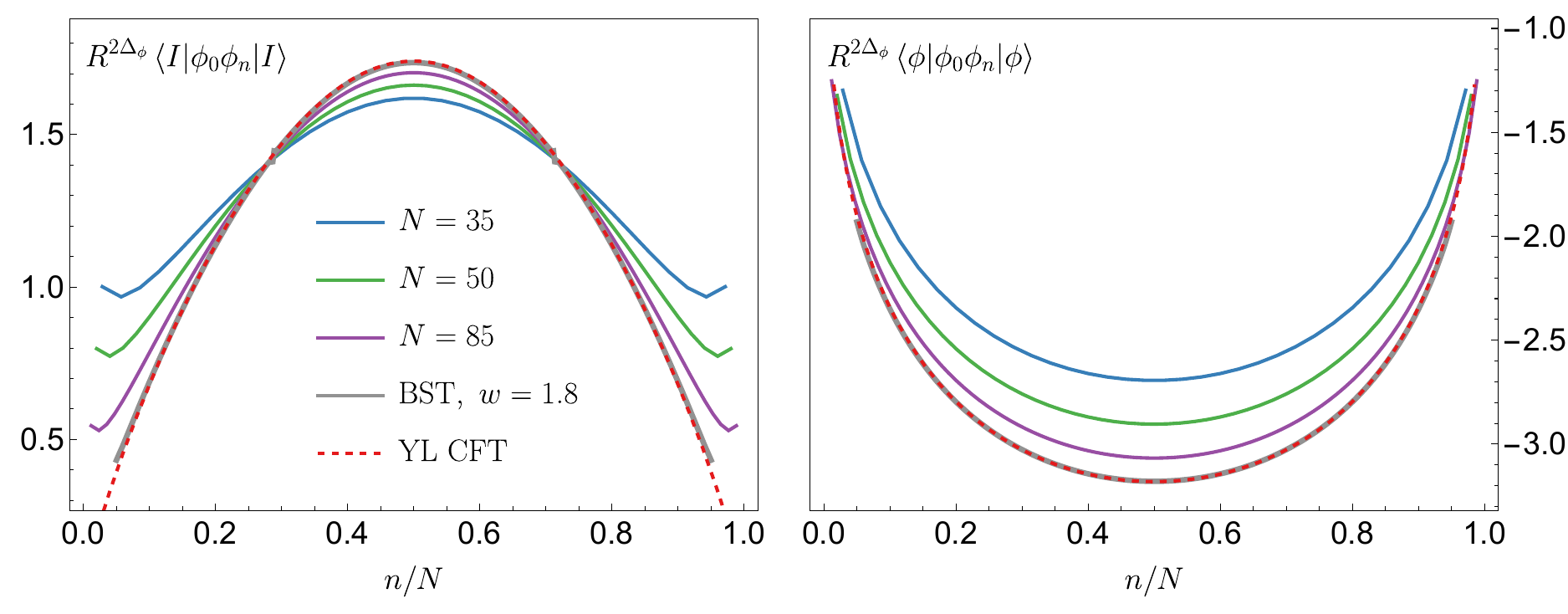}
    \caption{Two-point functions of  $\phi$ in the identity state $|I\rangle$ and the ground state $|\phi\rangle$ as functions of $n/N$ for $J=1$, $h_{x}=1.25$, and $h_{z}=h_{z,c}$. In gray, we show the thermodynamic extrapolation obtained using the BST algorithm with $w=1.8$ and with dashed red lines, we plot the CFT expressions \eqref{eq_XX_ES} and \eqref{eq_XX_GS}, respectively.}
    \label{fig:corrXX}
\end{figure}

Similarly, we calculate the two-point functions of  $\phi$ in the identity state $|I\rangle$ and in the state $|\phi\rangle$. The CFT expressions for these functions on a cylinder are \cite{Cardy:1985yy}
\begin{align}
\bra{I}\phi_{0} \phi_{n} \ket{I} &= R^{-2\Delta_{\phi}} \left|2\sin{\left( \frac{\pi}{N}n\right)} \right|^{-2\Delta_{\phi}} \,, \label{eq_XX_ES} \\
\bra{\phi}\phi_{0} \phi_{n} \ket{\phi} &= R^{-2\Delta_{\phi}} \left|2\sin{\left( \frac{\pi}{N}n\right)} \right|^{-2\Delta_{\phi}} \left[\left| F_{1}(e^{\frac{2\pi n}{N} i}) \right|^2 
 - |C_{\phi\phi\phi}|^2 \left| F_{2}(e^{\frac{2\pi n}{N} i}) \right|^2  \right]\,,
 \label{eq_XX_GS}
\end{align}
where $F_{1}(\chi)=\ _2F_{1}\left( \frac{3}{5},\frac{4}{5},\frac{6}{5};\chi \right)$ and $F_{2}(\chi)=\chi^{\Delta_{\phi}/2}\ _2F_{1}\left( \frac{3}{5},\frac{2}{5},\frac{4}{5};\chi \right)$.
Using the expression for the operator $Z$ in \eqref{eq_lat_CFT}, we can compute  these two-point functions numerically as: 
\begin{align}
R^{2\Delta_{\phi}} \bra{I}\phi_{0} \phi_{n} \ket{I} &  = \lim_{N \to \infty}\frac{\bra{I} Z_{0}Z_{n}\ket{I} -\langle I | Z|I\rangle^2}{\langle I | Z|\phi\rangle^2}\,,  \\
R^{2\Delta_{\phi}} \bra{\phi}\phi_{0} \phi_{n} \ket{\phi} 
&= \lim_{N \to \infty}\frac{\bra{\phi} Z_{0}Z_{n}\ket{\phi} -2\langle I | Z|I\rangle  \bra{\phi} Z \ket{\phi} + \langle I | Z|I\rangle^2 }{\langle I | Z|\phi\rangle^2}\,.
\end{align}
The results for different values of $N$ are shown in Figure \ref{fig:corrXX}. 
In gray, we show the extrapolation to the thermodynamic limit $N = \infty$, obtained using the BST algorithm with $w=1.8$. We see that the thermodynamic curves overlap with the theoretical expressions \eqref{eq_XX_ES} and \eqref{eq_XX_GS}, showing clear agreement with the CFT expectations.

\section{Ising criticality in antiferromagnetic non-Hermitian  Ising model}
\label{AFMIS}

\begin{figure}[h]
    \centering
    \includegraphics[width=8cm]{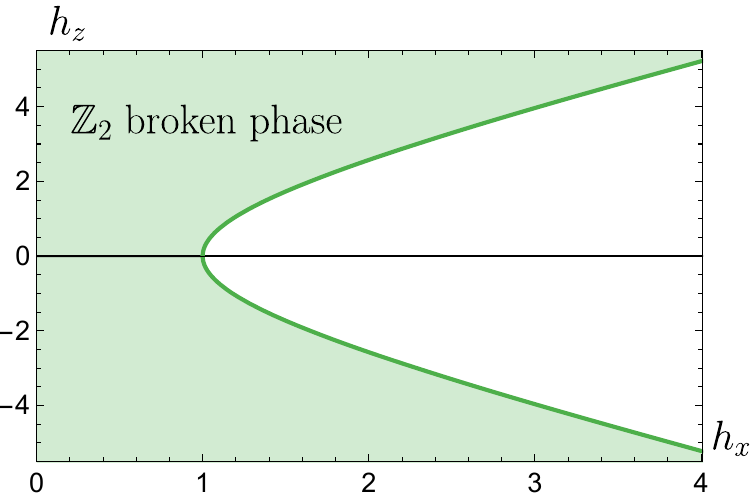}
    \caption{Phase diagram of the Hamiltonian (\ref{eq_AFHIS2}) for $J=1$. The green line represents the Ising critical points, and the shaded region represents the $\mathbbm{Z}_{2}$ broken phase.}
    \label{fig:PD_AFYL}
\end{figure}

The quantum Ising model in (\ref{eq_H_YL}) has a YL phase transition only when the exchange coupling is ferromagnetic, i.e. $J>0$. By contrast, the same model with antiferromagnetic coupling, $J<0$, does not exhibit YL criticality \cite{Starkov:2022pax}.   
By flipping the spins on every odd site, we map the antiferromagnetic Hamiltonian \eqref{eq_H_YL} to the ferromagnetic one:
\begin{align}
    H= -J \sum_{n=0}^{N-1} Z_{n}Z_{n+1} -h_{x} \sum_{n=0}^{N-1} X_{n} -ih_{z} \sum_{n=0}^{N-1}(-1)^{n} Z_{n}\,,
    \label{eq_AFHIS2}
\end{align}
where now $J>0$ and the imaginary magnetic field term is staggered. 
This Hamiltonian has the following global symmetries:
\begin{equation}
\begin{aligned}
    \mathbb{Z}_{2} &: \qquad Z_{n} \to -Z_{n+1}, \qquad X_{n} \to X_{n+1} \\
    \mathcal{PT}&: \qquad Z_{n} \to -Z_{n}, \qquad \quad i \to -i.
\end{aligned}
\end{equation}
It turns out that the low-lying energy levels of \eqref{eq_AFHIS2} are real for a range of values of $h_{x}$ and $h_{z}$ and this model does not have a YL phase transition \cite{Starkov:2022pax}. In fact, as we explain below using the Landau-Ginzburg-Wilson action, the low-energy behavior of this model belongs to the Ising universality class.  We find numerically that there is a line of  Ising phase transitions, and we present the phase diagram of \eqref{eq_AFHIS2} in Figure \ref{fig:PD_AFYL}. 

\begin{figure}[h!]
  \centering
\includegraphics[width=0.55\textwidth]{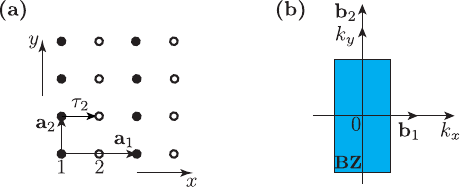}
\caption{\textbf{(a)} Layered square lattice with the primitive lattice vectors $\textbf{a}_{1} = 2a \hat{x}$ and $\textbf{a}_{2} = a\hat{y}$ and two sites per unit cell, with  basis position vectors $\tau_{1}=0$ and $\tau_{2}=a\hat{x}$. Sublattice $1$ is depicted by black circles and sublattice $2$ is depicted by white circles. \textbf{(b)} The lattice Brillouin zone (BZ). The reciprocal lattice vectors are $\textbf{b}_{1} = \frac{\pi}{a}\hat{x}$, $\textbf{b}_{2} = \frac{2\pi}{a}\hat{y}$.}
\label{bp_lat_fig2}
\end{figure}

In order to explain this behavior, we consider the classical Ising model on a square lattice in a layer-alternating imaginary magnetic field, with Hamiltonian 
\begin{align}
\mathcal{H} =  -\frac{1}{2}\sum_{\alpha,\beta}\sum_{\textbf{r}, \textbf{r}'  \in L}V_{\alpha\textbf{r}, \beta\textbf{r}'} \sigma_{\alpha \textbf{r}}\sigma_{\beta\textbf{r}'} - i h\sum_{\alpha, \textbf{r} \in L}  (-1)^{\alpha}\sigma_{\alpha \textbf{r}}, \quad \sigma_{\alpha \textbf{r}} = \pm 1\,, \label{clAFHIS}
\end{align} 
where $\alpha,\beta =1,2$ go over the sublattice sites, and the ferromagnetic couplings are
\begin{equation}
\begin{aligned}
&V_{11}(\textbf{r}-\textbf{r}') = V_{22}(\textbf{r}-\textbf{r}') =   J(\delta_{\textbf{r}-\textbf{r}', \textbf{a}_{2}}  + \delta_{\textbf{r}-\textbf{r}', -\textbf{a}_{2}}) \,, \\
&V_{12}(\textbf{r}-\textbf{r}') = J(\delta_{\textbf{r}-\textbf{r}',0} +\delta_{\textbf{r}-\textbf{r}',\textbf{a}_{1}})\,, \label{V11V12}
\end{aligned}
\end{equation}
where $\textbf{a}_{1}=2a\hat{x}$ and $\textbf{a}_{2}=a\hat{y}$ are primitive lattice vectors, and the lattice is depicted in Figure \ref{bp_lat_fig2} (a).
The quantum Hamiltonian (\ref{eq_AFHIS2}) can be obtained from the classical Hamiltonian (\ref{clAFHIS}) in the highly anisotropic limit, discussed in Section \ref{FMISandYL}.
Performing the Polyakov-Hubbard transformation discussed in Section \ref{PHtrans}, we obtain for the partition function 
\begin{equation}
\begin{aligned}
Z =&  \mathcal{N} \int_{-\infty}^{+\infty} \prod_{\alpha, \textbf{r} }d\phi_{\alpha\textbf{r}} \exp\Big(-\frac{1}{2\beta} \sum_{\alpha\textbf{r},\beta \textbf{r}'}   \phi_{\alpha\textbf{r}}(V^{-1})_{\alpha\textbf{r},\beta \textbf{r}'}  \phi_{\beta\textbf{r}'}  \\
&\qquad + \sum_{\alpha,\textbf{r} } \ln (2\cosh(\phi_{\alpha\textbf{r}}+i\beta h (-1)^{\alpha}))\Big) \,.
\end{aligned}
\end{equation}
We define the discrete Fourier-transformed fields as
\begin{align}
\phi_{\alpha \textbf{k}} = \frac{1}{\sqrt{N}}\sum_{\textbf{r}\in L} \phi_{\alpha \textbf{r}}e^{-i\textbf{k}(\textbf{r}+\tau_{\alpha})}\,, \quad 
\phi_{\alpha \textbf{r}} = \frac{1}{\sqrt{N}}\sum_{\textbf{k} \in \textrm{BZ}} \phi_{\alpha \textbf{k}}e^{i\textbf{k}(\textbf{r}+\tau_{\alpha})}\,,
\end{align} 
where $N$ is the total number of  unit cells in the lattice, and $\tau_{\alpha}$ are the basis position vectors. The Bloch vectors belong to the Brillouin zone, depicted in Figure \ref{bp_lat_fig2} (b). 
We then find for the partition function 
\begin{equation}
\begin{aligned}
Z =&  \mathcal{N} \int_{-\infty}^{+\infty} \prod_{\alpha, \textbf{r} }d\phi_{\alpha\textbf{r}} \exp\Big(-\frac{1}{2\beta} \sum_{\alpha,\beta,\textbf{k}}  V^{-1}_{\alpha\beta}(\textbf{k})  \phi_{\alpha\textbf{k}} \phi^{*}_{\beta\textbf{k}}  \\
&\qquad + \sum_{\alpha,\textbf{r} } \ln (2\cosh(\phi_{\alpha\textbf{r}}+i\beta h (-1)^{\alpha}))\Big) \,,
\end{aligned}
\end{equation}
where $V^{-1}_{\alpha \beta}(\textbf{k})  \equiv \sum_{\textbf{r}\in L} V^{-1}_{\alpha\beta}(\textbf{r}) e^{i\textbf{k}(\textbf{r}+\tau_{\alpha}-\tau_{\beta})}$, and 
$V_{\alpha \beta}(\textbf{k})  \equiv \sum_{\textbf{r}\in L} V_{\alpha\beta}(\textbf{r}) e^{i\textbf{k}(\textbf{r}+\tau_{\alpha}-\tau_{\beta})}$ and  $V_{\alpha \beta}(\textbf{k})V^{-1}_{\beta \gamma}(\textbf{k})=\delta_{\alpha\gamma}$.  Using (\ref{V11V12}) we obtain 
\begin{align}
&V(\textbf{k}) = \left(  \begin{array}{cc}
    2J\cos(\textbf{k}\textbf{a}_{2}) & 2J \cos(\textbf{k}\textbf{a}_{1}/2) \\ 
    2J\cos(\textbf{k}\textbf{a}_{1}/2) & 2J\cos(\textbf{k}\textbf{a}_{2}) \\ 
  \end{array}\right)\,.
\end{align}
As in the case of the classical Ising model on a square lattice, the matrix $V(\textbf{k})$ is not positive definite, but this can be easily fixed by adding the identity matrix $2c J \mathbbm{1}$ to $V(\textbf{k})$, which is a mere shift in energy. This matrix has two eigenvalues 
\begin{align}
v_{1}(\textbf{k}) = 2J(c+\cos(k_{x}a) + \cos(k_{y}a)), \quad 
v_{2}(\textbf{k}) = 2J(c-\cos(k_{x}a) + \cos(k_{y}a)) \label{v1v2eq}
\end{align}
with the eigenvectors $e_{1}=(1,1)/\sqrt{2}$ and $e_{2}=(1,-1)/\sqrt{2}$. Introducing the new variables  $\chi_{s\textbf{r}}=(\phi_{1\textbf{r}}+\phi_{2\textbf{r}})/\sqrt{2}$ and 
$\chi_{f\textbf{r}}=(\phi_{1\textbf{r}}-\phi_{2\textbf{r}})/\sqrt{2}$, we can write the partition function in the form
\begin{align}
Z =&  \mathcal{N} \int_{-\infty}^{+\infty} \prod_{\textbf{r} } d\chi_{s \textbf{r}}d\chi_{f \textbf{r}} \exp(-S[\chi_{s \textbf{r}},\chi_{f \textbf{r}}])\,,
\end{align} 
where the Landau-Ginzburg-Wilson action is 
\begin{equation}
\begin{aligned}
S[\chi_{s \textbf{r}},\chi_{f \textbf{r}}] = &
\frac{1}{2}\sum\limits_{\textbf{k}\in \textrm{BZ}} (\varepsilon_{1}(\textbf{k})|\chi_{s\textbf{k}}|^2 + 
\varepsilon_{2}(\textbf{k})|\chi_{f\textbf{k}}|^2) +\sum\limits_{\textbf{r}}\bigg( \frac{\chi_{s\textbf{r}}^2}{2\beta v_{1}(0)} + \frac{\chi_{f\textbf{r}}^2}{2\beta v_{2}(0)} \\
& -\ln\Big(4\cosh\big(\frac{\chi_{s\textbf{r}}+\chi_{f\textbf{r}}}{\sqrt{2}}-i\beta h\big)\cosh\big(\frac{\chi_{s\textbf{r}}-\chi_{f\textbf{r}}}{\sqrt{2}}+i\beta h\big)\Big) \bigg)\,, \label{LGWafi}
\end{aligned}
\end{equation}
and $\varepsilon_{\alpha}(\textbf{k})\equiv  (v_{\alpha}^{-1}(\textbf{k}) - v_{\alpha}^{-1}(0))/\beta$ with $\alpha = 1,2$. The field $\chi_{s\textbf{r}}$, which we call slow, has a usual kinetic energy $\varepsilon_{1}(\textbf{k})$, which is proportional to $\textbf{k}^2$ at small momenta.  The action has a $\mathbbm{Z}_{2}$ symmetry, $\chi_{s\textbf{r}} \to - \chi_{s\textbf{r}}$.  In contrast, the field $\chi_{f\textbf{r}}$, which we call fast, has a larger bare mass term and an unusual kinetic energy $\varepsilon_{2}(\textbf{k})$, which is linear at small momenta. The action has a $\mathcal{PT}$ symmetry, $\chi_{f\textbf{r}} \to - \chi_{f\textbf{r}}$ and $i\to -i$.

By integrating out the fast (high-energy) field $\chi_{f\textbf{r}}$, we obtain the effective action for the  slow (low-energy) field $\chi_{s\textbf{r}}$:
\begin{align}
e^{-S_{\textrm{eff}}[\chi_{s\textbf{r}}]} =  \int_{-\infty}^{+\infty} \prod_{\textbf{r}} d\chi_{f\textbf{r}} \, e^{-S[\chi_{s\textbf{r}}, \chi_{f\textbf{r}}]}\,.
\end{align} 
This effective action is $\mathbbm{Z}_{2}$ symmetric $S_{\textrm{eff}}[-\chi_{s\textbf{r}}]=S_{\textrm{eff}}[\chi_{s\textbf{r}}]$, and real, $S^{*}_{\textrm{eff}}[\chi_{s\textbf{r}}] = S_{\textrm{eff}}[\chi_{s\textbf{r}}]$. Thus, we should expect that the model undergoes the Ising phase transition at some critical temperature, as we indeed verify numerically in the corresponding quantum model (\ref{eq_AFHIS2}).

\section{Yang-Lee criticality in antiferromagnetic quantum Ising model}
\label{AFMYL}

\begin{figure}[h]
    \centering
    \includegraphics[width=8cm]{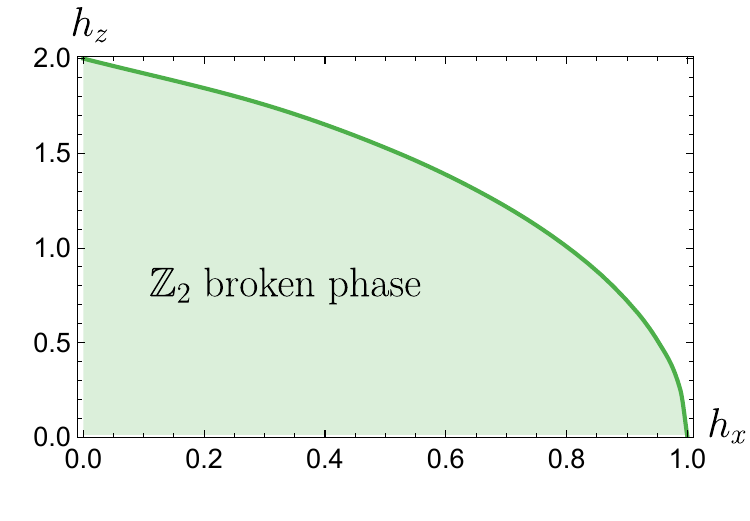}
    \caption{Phase diagram of the antiferromagnetic Ising model \eqref{eq_H_AFM} for different values of $h_{x}$ and $h_{z}$, with $J=1$. The green line corresponds to the Ising criticality, and the shaded region is the $\mathbb{Z}_{2}$ broken phase.}
    \label{fig:AFM_PD}
\end{figure}

In this section, we consider the Hamiltonian of the antiferromagnetic Ising model in real transverse and longitudinal  magnetic fields:
\begin{align}
    H_{\textrm{AFI}}= J \sum_{n=0}^{N-1} Z_{n}Z_{n+1} -h_{x} \sum_{n=0}^{N-1} X_{n} - h_{z} \sum_{n=0}^{N-1} Z_{n}\,,
    \label{eq_H_AFM}
\end{align}
where $J>0$, and the magnetic fields $h_{x}$ and $h_{z}$ are real.  It is known that this model has a line of critical points belonging to the Ising universality class \cite{NovotyJMM1986, sen2000quantum, ovchinnikov2003, Lauchli:2025fii}. 
The $\mathbb{Z}_{2}\cong \langle T\rangle/\langle T^2 \rangle$ symmetry in this model corresponds to  translation by one lattice site, $T$. 
The phase diagram of this model is shown in Figure \ref{fig:AFM_PD}. 

In order to construct a Hamiltonian that has the Yang-Lee critical point, we follow the original construction discussed in Section \ref{FMISandYL}. Namely, we add a term that (1) breaks the $\mathbb{Z}_{2}$ symmetry and (2) makes the Hamiltonian non-Hermitian, while preserving $\mathcal{PT}$ symmetry.  For the antiferromagnetic model (\ref{eq_H_AFM}), we can achieve this by adding an imaginary staggered longitudinal  magnetic field:
\begin{equation}
\begin{aligned}
H_{\textrm{AFYL}} &= H_{\textrm{AFI}} -ih_{z}^{\textrm{st}} \sum_{n=0}^{N-1}(-1)^{n}Z_{n} \,. \label{AYLHam1}
\end{aligned}
\end{equation}
Indeed, this Hamiltonian no longer has $\mathbbm{Z}_{2}$ symmetry, but it is invariant under $\mathcal{P}\mathcal{T}$ symmetry, which acts as 
$Z_{n}\to Z_{n+1}$, $X_{n}\to X_{n+1}$ and $i\to -i$.
Since the model (\ref{eq_H_AFM}) has a line of Ising critical points, we expect to find a surface of Yang-Lee critical points, i.e. a critical value  $h_{z,c}^{\textrm{st}}$ for each given pair  $(h_{x}, h_{z})$ when the model is in the paramagnetic phase. 

We can analyze the behavior of the model (\ref{AYLHam1}) through its corresponding classical version. Following steps similar to those in Section \ref{AFMIS} and using the Polyakov-Hubbard transformation, we arrive at the following LGW action:
\begin{equation}
\begin{aligned}
&S[\chi_{s \textbf{r}},\chi_{f \textbf{r}}] = 
\frac{1}{2}\sum\limits_{\textbf{k}\in \textrm{BZ}} (\varepsilon_{1}(\textbf{k})|\chi_{s\textbf{k}}|^2 + 
\varepsilon_{2}(\textbf{k})|\chi_{f\textbf{k}}|^2) +\sum\limits_{\textbf{r}}\bigg( \frac{\chi_{s\textbf{r}}^2}{2\beta v_{1}(0)} + \frac{\chi_{f\textbf{r}}^2}{2\beta v_{2}(0)} \\
&\qquad\quad -\ln\Big(4\cosh\big(\frac{\chi_{s\textbf{r}}+\chi_{f\textbf{r}}}{\sqrt{2}}-\beta h + i\beta h^{\textrm{st}}\big)\cosh\big(\frac{\chi_{s\textbf{r}}-\chi_{f\textbf{r}}}{\sqrt{2}}+\beta h+ i\beta h^{\textrm{st}}\big)\Big) \bigg)\,,
\end{aligned}
\end{equation}
where  $v_{\alpha}(\textbf{k})$ with $\alpha = 1,2$ are given in (\ref{v1v2eq}) and $\varepsilon_{\alpha}(\textbf{k})$ are defined below (\ref{LGWafi}). We see that, for $h^{\textrm{st}}=0$ the LGW action has a $\mathbb{Z}_{2}$ symmetry: $\chi_{s\textbf{r}} \to -\chi_{s\textbf{r}}$ for the low-energy (slow) field. Using  mean-field analysis, one can find the line of Ising critical points. 
Whereas at non-zero imaginary staggered  field $h^{\textrm{st}}$, the action has only $\mathcal{PT}$ symmetry: $\chi_{s\textbf{r}} \to -\chi_{s\textbf{r}}$, $i\to -i$. Using mean-field analysis, as in (\ref{YLmfeq}), one can find the Yang-Lee critical surface in this case.

\begin{figure}[h!]
    \centering
    \includegraphics[width=14cm]{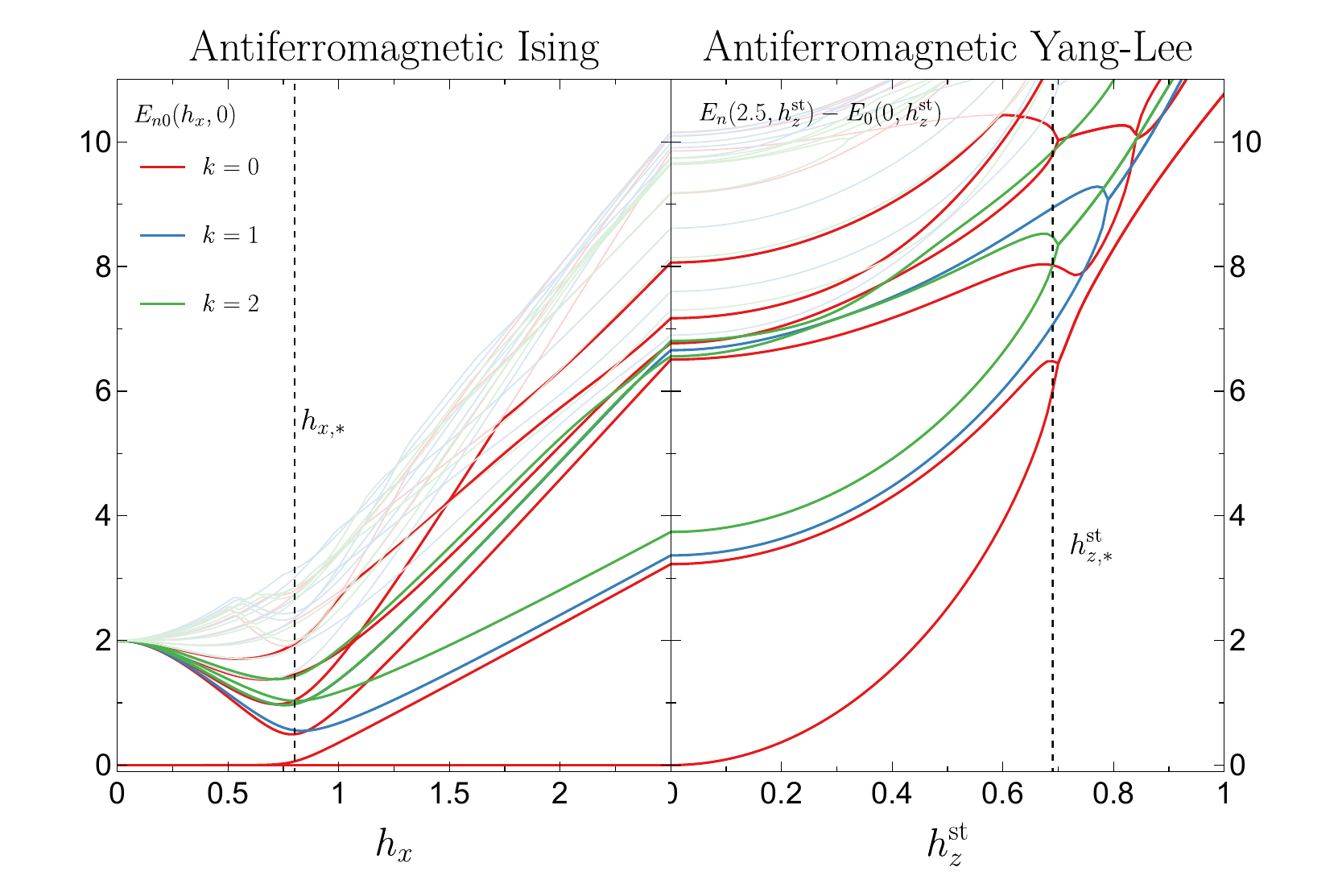}
    \caption{Spectrum of the Hamiltonian $H_\textrm{AFYL}$ in (\ref{AYLHam1}), for $N=20$, $J=1$, and $h_{z}=1$, and different spin sectors $k=0, 1, 2$. A pseudo-critical point that belongs to the Ising universality class at $h_{x,*} \approx  0.8$, and a pseudo-critical point that belongs to the Yang-Lee universality class at $h_{z,*}^{\textrm{st}} \approx 0.69$, for $h_{x}=2.5$ are denoted by vertical dashed lines.}
    \label{fig:YL2_Spectrum}
\end{figure}

To confirm our expectations numerically, we plot the real part of the low-energy spectrum $E_{n0}(h_{x}, h_{z}^{\textrm{st}})\equiv E_{n}-E_{0}$ of the Hamiltonian (\ref{AYLHam1}) in Figure \ref{fig:YL2_Spectrum} for the parameters $N=20$ and $h_{z}=1$, and along the path with $h_{z}^{\textrm{st}}=0$ and $h_{x}\in [0,2.5]$, and then the path with $h_{x}=2.5$ and $h_{z}^{\textrm{st}}\in [0, 1]$. 
We see that along this path we first cross the Ising pseudo-critical point at $h_{x,*} \approx  0.8$ and then cross the YL pseudo-critical point at $h_{z,*}^{\textrm{st}} \approx 0.69$. 
We note a striking similarity between this spectral flow and the one obtained for the ferromagnetic YL model (\ref{eq_H_YL}) in \cite{ArguelloCruz:2025zuq}. 

Below, we provide more details on finding pseudo-critical points and extracting scaling dimensions from the lattice Hamiltonians.  The finite-size scaling analysis \cite{Fisher.PRL.28.1516, Hamer:1981qt}  and the state-operator correspondence  \cite{CardyJPA17.1984, BlotePRL56.1986, CardyNPB.1986} imply that the energy gaps $E_{n0}\equiv E_{n}-E_{0}$ at the critical point and large $N$ behave as  
\begin{align}
    E_{n0} = \frac{2\pi v}{N}\ \Delta_{n} + \dots\,,
    \label{eq_gaps}
\end{align}
where $v$ is the model-dependent speed of light and $\Delta_{n}$ are the scaling dimensions of operators in the corresponding IR CFT.  For the Hamiltonian (\ref{AYLHam1}), we define the pseudo-critical point $h_{z,*}^{\textrm{st}}(N)$ as the point where the following condition is exactly satisfied \cite{Hamer:1981qt}:
\begin{align}
    \frac{(N+2) E^{k=0}_{10}(h_{z,*}^{\textrm{st}}(N), N+2)}{N E^{k=0}_{10}(h_{z,*}^{\textrm{st}}(N), N)}=1\,,
    \label{eq_FSS_AFM}
\end{align}
where $E_{n}^{k=0}$ are the energy levels in the $k=0$ momentum (spin) sector\footnote{The Hamiltonian (\ref{AYLHam1}) is translationally invariant under shifts by two lattice sites, therefore, the momentum $k$ can take values $k=0,1,2,\dots, N/2-1$. }. Conditions of this type are called criticality criteria, and they produce a sequence of pseudo-critical points  $h_{z,*}^{\textrm{st}}(N)$ that converges to the thermodynamic critical point $h_{z,c}^{\textrm{st}}$ in $N\to \infty$ limit.

Table \ref{tab::crit_pt_locations} shows the pseudo-critical points for different values of $N$, obtained using the criticality criterion (\ref{eq_FSS_AFM}). The thermodynamic  critical point is calculated via the BST algorithm with $w=2$, which gives $h^{\textrm{st}}_{z,c}=0.69092$ for $h_{x}=2.5$ and $h_{z}=1$.
\begin{table}[h]
  \centering
  \begin{tabular}{c}
      \hline\hline
  \begin{tabular}{c|c|c|c|c|c|c|c||c}
    $N$  & $8$ & $ 10$ & $12$ &  $14$ & $16$ & $18$ & $20$ & $\infty$  \\
    \hline
    $h_{z, *}^{\textrm{st}}$  & $ 0.69197 $ & $ 0.69138 $ & $ 0.69114 $ & $ 0.69104 $ & $ 0.69099 $ & $ 0.69096 $ & $ 0.69095 $ & $ 0.69092 $ \\ 
    \hline
   \end{tabular} 
  \end{tabular}
  \caption{Pseudo-critical points  $h_{z, *}^{\textrm{st}}$ of the Hamiltonian  (\ref{AYLHam1}) for various $N$, obtained using the criticality criterion \eqref{eq_FSS_AFM} with precision $10^{-7}$ for $J=1$, $h_{x}=2.5$, and $h_{z}=1$.}
  \label{tab::crit_pt_locations}
\end{table}
We calculate the scaling dimensions using the formula\footnote{Notice that we have assumed that the ground state corresponds to the $\phi$ operator and that the first excited state corresponds to the identity operator, as expected in 2D YL CFT.} 
\begin{align}
    \Delta^{k}_{n}(N)=\frac{E^{k}_{n}(N)-E_{1}^{k=0}(N)}{E^{k=1}_{0}(N)-E^{k=0}_{0}(N)} \underset{N\to \infty}{\longrightarrow} \frac{\Delta^{k}_{n}-\Delta_{I}}{\Delta_{\partial\phi}-\Delta_{\phi}}=\Delta^{k}_{n}\,,
    \label{eq_scaling_dim_YL}
\end{align}
where $E^{k}_{n}(N)$ is the $n$th energy level in the momentum sector $k$ for system size $N$, computed at the pseudo-critical point $h_{z, *}^{\textrm{st}}(N)$, and $\Delta^{k}_{n}$ is the $n$th scaling dimension of the spin-$k$ operator in the YL CFT. We list the numerical results in Table \ref{tab::scaling_dim} in Appendix \ref{AppAFMYL}. The extrapolated values in the $N\to \infty$ limit are $\Delta_{0}^{k=0}= -0.400304$, $\Delta_{2}^{k=0}= 1.6008443$, and $\Delta_{3}^{k=0}= 3.5714391$, $\Delta_{4}^{k=0}= 3.97963$, which are approximately $10^{-4}$ and $10^{-2}$ away from the exact theoretical scaling dimensions $\Delta_{\phi}=-2/5, \Delta_{\Box\phi}=8/5$, and $\Delta_{\Box^2\phi}=18/5,\ \Delta_{T\bar{T}}=4$, respectively.

As an additional check of  YL universality, we compute the effective central charge of the critical point, using the ground-state energy $E^{k=0}_0$: 
\begin{align}
    E^{k=0}_{0}(N) = - N a_{0} - v\frac{\pi}{6} \frac{c_{\textrm{eff}}}{N} + \dots \,,
    \label{eq_E0}
\end{align}
where $a_{0}$ is a bulk constant and $c_{\textrm{eff}} = c -12\Delta_{\textrm{min}}$ is the effective central charge in a non-unitary CFT, and $\Delta_{\textrm{min}}$ is the lowest scaling dimension \cite{Itzykson.EPL.1986}. For the 2D YL CFT, $c=-22/5$ and $\Delta_{\textrm{min}}= \Delta_{\phi}=-2/5$.
Thus, we can calculate $c_{\textrm{eff}}$ at a given $N$ as \cite{Gehlen.JMPB.1994}
\begin{align}
    c_{\textrm{eff}}(N)=\frac{3N(N-2)}{2(N-1)\pi v(N)} \left[(N-2)E^{k=0}_{0}(N) - NE^{k=0}_{0}(N-2) \right] \underset{N\to \infty}{\longrightarrow} c_{\textrm{eff}}\,.
    \label{eq_charge_AFM}
\end{align}
The speed of light $v(N)$ at a given $N$ can be obtained from the following energy gap:
\begin{align}
 v(N) = \frac{N}{2\pi} \left(\frac{E_{0}^{k=1} - E_{0}^{k=0}}{\Delta_{\partial \phi} - \Delta_{\phi}}\right)
    = \frac{N}{2\pi} (E_{0}^{k=1} - E_{0}^{k=0}) \,. \label{eq_v_AFM}
\end{align}
We first calculate $v(N)$ and then use it to obtain the effective central charge in \eqref{eq_charge_AFM}. 
The results are presented in Table \ref{tab::charge_AFM} in Appendix \ref{AppAFMYL}, and the extrapolated values are $v =3.237124 $ and $c_{\textrm{eff}}=0.399646$. The effective central charge is within $10^{-4}$ of the exact value $c_{\textrm{eff}}=2/5$ in 2D YL CFT. All these numerical results confirm that the critical point we find in the model (\ref{AYLHam1}) belongs to the 2D YL universality class.

\section{Yang-Lee criticality in the Schwinger model}
\label{YLSchwinger}

In this section, we show that the one-flavor Schwinger model \cite{SchwingerPR1962, LowensteinAnn1971, CasherPRD1974}  has a Yang-Lee critical point under deformation by an imaginary pseudo-scalar mass term.  
The Lagrangian density of the Schwinger model in the presence of a background electric field, represented by the $\theta$ angle, is given by \cite{Coleman:1975pw,Coleman:1976uz}
\begin{align}
    \mathcal{L}_{\textrm{S}} = -\frac{1}{4}F_{\mu\nu}F^{\mu\nu} -\frac{e\theta}{4\pi}\epsilon^{\mu\nu}F_{\mu\nu}+\bar{\psi}( i\slashed{\partial}- e\slashed{A} -m)\psi \,,
    \label{Schw_lagrange}
\end{align}
where $\psi$ is a two-component fermionic field, $\slashed{A} = \gamma^{\mu}A_{\mu}$, with $(\gamma^{0},\gamma^{1}, \gamma^{5})=(Z, iY, X)$ and $X,Y,Z$ are the Pauli matrices,  $F_{\mu\nu}=\partial_{\mu}A_{\nu}-\partial_{\nu}A_{\mu}$ is the  field strength,  $\epsilon^{01}=1$, and $e$ is the electric charge.
The Hamiltonian density of this model is given by:
\begin{align}
\mathcal{H}_{\textrm{S}} =  \frac{1}{2}\Big(E + \frac{e\theta}{2\pi} \Big)^2 +\bar{\psi}( m - i\gamma^1 (\partial_{1} +  i e A_{1}))\psi \,. \label{HSferm}
\end{align}
Using bosonization, the Hamiltonian density of the Schwinger model can be written as \cite{Coleman:1976uz}:
\begin{align}
\mathcal{H}_{\textrm{S}} = & N_{M}\Big[\frac{1}{2} \Pi^2 +\frac{1}{2}(\partial_{x}\phi )^2 +\frac{1}{2} M^2\phi^2 -c m M \cos(2\sqrt{\pi}\phi -\theta)\Big]\,,  \label{HSbos}
\end{align}
where $c=e^{\gamma}/(2\pi) \approx 0.2835 $,  $M=e/\sqrt{\pi}$ is the Schwinger boson mass and $N_{M}$ denotes normal ordering with respect to the mass $M$. 
It is known that, for the background electric field $\theta=\pi$, this model has a second-order phase transition that belongs to the 2D Ising universality class \cite{Byrnes_2002, byrnes2003density, PhysRevD.95.094509, Ohata:2023gru, Dempsey.PRR.2022,Dempsey.PRL.2023, ArguelloCruz:2024xzi, Cuomo:2026kmy} at $m_{c}/e=0.333561(4)$ \cite{ArguelloCruz:2024xzi, Fujii:2024reh}. 
In the Schwinger model at $\theta = \pi$, the Ising $\mathbb{Z}_2$ symmetry   is the charge conjugation symmetry, which acts as $\psi \to \gamma^{5}\psi^{*}$ and $A_{\mu}\to -A_{\mu}$ in the Hamiltonian (\ref{HSferm}) and as $\phi \to -\phi$ in the bosonic Hamiltonian (\ref{HSbos}).  To be consistent with the notation in this article, we denote the generator of this symmetry by $\mathcal{P}$ and it should not be confused with the spatial parity symmetry, which is also present in the Schwinger model.

In order to obtain the YL criticality, we deform the Schwinger model Lagrangian by a term that breaks the $\mathbb{Z}_{2}$ symmetry generated by $\mathcal{P}$ (charge conjugation), but preserves the $\mathcal{P}\mathcal{T}$ symmetry, where  $\mathcal{T}$ denotes the time-reversal operator and it acts as $E \to E$, $A_{1}\to -A_{1}$,  $\psi \to \gamma^{0}\psi$ and $i\to -i$. A deformation by an imaginary pseudo-scalar mass  satisfies these conditions\footnote{Note that the usual pseudo-scalar mass term includes a factor of $i$ in order for the operator to be Hermitian.} 
\begin{align}
    \mathcal{L}_{\textrm{SYL}}= \mathcal{L}_{\textrm{S}}  + i m_{5} (i\bar{\psi}\gamma^{5}\psi)\,, \label{LSYLeq}
\end{align}
where $m_{5}$ is a real parameter. This additional term is $\mathcal{P}$-odd and $\mathcal{T}$-odd and therefore the new Lagrangian is $\mathcal{PT}$ symmetric. Under bosonization, the pseudo-scalar mass  maps as \cite{Byrnes_2002} $:i\bar{\psi}\gamma^{5}\psi: \leftrightarrow -c M N_{M}\sin (2\sqrt{\pi}\phi)$  and therefore the  bosonized Hamiltonian density for the deformed Schwinger model at $\theta = \pi$ reads
\begin{align}
\mathcal{H}_{\textrm{SYL}} = & N_{M}\Big[\frac{1}{2} \Pi^2 +\frac{1}{2}(\partial_{x}\phi )^2 +\frac{1}{2} M^2\phi^2 +c m M \cos(2\sqrt{\pi}\phi) + i c m_{5} M \sin(2\sqrt{\pi}\phi)\Big]\,.
\label{eq_HSM_cont}
\end{align}
We can estimate the values of the Ising and Yang-Lee critical points using the mean-field analysis, discussed in Section \ref{PHtrans}. For the Ising critical point, when $m_{5}=0$ we find $m_{c} = M/(4\pi c)$, which gives $m_{c}/e = e^{-\gamma}/(2\sqrt{\pi})\approx 0.158$. For the Yang-Lee critical point, using (\ref{YLmfeq}), we obtain two equations that determine the field shift $\phi \to \phi + \frac{i\phi_{*}}{2\sqrt{\pi}}$ and the critical value $m_{5,c}$:
\begin{align}
m = \frac{M}{4\pi c} (\cosh(\phi_{*})- \phi_{*} \sinh(\phi_{*})), \quad m_{5,c}=
\frac{M}{4\pi c} (\sinh(\phi_{*})- \phi_{*} \cosh(\phi_{*})) \,.
\end{align}
These are transcendental equations that can be easily solved numerically. In the case $m=0$, we find
\begin{align}
\frac{m_{5,c}}{e} = \sqrt{\phi_{*}^2 -1}\frac{e^{-\gamma}}{2\sqrt{\pi}} \approx 0.105\,, \label{m5cMF}
\end{align}
where $\phi_{*}\approx -1.200$ is found from the equation $\phi_{*}\tanh(\phi_{*})=1$. Below, we analyze the Schwinger model numerically by discretizing it on a lattice. We indeed find the Yang-Lee critical point after deforming the model by the imaginary pseudo-scalar mass term, and the critical value of $m_{5,c}$ turns out to be close to the mean-field estimate.

\subsection{Lattice description of the Schwinger model and numerical results}
The Schwinger model Hamiltonian (\ref{HSferm}) on a lattice, using the staggered-fermion formulation of \cite{kogutPRD.1975}, has the form \cite{BanksPRD.13.1043}
\begin{equation}
\begin{aligned}
H_{\textrm{S}} =& \frac{e^{2}a}{2} \sum_{n=0}^{N-1}\left(L_{n}+\frac{\theta}{2\pi}\right)^{2} + m_{\textrm{lat}} \sum_{n=0}^{N-1}(-1)^{n}c_{n}^{\dag}c_{n} \\
&\qquad -\frac{i}{2a}\sum_{n=0}^{N-1}\left(c_{n}^{\dag}U_{n} c_{n+1} - c_{n+1}^{\dag} U_{n}^{\dag}c_{n}\right)\,, \label{SchLatt}
\end{aligned}
\end{equation}
where $a$ is the lattice spacing, $U_{n}$ and $L_{n}$ are the gauge-field variables that live on the links between sites $n$ and $n+1$ and obey $[L_{n},U_{m}]=\delta_{nm}U_{m}$, and $c_{n}, c^{\dagger}_{n}$ are the fermion annihilation and creation operators at site $n$, which satisfy $\{c_{n},c_{m}^{\dag}\} = \delta_{nm}$. In order to preserve the discrete remnant of  chiral symmetry on the lattice, the lattice mass $m_{\textrm{lat}}=m-e^2a/8$ has a shift \cite{Dempsey.PRR.2022,Dempsey.PRL.2023, Dempsey:2025wia, Cuomo:2026kmy}, which vanishes in the continuum limit $a\to 0$, but significantly improves the numerical convergence of the model.  The Gauss law relates gauge and fermionic degrees of freedom and reads
\begin{align}
L_{n}-L_{n-1} = Q_{n}, \quad Q_{n} = c^{\dag}_{n}c_{n} - \delta_{n,\textrm{odd}}\,. \label{Glaw}
\end{align}
We assume periodic boundary conditions (PBC), so $L_{-1}\equiv L_{N-1}$ and $c_{N}\equiv c_{0}$, and  Gauss's law  implies that the total charge must be zero $\sum_{n} Q_{n}=0$.  The Hilbert space of the lattice Schwinger model is 
\begin{align}
\mathcal{H} = \{ |n_{0}n_{1}\dots n_{N-1}\rangle_{F} \otimes |\ell_{0}\ell_{1}\dots \ell_{N-1}\rangle_{G},\; n_{n}=0,1\,,\; \ell_{n}\in \mathbbm{Z} \} \,,
\end{align}
where $L_{n}|\ell_{n}\rangle_{G}=\ell_{n}|\ell_{n}\rangle_{G}$ and $Q_{n}|n_{n}\rangle_{F}  = (n_{n}- \delta_{n,\textrm{odd}})|n_{n}\rangle_{F}$.
Using Gauss's law  (\ref{Glaw}), one can write the Hamiltonian (\ref{SchLatt}) in the form (see eq. (B12) of \cite{Dempsey.PRR.2022}):
\begin{equation}
\begin{aligned}
    \frac{2H_\textrm{S}}{e^{2}a} =& N\left( \mathcal{E} +\frac{\theta}{2\pi}\right)^2  -\frac{1}{2N} \sum_{n,n'=0}^{N-1}|n-n'|\left( N- |n-n'|\right)  Q_{n}Q_{n'}  \\
    &- ix \sum_{n=0}^{N-1} \left( c^{\dagger}_{n}Uc_{n+1} - c^{\dagger}_{n+1} U^{\dag}c_{n} \right) + \mu\sum_{n=0}^{N-1} (-1)^{n}c_{n}^{\dagger}c_{n} \,, \label{HSlatt2}
\end{aligned} 
\end{equation}
where $x \equiv 1/(e^{2} a^{2})$ and $\mu  \equiv 2m_{\textrm{lat}}/(e^2 a)$ and we have introduced the average electric field $\mathcal{E} \equiv (L_{0}+L_{1}+\dots +L_{N-1})/N$ and the link variable $U$, which is conjugate to $\mathcal{E}$ and obeys $[\mathcal{E},U]=U/N$.  The advantage of the Hamiltonian (\ref{HSlatt2}) is that it is explicitly translationally invariant and acts on the reduced Hilbert space\footnote{We note that, in the case of open boundary conditions (OBC), Gauss’s law allows one to eliminate all gauge degrees of freedom from the lattice Hamiltonian, whereas for PBC one is left with a single gauge degree of freedom, the holonomy, represented by the average electric field $\mathcal{E}$. }
\begin{align}
\mathcal{H} = \{ |n_{0}n_{1}\dots n_{N-1}\rangle_{F} \otimes |\mathscr{E}\rangle_{G},\; n_{n}=0,1\,,\; \mathscr{E}\in \mathbbm{Z}/N \} \,,
\end{align}
where $\mathcal{E}|\mathscr{E}\rangle_{G} = \mathscr{E}|\mathscr{E}\rangle_{G}$ and $U|\mathscr{E}\rangle_{G}=|\mathscr{E}+1/N\rangle_{G}$.
The gauge variables $\ell_{n}$ are integers and can be expressed in terms of $\mathscr{E}$ and the charges $q_{n}=n_{n}-\delta_{n,\textrm{odd}}$ via the formula 
\begin{align}
\ell_{n} =\mathscr{E} + \frac{1}{N}\sum_{m=1}^{N} m q_{m} - \sum_{m=n+1}^{N-1}q_{m} \,.
\end{align}
Therefore, $\mathscr{E}$ can not take arbitrary values in $\mathbbm{Z}/N$, but only those for which $\mathscr{E} + \frac{1}{N}\sum_{m=1}^{N} m q_{m}$ is an integer. Let us define $\mathscr{E}_{0}\in[-1/2, 1/2)$ as \cite{Dempsey.PRR.2022}
\begin{align}
\mathscr{E}_{0}  \equiv -\frac{1}{N}\sum_{m=1}^{N} m q_{m} \; \textrm{mod}\; 1\,. \label{E0def}
\end{align}
Then $\mathscr{E}$ can take values $\mathscr{E}=\mathscr{E}_{0}+h$, where the holonomy number $h\in \mathbbm{Z}$. For a given distribution of occupation numbers $\{n_{n}\} \equiv \{n_{0},n_{1},\dots n_{N-1}\}$, one unambiguously finds $\mathscr{E}_{0}=\mathscr{E}_{0}(\{n_{n}\})$ in the interval $[-1/2, 1/2)$ using (\ref{E0def}). Then the model states $|\{n_{n}\}\rangle_{F}\otimes |\mathscr{E}\rangle_{G}$ can be written as  $|\{n_{n}\}\rangle_{F} \otimes |h\rangle_{G}$, with the holonomy number $h\in \mathbbm{Z}$, and they satisfy
\begin{align}
c_{n}^{\dag}c_{n+1}|\{n_{n}\}\rangle_{F}\otimes U|h\rangle_{G} = 
\begin{cases}
c_{n}^{\dag}c_{n+1}|\{n_{n}\}\rangle_{F}\otimes |h\rangle_{G}, \qquad\, \textrm{if}\;\; \mathscr{E}_{0}+1/N < 1/2 \\
c_{n}^{\dag}c_{n+1}|\{n_{n}\}\rangle_{F}\otimes |h+1\rangle_{G}, \;\;\textrm{if}\;\;  \mathscr{E}_{0}+1/N \geq  1/2 
\end{cases}\,,
\end{align}
and similarly the operator $c^{\dagger}_{n+1}c_{n} U^{\dag}$ does not change the holonomy number $h$ if $\mathscr{E}_{0}-1/N \geq -1/2$ and lowers the holonomy by one if $\mathscr{E}_{0}-1/N < -1/2$. 
The pseudo-scalar mass  operator on the staggered lattice is  \cite{Sussking.PRD.1997}:
\begin{align}
    i\bar{\psi}\gamma^{5}\psi \to \frac{1}{a}i(-1)^{n}\left(c^{\dagger}_{n} c_{n+1} - c^{\dagger}_{n+1} c_{n} \right).
\end{align}
Therefore, the lattice Hamiltonian for the deformed Schwinger model (\ref{LSYLeq}) has the form
\begin{align}
    H_\textrm{SYL} &= H_\textrm{S} - i m_{5} \sum_{n=0}^{N-1} i(-1)^{n}\left( c^{\dagger}_{n}Uc_{n+1} - c^{\dagger}_{n+1} U^{\dag}c_{n} \right)\,. \label{HSYLlatt}
\end{align}
The charge-conjugation operator $\mathcal{P}$, which represents the Ising $\mathbbm{Z}_{2}$ symmetry of the Schwinger model at $\theta = \pi$, and the time-reversal operator $\mathcal{T}$ act on the lattice operators as follows:
\begin{equation}
\begin{aligned}
&\mathcal{P}:\quad \mathcal{E} \to -\mathcal{E}-1, \quad U \to U^{\dag}, \quad c_{n}  \to c^{\dagger}_{n+1}, \quad   c^{\dagger}_{n} \to c_{n+1}\,, \\
&\mathcal{T}:\quad \mathcal{E} \to \mathcal{E}, \quad U \to U, \quad c_{n}  \to (-1)^{n}c_{n}, \quad   c^{\dagger}_{n} \to (-1)^{n}c^{\dag}_{n}, \quad i \to -i\,.
\end{aligned}
\end{equation}
The Hamiltonian (\ref{HSlatt2}) is invariant under the $\mathcal{P}$ and $\mathcal{T}$ transformations separately, whereas the deformation term in (\ref{HSYLlatt}) is invariant only under the combined $\mathcal{PT}$ transformation.

We can rewrite the Hamiltonian (\ref{HSYLlatt}) in terms of spin variables using the Jordan-Wigner transformation 
\begin{align}
c_{n} = \prod_{m=0}^{n-1}(-iZ_{m}) \otimes  \sigma^{-}_{n}, \quad c^{\dag}_{n} = \prod_{m=0}^{n-1}(iZ_{m}) \otimes  \sigma^{+}_{n}\,, \label{JWtrans}
\end{align}
where $\sigma^{\pm}_{n}=\frac{1}{2}(X_{n} \pm iY_{n})$. We arrive at the following spin representation:
\begin{equation}
\begin{aligned}
    \frac{ 2 H_{\textrm{SYL}}}{e^2a} =& N\left( \mathcal{E} +\frac{\theta}{2\pi}\right)^2 - \frac{1}{2N} \sum_{n,n'=0}^{N-1}|n-n'|\left( N- |n-n'|\right)  Q_{n}Q_{n'} \\
    &+x \sum_{n=0}^{N-1} \left( \sigma^{+}_{n}\sigma^{-}_{n+1} U +\sigma^{-}_{n}\sigma^{+}_{n+1} U^{\dag} \right) + \frac{\mu}{2} \sum_{n=0}^{N-1} (-1)^{n}Z_{n}   \\
      &+ i\mu_{5} \sum_{n=0}^{N-1} (-1)^{n}\left( \sigma^{+}_{n}\sigma^{-}_{n+1} U +\sigma^{-}_{n}\sigma^{+}_{n+1} U^{\dag}\right)\,, 
    \label{spinHSYM}
\end{aligned}
\end{equation}
where $Q_{n} = \frac{1}{2}(1+Z_{n}) - \delta_{n,\textrm{odd}}$, $\mu_{5}\equiv 2\sqrt{x}\ m_{5}/e$ and $\sigma^{\pm}_{N}\equiv \sigma^{\pm}_{0}$. We note that the Hamiltonian (\ref{spinHSYM}) is explicitly invariant under translations by two lattice sites. This invariance is not automatically preserved after the transformation (\ref{JWtrans}), since the boundary terms $\sigma_{N-1}^{+}\sigma_{0}^{-} U_{N-1}$ and $\sigma_{N-1}^{-}\sigma_{0}^{+} U_{N-1}^{\dagger}$ appear with opposite signs. To fix this, we make a canonical transformation $U_{N-1} \to -U_{N-1}$ and $L_{N-1}\to L_{N-1}$ on the boundary link.

\begin{figure}[h!]
    \centering
    \includegraphics[width=14cm]{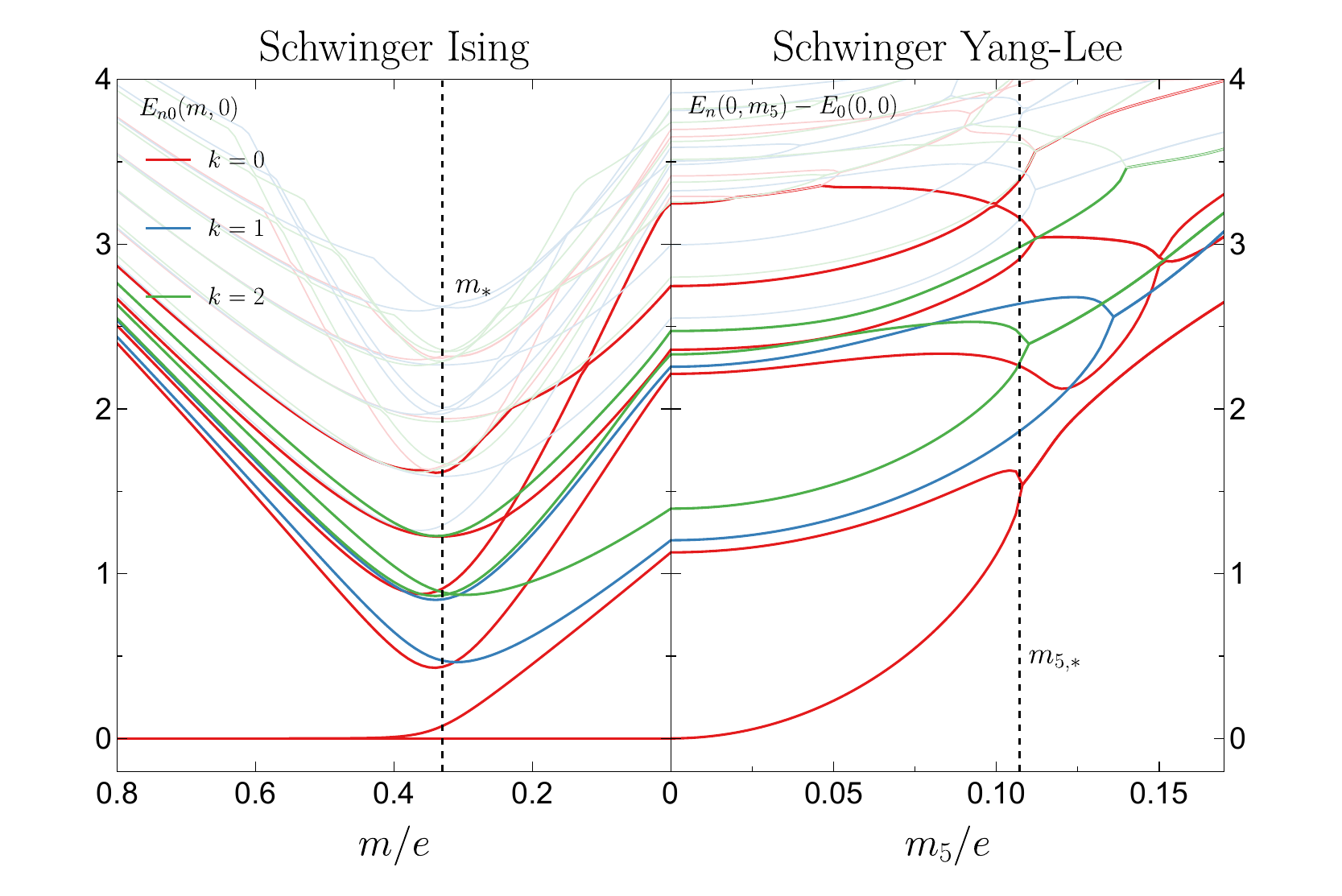}
    \caption{Spectrum of the Hamiltonian $H_\textrm{SYL}$ in (\ref{spinHSYM}), using $N=28$, $x=1$, and $h_{\textrm{max}}=5$, for different spin sectors $k=0, 1, 2$. A pseudo-critical point that belongs to the Ising universality class at $m_{*}/e \approx  0.33$, and a pseudo-critical point that belongs to the Yang-Lee universality class at $m_{5,*}/e \approx 0.107$, for $m/e=0$, are denoted by vertical dashed lines.}
    \label{fig:Spec_SYL}
\end{figure}

For numerical computations, we work with the Hamiltonian (\ref{spinHSYM}) at $\theta = \pi$ and truncate the holonomy space. Thus, we assume that $h$  takes only a finite number of values $h = -h_{\textrm{max}},\dots, h_{\textrm{max}}$, and in this work we  use $h_{\textrm{max}}=5$. We plot the real part of the low-energy spectrum $E_{n0}(m/e, m_{5}/e)\equiv E_{n}-E_{0}$ of the Hamiltonian (\ref{spinHSYM}) for momenta $k=0,1,2$\;\footnote{We note that, for the fermionic representation of the Schwinger Hamiltonian (\ref{HSYLlatt}), the eigenstates of the translation operator differ from those of the spin Hamiltonian (\ref{spinHSYM}). In particular, in the fermionic representation, for $N = 0\, \textrm{mod}\, 4$, the ground-state energy belongs to the $k=N/4$ sector, whereas for $N = 2 \, \textrm{mod}\, 4$, it belongs to the $k=0$ sector.} in Figure \ref{fig:Spec_SYL} for the parameters $N=28$ and $x=1$, and along the path with $m_{5}/e=0$ and $m/e \in [0.8,0]$, followed by the path with $m/e=0$ and $m_{5}/e\in [0, 0.16]$.  
We see that along this path we first cross the Ising pseudo-critical point at $m_{*}/e \approx  0.33$ and then we cross the YL pseudo-critical point at $m_{5,*}/e \approx 0.107$. 
We note the similarity with Figure \ref{fig:YL2_Spectrum}.

To find the YL pseudo-critical point, we use the same method as in the previous section and 
define the pseudo-critical point $m_{5,*}(N)/e$ as
\begin{align}
    \frac{(N+2) E_{10}^{k=0}(m_{5,*}(N)/e,N+2)}{NE_{10}^{k=0}(m_{5,*}(N)/e, N)}=1 \,.
    \label{eq_FSS_SM}
\end{align}
Using this expression, we present in Table \ref{tab::crit_pt_locations_SM} in Appendix \ref{AppSYL} the results for the pseudo-critical points $m_{5,*}/e$ for different values of $N$ and $x$. Extrapolating to the thermodynamic limit using the BST algorithm with $w=2$, we find $m_{5,c}/e \approx 0.107,\, 0.102,\, 0.097$ for $x=1,\, 1.5,\, 4$, respectively. These results are quite close to the mean-field estimate in (\ref{m5cMF}). We  also calculate the scaling dimension of the first gap in the $k=0$ sector, the speed of light $v$ and the effective central charge $c_{\textrm{eff}}$, for different values of $N$ and $x$ using (\ref{eq_scaling_dim_YL}), \eqref{eq_charge_AFM} and \eqref{eq_v_AFM}. The results are shown in Tables \ref{tab::scaling_dims_SM} and \ref{tab::charge_SM} in Appendix \ref{AppSYL}.  Extrapolating to $N=\infty$ using the BST algorithm with $w=2$, we find  $\Delta_{\phi} \approx  -0.4002, -0.4002, -0.3996$,  $c_{\textrm{eff}} \approx 0.3985, 0.3975, 0.3975$ for $x=1,1.5, 4$ respectively. We see that these values are within $10^{-3}$ of the exact values $\Delta_{\phi}=-2/5$ and $c_{\textrm{eff}}=2/5$ of the Yang-Lee CFT.

\section{Yang-Lee criticality in the Blume-Capel model}
\label{secYLinBC}

In this section, we revisit Yang-Lee criticality in the spin-1 ferromagnetic Blume-Capel (BC) quantum model in an imaginary longitudinal magnetic field, first discussed in \cite{Becker:1991nq, Becker:1991nr, Gehlen.JMPB.1994}. Without the imaginary magnetic field, the Blume-Capel model Hamiltonian reads \cite{Gefen.PRB.1981, Alcaraz.PRB.1985, Balbao.JPA.1987}: 
\begin{align}
    H_{\textrm{BC}}= - J \sum_{n=0}^{N-1} S_{n}^{z}S_{n+1}^{z} - h_{x} \sum_{n=0}^{N-1} S_{n}^{x}  + \alpha \sum_{n=0}^{N-1}(S_{n}^{z})^2 \,,
    \label{eq_H_BC}
\end{align}
where $S_{N}^{z}\equiv S_{0}^{z}$ (PBC) and $J>0$, $h_{x}$, and $\alpha$ are real parameters.   
Here  $S^{x}, S^{y}, S^{z}$ are  $3\times3 $ spin-1 matrices, represented as
\begin{align}
    S^{x}= \frac{1}{\sqrt{2}} \left(
    \begin{matrix}
        0 & 1 & 0 \\
        1 & 0 & 1 \\
        0 & 1 & 0
    \end{matrix}
    \right), \qquad 
    S^{y}= \frac{1}{\sqrt{2}} \left(
    \begin{matrix}
        0 & -i & 0 \\
        i & 0 & -i \\
        0 & i & 0
    \end{matrix}
    \right), \qquad 
    S^{z}=  \left(
    \begin{matrix}
        1 & 0 & 0 \\
        0 & 0 & 0 \\
        0 & 0 & -1
    \end{matrix}
    \right).
\end{align}
This model has a global $\mathbb{Z}_{2}$ symmetry generated by the operator $\mathcal{P}$:
\begin{align}
    \mathcal{P}=\prod_{n=0}^{N-1} (2 \left(S_{n}^{x}\right)^2 - 1) \,,
\end{align}
and one can check that $[\mathcal{P}, H_{\textrm{BC}}]=0$. It is well known that \eqref{eq_H_BC} has a line of Ising critical points in  the $(\alpha, h_{x})$ plane, which ends at the tricritical Ising fixed point at $\alpha_{tc} = 0.910207$ and $h_{x,tc}= 0.415685$ \cite{Alcaraz.PRB.1985, Balbao.JPA.1987}.

As in the Ising spin-1/2 chain, we deform the Hamiltonian (\ref{eq_H_BC}) by an imaginary longitudinal  magnetic field:
\begin{align}
    H_{\textrm{BCYL}}=  H_{\textrm{BC}} -ih_{z} \sum_{n=0}^{N-1}S_{n}^{z} 
    \label{eq_H_BCYL}\,.
\end{align}
This deformation breaks the $\mathbb{Z}_{2}$ symmetry but preserves  $\mathcal{P}\mathcal{T}$ invariance. The spectrum of \eqref{eq_H_BCYL} is either real or consists of complex conjugate pairs.  Following \cite{Gehlen.JMPB.1994}, we can find Yang-Lee criticality for $\alpha < \alpha_{tc}$ and $h_{x} > h_{x,c}$, where $h_{x,c}$ corresponds to the Ising fixed point at a given $\alpha$.   

To better explain the expected phase diagram of the quantum model  (\ref{eq_H_BCYL}), we consider the classical model, from which the quantum model
can be obtained in the highly anisotropic limit. The classical Blume-Capel Hamiltonian in an imaginary magnetic field reads 
\begin{align}
\mathcal{H} = -\frac{1}{2}\sum_{\textbf{r},\textbf{r}'}V_{\textbf{r}\textbf{r}'} \sigma_{\textbf{r}}\sigma_{\textbf{r}'} + \alpha \sum_{\textbf{r}} \sigma_{\textbf{r}}^2 -  i h\sum_{\textbf{r}} \sigma_{\textbf{r}}, \quad \sigma_{\textbf{r}} = 0,\pm 1\,,
\end{align}
where $V_{\textbf{r},\textbf{r}'}$ is the nearest-neighbor potential on a simple square lattice,  
\begin{align}
V_{\textbf{r},\textbf{r}'}= J(\delta_{\textbf{r},\textbf{r}'+a\hat{x}} +
\delta_{\textbf{r},\textbf{r}'-a\hat{x}}+ \delta_{\textbf{r},\textbf{r}'+a\hat{y}} +
\delta_{\textbf{r},\textbf{r}'-a\hat{y}})\,.
\end{align}
Following  steps similar to those in Section \ref{PHtrans}, we arrive at the following LGW action: 
\begin{align}
S[\phi_{\textbf{r}}]=  \frac{1}{2}\sum\limits_{\textbf{k} \in \textrm{BZ}} \varepsilon(\textbf{k}) |\phi_{\textbf{k}}|^2 +\sum_{\textbf{r}}\left(\frac{\phi_{\textbf{r}}^2}{2\beta V(0)} -\ln (1+2e^{-\beta(\alpha+cJ)}\cosh(\phi_{\textbf{r}} + i\beta h))\right)\,,
\end{align}
where $\varepsilon(\textbf{k})\equiv  (V^{-1}(\textbf{k}) - V^{-1}(0))/\beta$ with $V(\textbf{k})= 2J(c+\cos(k_{x}a) + \cos(k_{y}a))$ and $\textrm{BZ}= \{-\pi/a \leq  k_{x},k_{y} \leq \pi/a\}$. We note that, in the BC model, the shift $c$ in the exchange coupling $V_{\textbf{r},\textbf{r}'}$ introduced to make it positive definite has to be compensated by a shift of the parameter $\alpha$. Using the mean-field approach in the case of zero imaginary magnetic field, we find the Ising critical line $T_{c}(\alpha)$ given by the equation 
\begin{align}
\beta V(0) = 1 + \frac{1}{2}e^{\beta(\alpha + c J)}\,.
\end{align}
The Ising line ends at the tricritical Ising point $(\alpha_{tc},T_{tc})$, determined by the equations $\beta V(0)  = 3$ and 
$2e^{-\beta(\alpha+cJ)}=1/2$. There is a line of first-order phase transitions for $\alpha > \alpha_{tc}$. For a non-zero imaginary magnetic field $ih$, the mean-field equations (\ref{YLmfeq}) always have a solution when the temperature satisfies $T >T_{c}(\alpha)$ and $\alpha < \alpha_{tc}$ and the YL critical points form surfaces which are attached to the Ising critical line.
One can also apply the mean-field analysis directly to the quantum model (\ref{eq_H_BCYL}) and the results are similar to those in the classical case \cite{Gefen.PRB.1981, Becker:1991nq, Becker:1991nr, Gehlen.JMPB.1994}.

The mean-field analysis also indicates that there are two lines in $(\alpha,T,h)$ space, emanating from the tricritical Ising point $(\alpha_{tc},T_{tc},0)$, along which $\delta^{n}S/\delta \phi^n|_{\phi = i\phi_{*}, h = h_{c}} = 0$ for $n=1,2,$ and $4$. It was argued in \cite{Gehlen.JMPB.1994} that these lines are the edges of the YL critical surfaces and consist of critical points described by the $M(2,7)$ minimal model. Beyond these lines, the model is not critical. Since the $M(2,7)$ minimal model has a relevant spin-$2$ operator \cite{Katsevich:2025ojk}, one should expect to find only a single $M(2,7)$ point on each edge line, because spinful operators are allowed to be generated in the Hamiltonian formalism. Moreover, the presence of the relevant spin-$2$ operator  indicates that the YL theory may flow to a theory with broken Lorentz symmetry. It would be interesting to analyze this possibility further with higher numerical precision.

\begin{figure}[h]
    \centering
    \includegraphics[width=7cm]{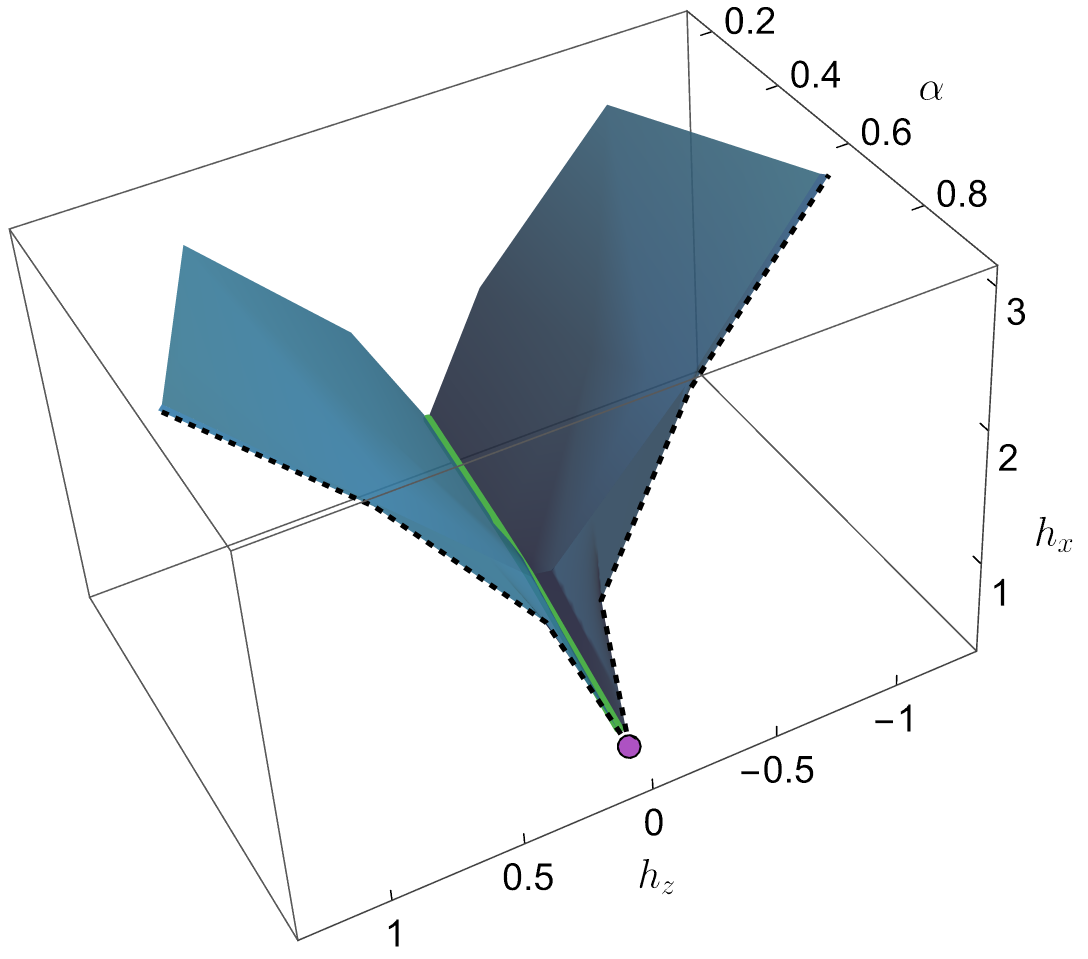}
    \caption{Phase diagram of the Hamiltonian $H_\textrm{BCYL}$ in (\ref{eq_H_BCYL}) for different values of $\alpha$, $h_{x}$, $h_{z}$, showing the YL critical surfaces (``wings'') in blue. The Ising critical line is shown in green, and the purple dot marks the tricritical Ising point. The YL edge lines are shown as black dashed lines.}
    \label{fig:BC_phase_diag}
\end{figure}

In \cite{Gehlen.JMPB.1994}, the YL edge lines were found by identifying the points at which the three lowest-energy states merge with one another. Although it remains unclear whether these lines describe $M(2,7)$ criticality, below we confirm the existence of these YL critical edge lines using the standard criticality criterion:
\begin{align}
\frac{(N+1)E_{10}^{k=0}(h_{z,*}(N),N+1)}{N E_{10}^{k=0}(h_{z,*}(N), N)}=1\,, \label{eq_FSS_BC}
\end{align}
which defines the pseudo-critical point $h_{z,*}(N)$. Some results for the pseudo-critical points are shown in Table \ref{tab::crit_pt_locations_BC} in Appendix \ref{AppBCYL}, and the corresponding phase diagram is presented in Figure \ref{fig:BC_phase_diag}, where the green line describes the Ising criticality ($h_{z}=0$), which ends at the tricritical Ising point, denoted by the purple dot.  The $\mathcal{P}\mathcal{T}$-broken phase lies below the YL critical surface, shown in blue.
The black dashed lines show the edges of the YL surfaces: beyond these edges, it is not possible to find pseudo-critical points that satisfy (\ref{eq_FSS_BC}). This phase diagram is in agreement with the previous results of \cite{Gehlen.JMPB.1994}.

\begin{figure}[h]
    \centering
    \includegraphics[width=14cm]{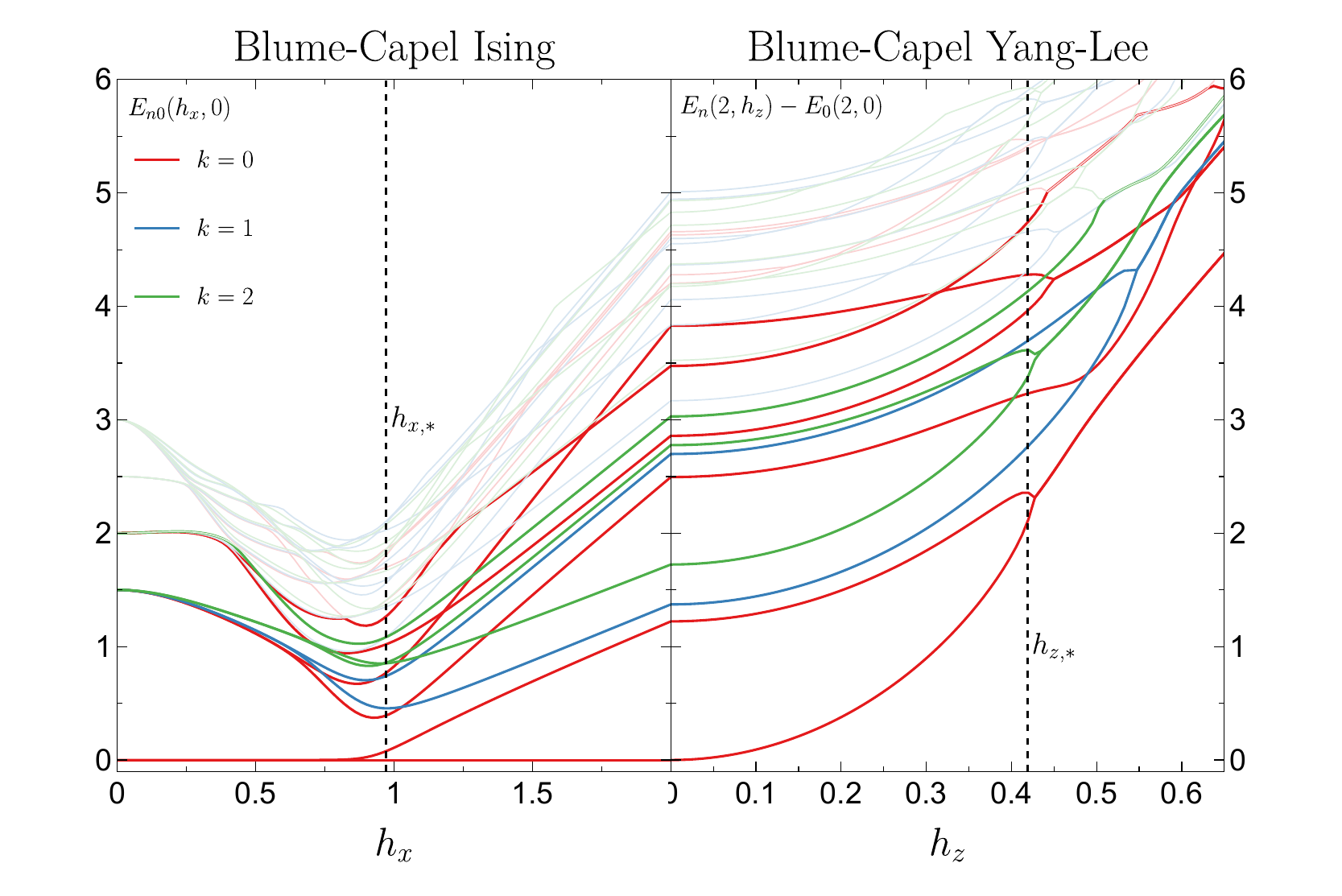}
    \caption{Spectrum of the Hamiltonian $H_\textrm{BCYL}$ in (\ref{eq_H_BCYL}), using $N=14$, $J=1$, and $\alpha=0.5$, for different spin sectors $k=0, 1, 2$. A pseudo-critical point that belongs to the Ising universality class at $h_{x,*}\approx 1.0$, and a pseudo-critical point that belongs to the Yang-Lee universality class at $h_{z,*} \approx 0.42$, for $h_{x}=2.0$ are denoted by vertical dashed lines.}
    \label{fig:BC_Spectrum}
\end{figure}

Next, we plot the real part of the low-energy spectrum $E_{n0}(h_{x}, h_{z})\equiv E_{n}-E_{0}$ of the Hamiltonian (\ref{eq_H_BCYL}) for momenta $k=0,1,2$ in Figure \ref{fig:BC_Spectrum} for the parameters $N=14$, $J=1$ and $\alpha =0.5$,  along the path with $h_{z}=0$ and $h_{x} \in [0,2]$ and then the path with $h_{x}=2$ and $h_{z}\in [0, 0.65]$.  

We see that along this path, we first cross the Ising pseudo-critical point at $h_{x,*} \approx  1.0$ and then we cross the YL pseudo-critical point at $h_{z,*} \approx 0.42$. 
Note the similarity with Figures \ref{fig:YL2_Spectrum} and \ref{fig:Spec_SYL}. Using the formulas \eqref{eq_scaling_dim_YL}, (\ref{eq_v_AFM}) and 
\begin{align}
c_{\textrm{eff}}(N)=\frac{6N(N-1)}{(2N-1)\pi v(N)} \left[(N-1)E^{k=0}_{0}(N) - NE^{k=0}_{0}(N-1) \right] \underset{N\to \infty}{\longrightarrow} c_{\textrm{eff}}\,,
    \label{eq_charge_BC}
\end{align}
we calculate the lowest scaling dimensions $\Delta_{\phi}$ and $\Delta_{\Box\phi}$, speed of light $v$ and the effective central charge $c_{\textrm{eff}}$ for fixed system sizes, together with their corresponding $N=\infty$ extrapolations using the BST algorithm with $w=2$. The results are summarized in Tables \ref{tab::scaling_dim_BC}, \ref{tab::scaling_dim_BC2} and \ref{tab::charge_BC} in Appendix \ref{AppBCYL}, where we see that the first two scaling dimensions are within $10^{-4}$ of the exact results $\Delta_{\phi}=-2/5$ and $\Delta_{\Box\phi}=8/5$, while the effective central charge is within $10^{-5}$--$10^{-3}$ of the exact value $c_{\textrm{eff}}=2/5$.

\section{Yang-Lee criticality in the three-state quantum clock model}
\label{YLCMcrit}

In this section, we consider the Yang-Lee criticality that can be reached by deforming the three-state quantum clock model (CM) \cite{Horn.PRD.1979, Ostlund.PRB.1981, Huse.PRB.1981, Ortiz:2012zz,Fendley:2012vv,Whitsitt.PRB.2018,Samajdar.PRA.2018}, defined by the Hamiltonian
\begin{align}
    H_{\textrm{CM}} = -J\sum_{n=0}^{N-1} (\sigma_{n}^{\dagger} \sigma_{n+1} +\sigma_{n} \sigma_{n+1}^{\dagger} ) -f\sum_{n=0}^{N-1}(\tau_{n}+\tau^{\dagger}_{n}) \,, \label{eq_H_CM}
\end{align}
where the  matrices $\sigma$ and $\tau$ are
\begin{align}
    \sigma = \left( 
    \begin{matrix}
        1 & 0 & 0 \\
        0 & \omega & 0 \\
        0 & 0 & \omega^2
    \end{matrix}
    \right), \quad \tau = \left( 
    \begin{matrix}
        0 & 0 & 1 \\
        1 & 0 & 0 \\
        0 & 1 & 0
    \end{matrix}
    \right), \quad \omega =\exp{(2\pi i/3)} \,,
\end{align}
with $\sigma_{N}\equiv \sigma_{0}$. These matrices satisfy the relations 
\begin{align}
    \sigma^3= \tau^3  = 1, \quad \sigma \tau = \omega \tau \sigma, \quad \sigma^{\dagger} \tau = \omega^2 \tau \sigma^{\dagger}\,.
\end{align}
We note that this model is also called the three-state Potts quantum model \cite{DotsenkoNPB.1984, Gehlen.JPA.1986, Gehlen.JPA.1987}.  It is known that this Hamiltonian is self-dual and, at $f_{c}/J=1$, has the three-state Potts critical point, which is described by the $D$-series of the $M(5,6)$ minimal model \cite{DotsenkoNPB.1984}. The Hamiltonian \eqref{eq_H_CM} has $\mathbb{Z}_{3}$ and $\mathbb{Z}_{2}$ global symmetries generated by the operators \cite{Zhuang.PRB.2015, Whitsitt.PRB.2018}:
\begin{align}
    \mathcal{Q}=\prod _{n=0}^{N-1}\tau_{n}, \quad \mathcal{C}=\prod _{n=0}^{N-1} \mathcal{C}_{n},\quad 
    \mathcal{C}_{n}= \left( 
    \begin{matrix}
        1 & 0 & 0 \\
        0 & 0 & 1 \\
        0 & 1 & 0
    \end{matrix}
    \right)\,,  \label{QCops}
\end{align}
which act on $\sigma$ and $\tau$ matrices as 
\begin{equation}
\begin{aligned}
&\mathcal{Q}^{\dag} \sigma_{n} \mathcal{Q} = \omega\sigma_{n}, \quad \mathcal{Q}^{\dag} \tau_{n} \mathcal{Q} = \tau_{n}, \quad \mathcal{Q}^3 = 1 \,, \\
&\mathcal{C} \sigma_{n} \mathcal{C} = \sigma_{n}^{\dagger}, \quad \mathcal{C} \tau_{n} \mathcal{C} = \tau_{n}^{\dagger}, \quad \mathcal{C}^2 = 1 \,.
\end{aligned}
\end{equation}
One can check that $[\mathcal{Q},H_{\textrm{CM}}]=0$, $[\mathcal{C},H_{\textrm{CM}}]=0$ and $\mathcal{C}\mathcal{Q}=\mathcal{Q}^{\dag}\mathcal{C}$. The charge conjugation symmetry $\mathcal{C}$ implies that, for every eigenstate $\ket{\psi}$ of $H_{\textrm{CM}}$ with $\mathbb{Z}_{3}$ charge $\omega$, the state $\mathcal{C}\ket{\psi}$ has the same energy and $\mathbb{Z}_{3}$ charge $\omega^{*}$.

To find the YL critical point, we deform the clock model Hamiltonian \eqref{eq_H_CM} as
\begin{align}
H_{\textrm{CMYL}} = H_{\textrm{CM}} + \frac{1}{2}h(\lambda+1)\sum_{n=0}^{N-1}\sigma_{n} + \frac{1}{2}h(\lambda-1)\sum_{n=0}^{N-1}\sigma^{\dag}_{n}\,, \label{eq_H_CM_YL}
\end{align}
where $h$ and $\lambda$ are real parameters.
Such a deformation breaks the $\mathbb{Z}_{3}$ and $\mathcal{C}$ symmetries, but preserves the combination $\mathcal{C}\mathcal{T}$, where $\mathcal{T}$ is the time-reversal operator, which acts as complex conjugation $\mathcal{T} i \mathcal{T}= -i$, $\mathcal{T} \sigma\mathcal{T} = \sigma^{\dagger}$ and $\mathcal{T} \tau \mathcal{T}  = \tau$. This symmetry is the Yang-Lee $\mathcal{PT}$ symmetry, where $\mathcal{P}$ is identified with $\mathcal{C}\,$\footnote{We note that in \cite{Wydro.IJMB.2005, Wydro.JPCS.2006}, it was argued that the YL critical point is present in the three-state clock model deformed by the term $h \sum_{n}(\sigma_{n} + \sigma_{n}^{\dag})$ with complex $h$. Such a deformation does not preserve any $\mathcal{PT}$ symmetry, since the energy levels acquire imaginary parts for arbitrarily small complex $h$.}.  We argue below, using the Landau-Ginzburg-Wilson action obtained via the Polyakov-Hubbard transformation, that for $f>f_{c}$ and $\lambda > 0$, the Hamiltonian (\ref{eq_H_CM_YL}) has a YL critical point at some value of $h > 0$.

The classical three-state clock model Hamiltonian related to (\ref{eq_H_CM_YL})  reads 
\begin{align}
 \mathcal{H} = -J \sum_{\textbf{r},\bm{\delta}}\cos\Big(\frac{2\pi}{3}(\sigma_{\textbf{r}}-\sigma_{\textbf{r}+\bm{\delta}})\Big) + \frac{1}{2}h(\lambda+1) \sum_{\textbf{r}} e^{\frac{2\pi i}{3}\sigma_{\textbf{r}}}+ \frac{1}{2}h(\lambda-1) \sum_{\textbf{r}} e^{-\frac{2\pi i}{3}\sigma_{\textbf{r}}}\,, 
\end{align}
where $\bm{\delta} = a\hat{x}, a \hat{y}$ and $\sigma_{\textbf{r}}=0,1,2$. Using the Polyakov-Hubbard transformation and  steps similar to those in Section \ref{PHtrans}, we obtain the following LGW action \cite{Zia.JPA.1975,Amit.JPA.1979, Whitsitt.PRB.2018}:
\begin{equation}
\begin{aligned}
S[\Phi_{\textbf{r}},\Phi_{\textbf{r}}^{*}] =& \frac{1}{2} \sum_{\textbf{k}\in \textrm{BZ}} \varepsilon(\textbf{k}) |\Phi_{\textbf{k}}|^2  + \sum_{\textbf{r}}\bigg(\frac{|\Phi_{\textbf{r}}|^2}{2\beta V(0)} +U[\Phi_{\textbf{r}},\Phi^{*}_{\textbf{r}}]\bigg)\,, \label{LGWCM}
\end{aligned}
\end{equation}
where $\Phi_{\textbf{r}}=\phi_{1\textbf{r}}+i \phi_{2\textbf{r}}$ is a complex field, $\varepsilon(\textbf{k})\equiv  (V^{-1}(\textbf{k}) - V^{-1}(0))/\beta$ with $V(\textbf{k})= 2J(c + \cos(k_{x}a) + \cos(k_{y}a))$, $k_{x},k_{y} \in [-\pi/a,  \pi/a]$, and the potential $U$ has the form
\begin{align}
U[\Phi,\Phi^{*}] =  - \ln \Big(\sum_{\sigma=0}^{2}\exp\Big(\frac{1}{2}(\Phi^{*}-\beta h(\lambda+1))\omega^{\sigma}+\frac{1}{2}(\Phi-\beta h(\lambda-1))\omega^{* \sigma}\Big)\Big) \,.
\end{align}
For $h=0$, the action (\ref{LGWCM}) is invariant under the $\mathbb{Z}_{3}$ transformation $\mathcal{Q}$: $\Phi \to \omega \Phi$ and charge conjugation  $\mathcal{C}:$  $\Phi \to \Phi^{*}$. Writing the potential $U$ in terms of the real fields $\phi_{1}$ and $\phi_{2}$, we find 
\begin{equation}
\begin{aligned}
U[\phi_{1},\phi_{2}] &=  - \ln \Big(e^{\phi_{1}-\beta h \lambda} + 2 e^{\frac{1}{2}(\beta h \lambda -\phi_{1})}\cosh\Big(\frac{\sqrt{3}}{2}(\phi_{2}-i\beta h)\Big)\Big) \,,
\end{aligned}
\end{equation}
and we see that, for $h >0$, the action (\ref{LGWCM}) has the $\mathcal{PT}$ symmetry: $\phi_{2} \to -\phi_{2}$ and $i \to -i$.
For $\beta h \lambda > 0$ and sufficiently large $h$,  we can drop the first term under the logarithm and obtain  
\begin{equation}
\begin{aligned}
U[\phi_{1},\phi_{2}] &\approx \frac{1}{2}\phi_{1}  - \ln \Big(\cosh\Big(\frac{\sqrt{3}}{2}(\phi_{2}-i\beta h)\Big)\Big) \,,\label{Vappr1}
\end{aligned} 
\end{equation}
where we have omitted unimportant constants.  In this case, the YL field $\phi_{2}$ fully decouples from the massive field $\phi_{1}$, and its action coincides with the action (\ref{Sphi3micro}) for the classical Ising model in an imaginary magnetic field.  We show below that, in the numerical spectrum of the Hamiltonian (\ref{eq_H_CM_YL}), we indeed find massless YL states together with massive states when $h \lambda > 0$.  We note that,  for $\beta h \lambda < 0$ and sufficiently large $h$, the potential is simply
\begin{equation}
\begin{aligned}
U[\phi_{1},\phi_{2}] &=  - \phi_{1}\,,
\end{aligned}
\end{equation}
and the YL field $\phi_{2}$ is suppressed. In this case, we do not observe a YL critical point numerically. In general, the larger $\lambda$ is, the easier it is to locate the YL critical point, since the approximation (\ref{Vappr1}) becomes more accurate.

\begin{figure}[h]
    \centering
    \includegraphics[width=15cm]{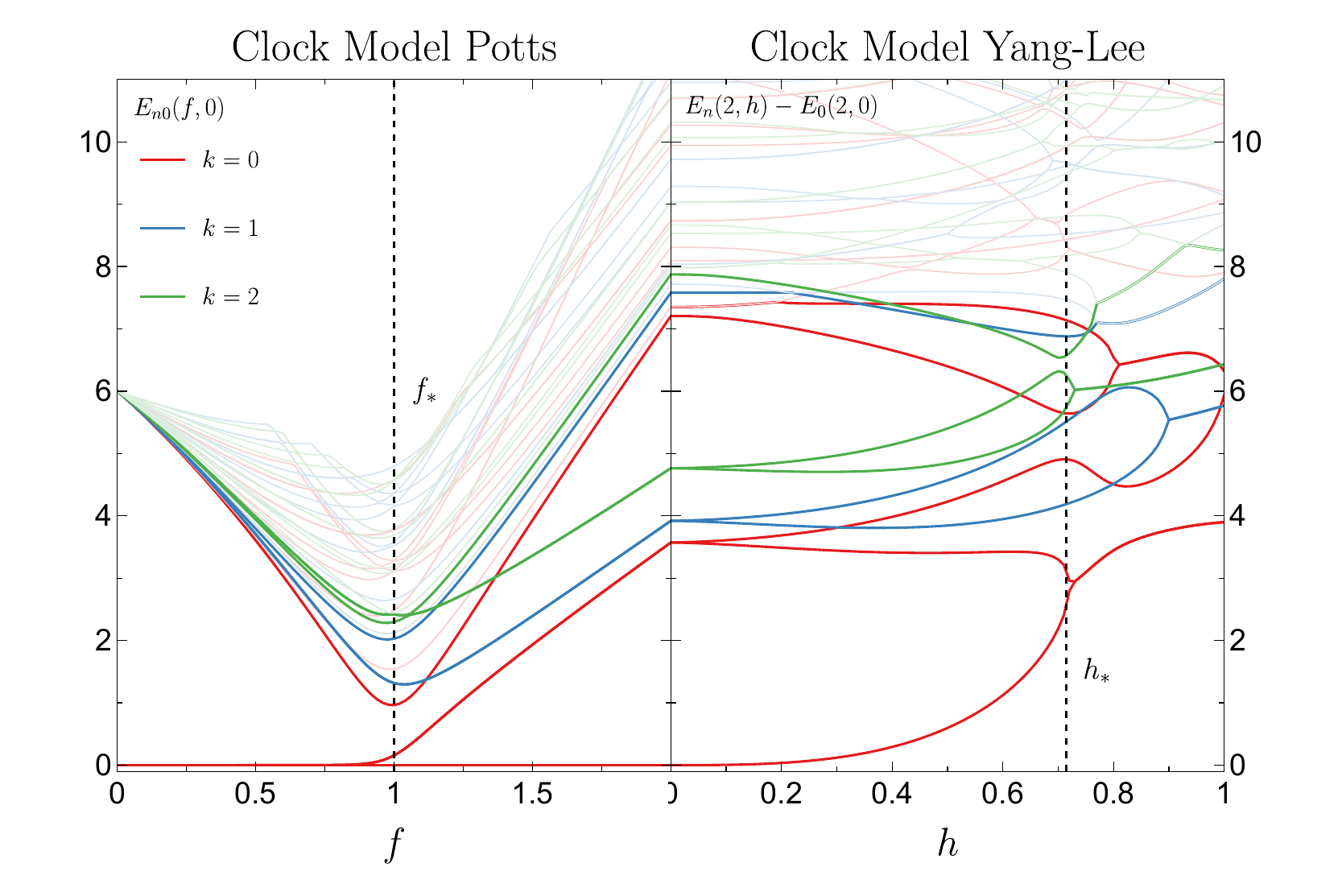}
    \caption{Spectrum of the Hamiltonian $H_\textrm{CMYL}$ in (\ref{eq_H_CM_YL}), computed for $N=14$, $J=1$, and $\lambda=1$, in different spin sectors $k=0, 1, 2$. A pseudo-critical point in the three-state Potts universality class at $f_{*}=1.0$, and a pseudo-critical point in the Yang-Lee universality class at $h_{*} \approx 0.71$ for $f=2.0$ are shown by vertical dashed lines. In addition to the Yang-Lee states,
    we observe massive states.
    }
    \label{fig:w2_breaking}
\end{figure}

Below, we confirm the above expectations numerically. 
We start by plotting the real part of the low-energy spectrum $E_{n0}(f, h)\equiv E_{n}-E_{0}$ of the Hamiltonian (\ref{eq_H_CM_YL}) for momenta $k=0,1,2$ in Figure \ref{fig:w2_breaking}. We use the parameters $N=14$, $J=1$ and $\lambda =1$, and follow the path with $h=0$ and $f \in [0,2]$ and then the path with $f=2$ and $h\in [0, 1]$.  
We see that along this path, we first cross the three-state Potts pseudo-critical point at $f_{*} =  1.0$ and then we cross the YL pseudo-critical point at $h_{*} \approx 0.7144$. In addition to the Yang-Lee states along the path with $f=2$ and $h\in [0,1]$, we observe massive states. We analyze these states  more carefully below. In Table \ref{table:PottsYLflow}, we summarize the flow of operators from the $D$-series of $M(5,6)$  to $M(2,5)$, motivated by Figure \ref{fig:w2_breaking}. 
\begin{table}[ht]
\centering
\begin{tabular}{|c|c|}
\hline
Operators in $M(5,6)$ & Operators in $M(2,5)$ \\
\hline
$I$ ($\Delta=0$) & $\phi$ ($\Delta=-2/5$) \\
$\sigma-\sigma^\dagger$ ($\Delta=2/15$) & $I$ ($\Delta=0$) \\
$\epsilon$ ($\Delta=4/5$) &  $\Box\phi$ ($\Delta=8/5$) \\
\hline
\end{tabular}
\caption{A flow of the lowest operators from the three-state Potts critical point, described by the $D$-series of the $M(5,6)$ minimal model,  to the Yang-Lee critical point, described by the $M(2,5)$ minimal model.}
\label{table:PottsYLflow}
\end{table}

\begin{figure}[h]
    \centering
    \includegraphics[width=4.5cm]{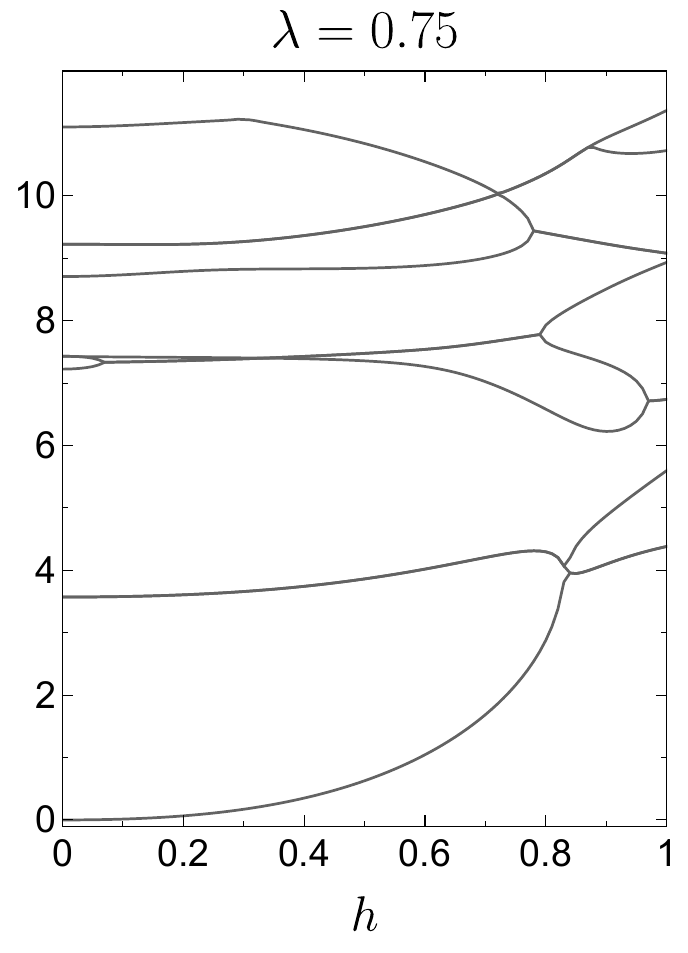}
    \includegraphics[width=4.5cm]{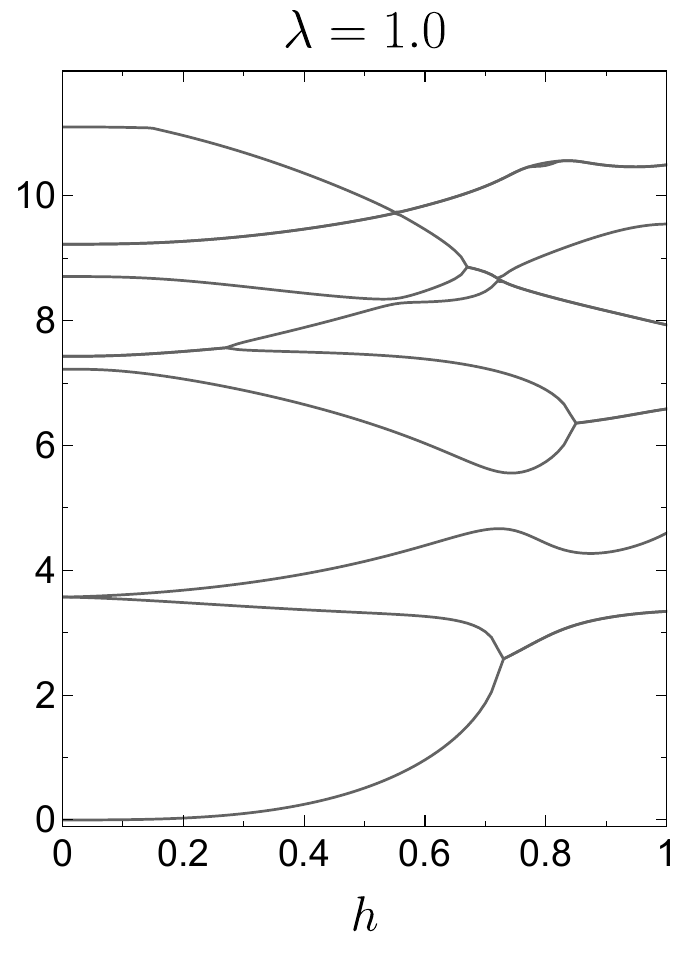}
    \includegraphics[width=4.5cm]{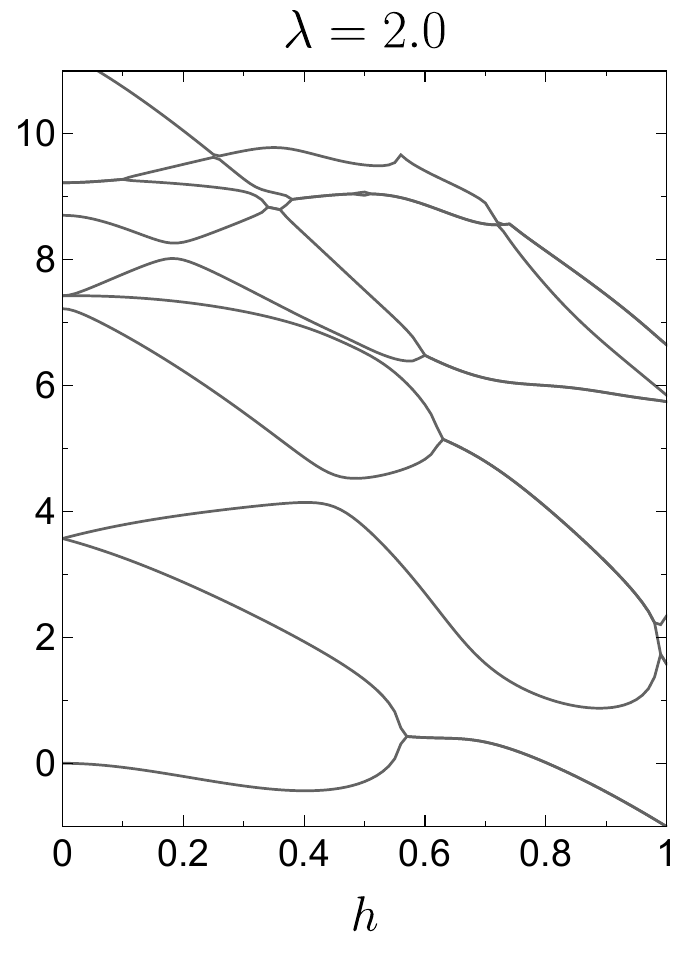}
    \caption{Spectrum of  the Hamiltonian $H_\textrm{CMYL}$ in (\ref{eq_H_CM_YL}) as a function of $h$ in the spin sector $k=0$ for  $\lambda =0.75, 1, 2$ and $N=14$, $J=1$, $f=2$. 
    }
    \label{fig:gamma_deform}
\end{figure}

We also plot the spectrum of the Hamiltonian $H_\textrm{CMYL}$ in (\ref{eq_H_CM_YL}) along the path $h\in[0,1]$ for $N=14$, $J=1$, $f=2$, and different values of $\lambda$ in Figure \ref{fig:gamma_deform}.

We define the YL pseudo-critical point $h_{*}(N)$ as the point at which the condition 
\begin{align}
\frac{(N+1)E_{10}^{k=0}(h_{*}(N),N+1)}{N E_{10}^{k=0}(h_{*}(N), N)}=1\,. \label{eq_FSS_CM}
\end{align}
is satisfied. We present the results of the pseudo-critical point $h_{*}(N)$ for different values of $N$ using $J=1$, $f=2$, and $\lambda=1$, in Table \ref{tab::crit_pt_locations_CM} in Appendix \ref{AppCMYL}. We solve (\ref{eq_FSS_CM}) with accuracy $10^{-7}$. Using the BST algorithm with $w=2$, we obtain the thermodynamic value of the critical point,  $h_{c}= 0.71406$.

\begin{figure}[h]
    \centering
    \includegraphics[width=10cm]{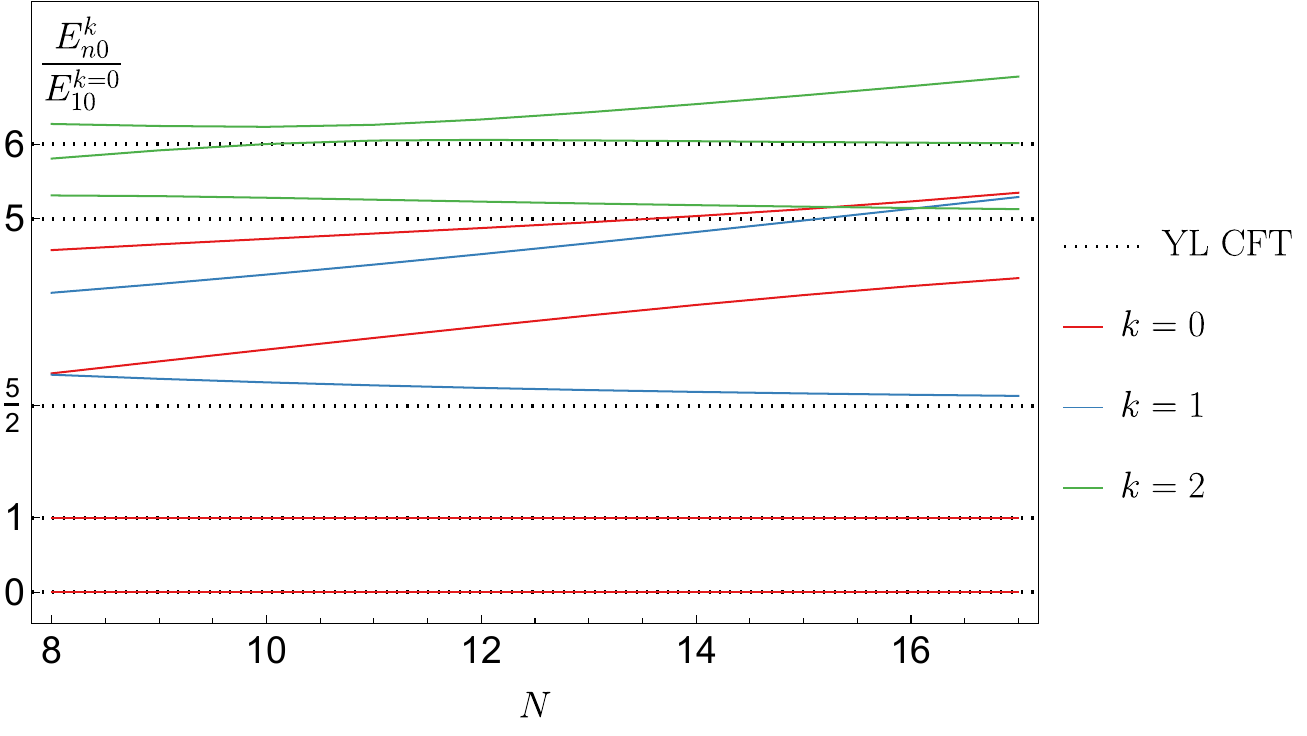}
    \caption{Dependence of the energy-gap ratios $E^{k}_{n0}/E^{k=0}_{10}$ on $N$ for the first five excited states of the Hamiltonian $H_\textrm{CMYL}$ in (\ref{eq_H_CM_YL}) at the pseudo-critical points listed in Table \ref{tab::crit_pt_locations_CM}, with $J=1$, $f=2$, and $\lambda=1$. Black dotted horizontal lines mark the YL CFT energy ratios.}
    \label{fig:gaps_vs_N}
\end{figure}

To see how the massive states flow in the thermodynamic limit, we study the behavior of the energy gaps $E_{n0}$ as a function of $N$. In particular, if a state belongs to the YL CFT, we expect that $N E_{n0} \to \textrm{const.}$, whereas the massive states will grow with $N$. We show the ratios $E_{n0}^{k}/E^{k=0}_{10}$ as functions of $N$ in Figure \ref{fig:gaps_vs_N} for the first five excited states at the pseudo-critical points listed in Table \ref{tab::crit_pt_locations_CM}. 
We indeed observe that, whereas some of the lines converge to the constant values predicted by the YL CFT, the remaining lines grow with $N$ and thus represent massive states.

Using the formulas \eqref{eq_scaling_dim_YL}, (\ref{eq_v_AFM}) and (\ref{eq_charge_BC}) for the  gapless states, we calculate the lowest scaling dimensions $\Delta_{\phi}$ and $\Delta_{\partial^2\phi}$, the speed of light $v$, and the effective central charge $c_{\textrm{eff}}$ for fixed system sizes, together with their corresponding $N=\infty$ extrapolations using the BST algorithm with $w=2$. The results are summarized in Tables \ref{tab::scaling_dim_CM}, \ref{tab::charge_CM} in Appendix  \ref{AppCMYL}. 

We see that the first spin-$0$ scaling dimension corresponds to $\Delta_{\phi}$ in the YL CFT, with a difference of $10^{-4}$ from the exact value $\Delta_{\phi}=-2/5$. The next spin-2 scaling dimension corresponds to  $\Delta_{\partial^2\phi}$, with an error of $10^{-3}$  from the exact result $\Delta_{\partial^2\phi}=8/5$. The accuracy of the higher scaling dimensions is lower because we use a small window of system sizes. To obtain more precise values, one has to increase the range of $N$. The result for the effective central charge in Table \ref{tab::charge_CM} is $10^{-3}$ away from the exact value $c_{\textrm{eff}}=2/5$. Finally, in Table \ref{tab::scaling_dims_lambda}, we show numerical results for the first scaling dimension $\Delta_{0}^{k=0}$ for $\lambda \in [0.9,3]$ extrapolated to $N=\infty$\;\footnote{For $\lambda \neq 1$, we locate the pseudo-critical points using  \eqref{eq_FSS_BC} with a precision of $10^{-5}$.}. We observe a good agreement with the exact result $\Delta_{\phi}=-2/5$ in the YL CFT.
\begin{table}[h]
  \centering
  \begin{tabular}{c}
      \hline\hline
  \begin{tabular}{c|c|c|c|c|c}
         $\lambda$& $0.9$ & $1.0$ & $1.5$ & $ 2.0$ & $3.0$\\ 
        \hline 
        $\Delta_{0}^{k=0}$ & $-0.39966$ & $-0.40099 $ & $-0.40033$ & $-0.40029$ & $-0.40033$ \\
        \hline
  \end{tabular}
  \end{tabular}
  \caption{Extrapolated to $N=\infty$ values of $\Delta_{0}^{k=0}$ for various $\lambda$ and $J=1$, $f=2$.}
  \label{tab::scaling_dims_lambda}
\end{table}

\section{Ising criticality in the three-state quantum clock model}
\label{IsCMcrit}

In this section, we consider a deformation of the three-state quantum clock model that 
has an Ising critical point:
\begin{align}
H_{\textrm{CMI}} = H_{\textrm{CM}} + h \sum_{n=1}^{N} \Big(\sigma_{n}\sigma_{n+1} +\sigma^{\dagger}_{n}\sigma^{\dagger}_{n+1} \Big) \,, \label{eq_H_CM_Ising}
\end{align}
where $H_{\textrm{CM}}$ is given in (\ref{eq_H_CM}) and  $h > 0$ is a real parameter. This deformation breaks the $\mathbb{Z}_{3}$ symmetry generated by the operator $\mathcal{Q}$, but preserves $\mathbbm{Z}_{2}$ symmetry generated by the charge conjugation operator $\mathcal{C}$ in (\ref{QCops}).

To gain a better picture of the behavior of the quantum model (\ref{eq_H_CM_Ising}), we consider the related classical Hamiltonian:
\begin{align}
 \mathcal{H} = -J \sum_{\textbf{r},\bm{\delta}}\cos\Big(\frac{2\pi}{3}(\sigma_{\textbf{r}}-\sigma_{\textbf{r}+\bm{\delta}})\Big) + h \sum_{\textbf{r}}\cos\Big(\frac{2\pi}{3}(\sigma_{\textbf{r}}+\sigma_{\textbf{r}+a\hat{x}})\Big)\,, 
\end{align}
where $\bm{\delta} = a\hat{x}, a\hat{y}$ and $\sigma_{\textbf{r}}=0,1,2$. The  partition function can be written as 
\begin{align}
Z = \sum_{\{ \sigma_{\textbf{r}} = 0,1,2\}} \exp\Big(\frac{1}{2}\sum_{\textbf{r},\textbf{r}' }\beta (V_{1,\textbf{r}\textbf{r}'} n_{1\textbf{r}}n_{1\textbf{r}'}+ V_{2,\textbf{r}\textbf{r}'} n_{2\textbf{r}}n_{2\textbf{r}'})\Big)\,,
\end{align}
where $n_{1\textbf{r}} \equiv \cos(\frac{2\pi}{3}\sigma_{\textbf{r}})$,  $n_{2\textbf{r}}\equiv \sin(\frac{2\pi}{3}\sigma_{\textbf{r}})$ and the potentials are 
\begin{equation}
\begin{aligned}
&V_{1,\textbf{r}\textbf{r}'} = (J-2h)(\delta_{\textbf{r},\textbf{r}'+a\hat{x}} +
\delta_{\textbf{r},\textbf{r}'-a\hat{x}}) + J(\delta_{\textbf{r},\textbf{r}'+a\hat{y}} +
\delta_{\textbf{r},\textbf{r}'-a\hat{y}})\,, \\
&V_{2,\textbf{r}\textbf{r}'} = (J+2h)(\delta_{\textbf{r},\textbf{r}'+a\hat{x}} +
\delta_{\textbf{r},\textbf{r}'-a\hat{x}}) + J(\delta_{\textbf{r},\textbf{r}'+a\hat{y}} +
\delta_{\textbf{r},\textbf{r}'-a\hat{y}})\,.
\end{aligned}
\end{equation}
Using the Polyakov-Hubbard transformation and  steps similar to those in Sections \ref{PHtrans} and \ref{YLCMcrit}, we obtain the following LGW action:
\begin{equation}
\begin{aligned}
S[\phi_{1\textbf{r}},\phi_{2\textbf{r}}] =& \frac{1}{2} \sum_{\textbf{k}\in \textrm{BZ}} (\varepsilon_{1}(\textbf{k}) |\phi_{1\textbf{k}}|^2 + \varepsilon_{2}(\textbf{k}) |\phi_{2\textbf{k}}|^2)  + \sum_{\textbf{r}}\bigg(\frac{\phi_{1\textbf{r}}^2}{2\beta V_{1}(0)} +
\frac{\phi_{2\textbf{r}}^2}{2\beta V_{2}(0)} \\
&- \ln \Big(e^{\phi_{1\textbf{r}}} + 2 e^{-\frac{1}{2}\phi_{1\textbf{r}}}\cosh\Big(\frac{\sqrt{3}}{2}\phi_{2\textbf{r}}\Big)\Big)\bigg)\,, \label{LGWCMI}
\end{aligned}
\end{equation}
where $\varepsilon_{\alpha}(\textbf{k})\equiv  (V_{\alpha}^{-1}(\textbf{k}) - V_{\alpha}^{-1}(0))/\beta$ and $V_{\alpha}(\textbf{k})= 2(cJ+ (J +(-1)^{\alpha}h)\cos(k_{x}a) + J\cos(k_{y}a))$ with $\alpha = 1,2$.
Because of the unequal $\phi_{1\textbf{r}}^2$ and $\phi_{2\textbf{r}}^2$ terms, the action (\ref{LGWCMI}) does not have $\mathbbm{Z}_{3}$ symmetry $\Phi_{\textbf{r}} \to \omega \Phi_{\textbf{r}}$, where $\Phi_{\textbf{r}}=\phi_{1\textbf{r}}+i \phi_{2\textbf{r}}$, but it has a $\mathbbm{Z}_{2}$ symmetry $\phi_{2\textbf{r}} \to -\phi_{2\textbf{r}}$. As the positive deformation parameter $h$ in (\ref{LGWCMI}) is increased, the UV mass of the field $\phi_{2\textbf{r}}$ decreases, while that of $\phi_{1\textbf{r}}$ increases.
Therefore, we expect that, at some value of $h$ we reach Ising criticality for the field $\phi_{2\textbf{r}}$, while the field $\phi_{1\textbf{r}}$ remains massive. Below, we confirm this expectation numerically by analyzing the spectrum of the Hamiltonian (\ref{eq_H_CM_Ising}). At a critical value of $h$, we indeed find massless states corresponding to the Ising critical point, together with massive states.

\begin{figure}[h]
    \centering
    \includegraphics[width=14cm]{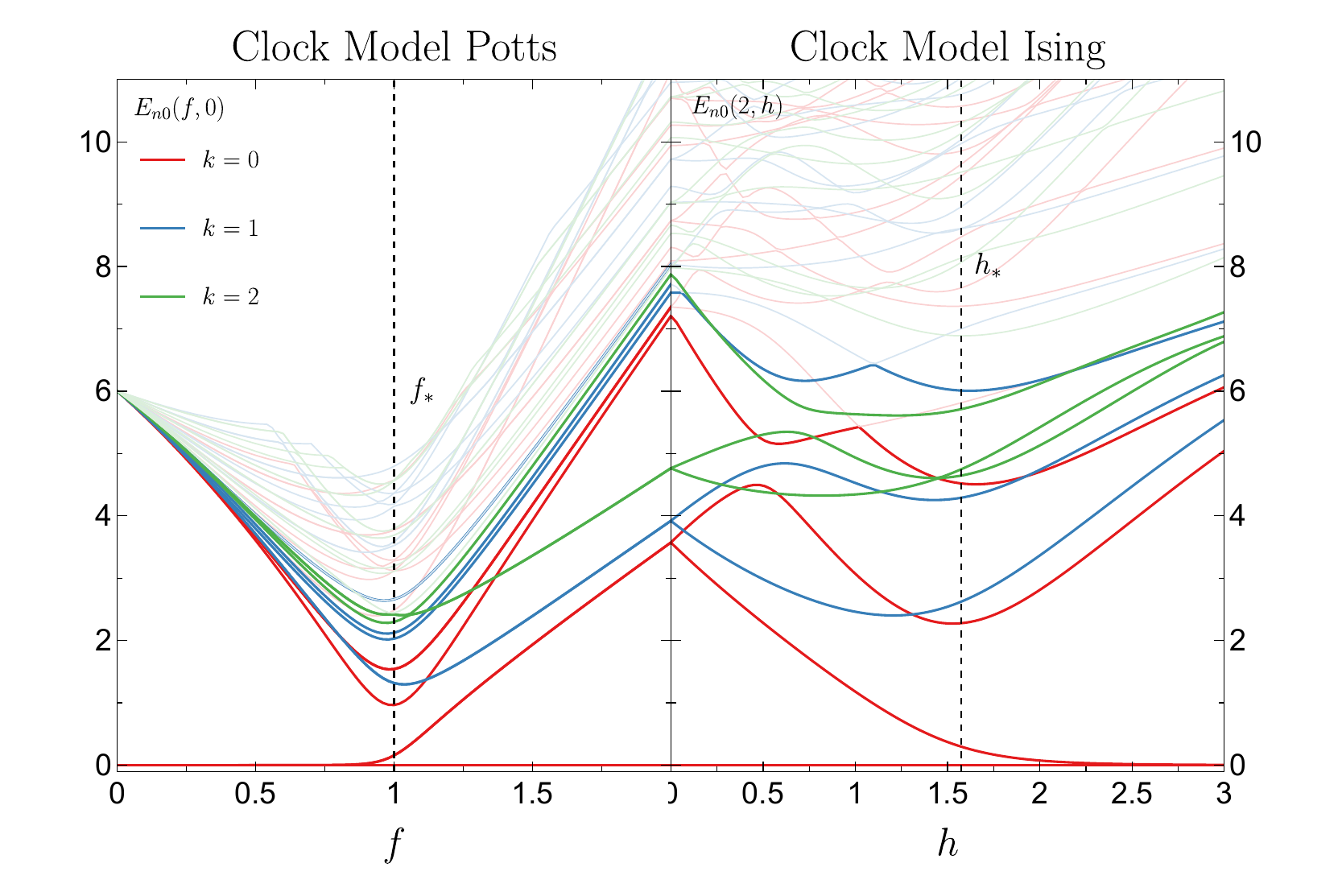}
    \caption{Spectrum of the Hamiltonian $H_\textrm{CMI}$ in (\ref{eq_H_CM_Ising}), computed for $N=14$, $J=1$, in different spin sectors $k=0, 1, 2$. A pseudo-critical point in the three-state Potts universality class at $f_{*}=1.0$ for $h=0$, and a pseudo-critical point in the Ising universality class at $h_{*} \approx 1.57$ for $f=2.0$ are shown by vertical dashed lines. In addition to the Ising states,
    we observe massive states.}
    \label{fig:h_breaking_Ising}
\end{figure}

We present the low-energy spectrum $E_{n0}(f, h)\equiv E_{n}-E_{0}$ of the Hamiltonian (\ref{eq_H_CM_Ising}) for momenta $k=0,1,2$ in Figure \ref{fig:h_breaking_Ising}, using the parameters $N=14$, $J=1$, and along the path with $h=0$ and $f \in [0,2]$, followed by the path with $f=2$ and $h\in [0, 3]$.  
We see that along this path, we first cross the three-state Potts pseudo-critical point at $f_{*} =  1.0$ and then we cross the Ising pseudo-critical point at $h_{*} \approx 1.57$. In addition to the Ising states along the path with $f=2$ and $h\in [0,3]$, we observe massive states. We further analyze these states more closely below.

We first locate the Ising critical point using \eqref{eq_FSS_CM} for different values of $N$. We present the results in Table \ref{tab::crit_pt_locations_Ising} in Appendix \ref{AppCMI}.
Then, we can calculate the scaling dimensions using the formula 
\begin{align}
    \Delta^{k}_{n}(N)=\frac{E^{k}_{n}(N)-E_{0}^{k=0}(N)}{E^{k=1}_{0}(N)-E^{k=0}_{1}(N)} \underset{N\to \infty}{\longrightarrow} \frac{\Delta^{k}_{n}-\Delta_{I}}{\Delta_{\partial\sigma}-\Delta_{\sigma}}=\Delta^{k}_{n}\,,
    \label{eq_scaling_dim_Ising}
\end{align}
where $E^{k}_{n}(N)$ is the $n$th  gapless energy level with momentum $k$ for system size $N$, and $\Delta^{k}_{n}$ is the $n$th scaling dimension of the Ising CFT  with spin $k$. The results are shown in Table \ref{tab::scaling_dim_Ising} in Appendix \ref{AppCMI}.  From this table, we see that the first two scaling dimensions correspond to those of the Ising operators $\sigma$ and $\varepsilon$, with errors of $10^{-6}$ and $10^{-5}$  from the exact Ising CFT ($M(3,4)$ minimal model) values $\Delta_{\sigma}=1/8$ and $\Delta_{\varepsilon} = 1$, respectively. The other scaling dimensions correspond to the operators $\partial\varepsilon$, $\Box \sigma$, $T$, and $\partial^2 \sigma$, with errors of $10^{-5}-10^{-4}$ from the theoretical values $\Delta_{\partial\varepsilon}=2$, $\Delta_{\Box\sigma}=17/8$, $\Delta_{T}=2$, and $\Delta_{\partial^2\sigma} = 17/8$, respectively. Similarly, we calculate the speed of light using
\begin{align}
v(N) = \frac{N}{2\pi} (E_{0}^{k=1} - E_{1}^{k=0})\,, \label{eq_v_Ising}
\end{align}
and the central charge using  \eqref{eq_charge_BC}.
The results are given in Table \ref{tab::charge_CM_Ising} in Appendix \ref{AppCMI}. We  confirm that the gapless sector has a critical point that belongs to the Ising universality class, with $c$ within $10^{-4}$ of the theoretical value $c=1/2$.

\begin{figure}[h]
    \centering
    \includegraphics[width=11cm]{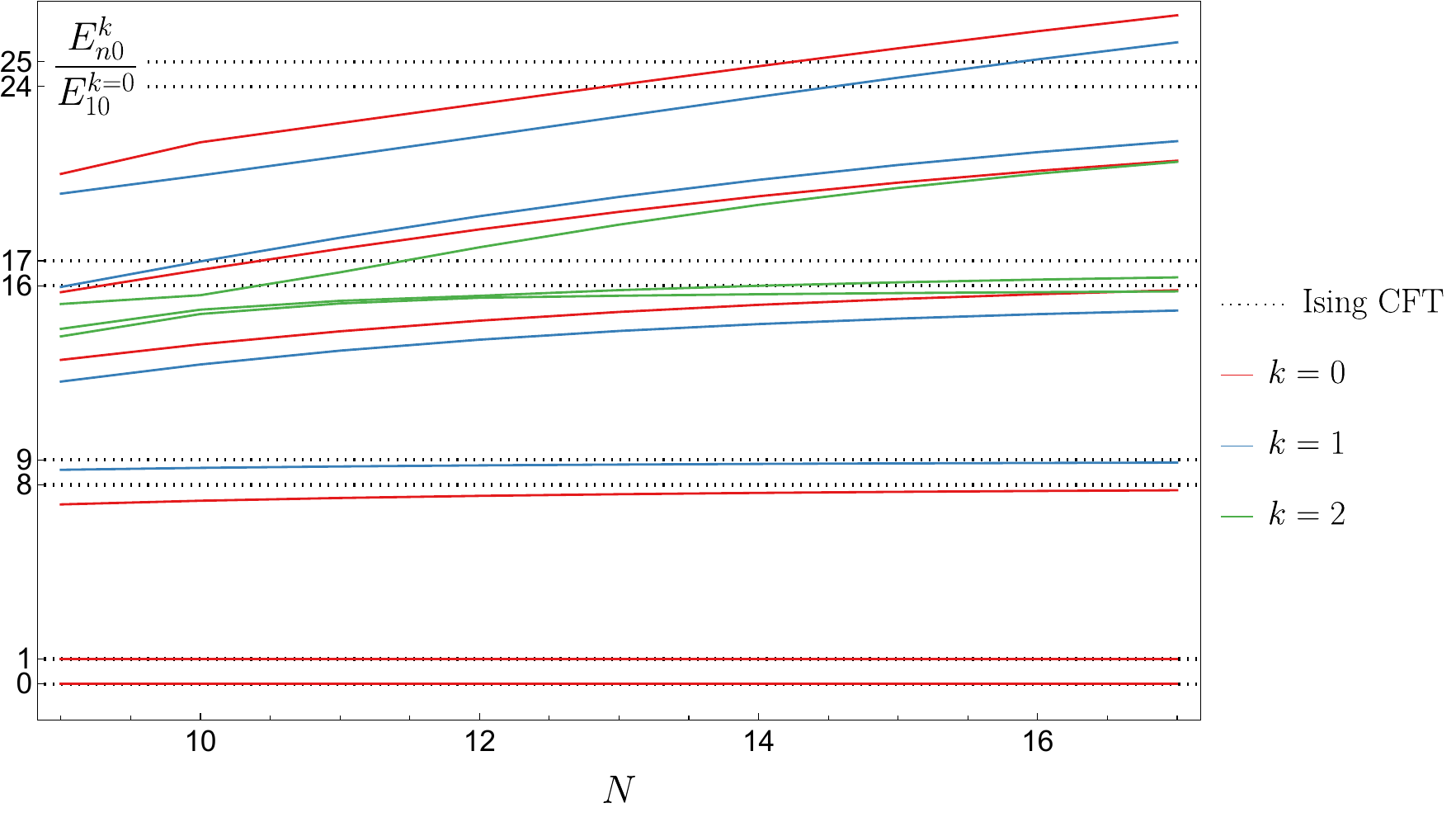}
    \caption{Dependence of the energy-gap ratios $E^{k}_{n0}/E^{k=0}_{10}$ on $N$ for the first seven excited states of the Hamiltonian $H_\textrm{CMI}$ in \eqref{eq_H_CM_Ising} at the pseudo-critical points listed in Table \ref{tab::crit_pt_locations_Ising}, with $J=1$ and $f=2$. Black dotted horizontal lines mark the Ising CFT energy ratios.}
    \label{fig:gaps_vs_N_Ising}
\end{figure}

As expected from the Landau-Ginzburg-Wilson description in (\ref{LGWCMI}), we should encounter additional states that are gapped in the thermodynamic limit and grow linearly with the system size $N$.
In Figure \ref{fig:gaps_vs_N_Ising}, we plot the dependence of the energy ratios $E_{n0}^{k}/E_{10}^{k=0}$ on the system size $N$ for the lowest energy states. 
We see that some states indeed grow linearly with $N$, while the others converge to the expected ratios of the Ising CFT scaling dimensions.  Since there are no symmetries that distinguish the massive states from the Ising CFT states, they can not cross each other. Therefore, the massive states collide with the CFT states through avoided crossings.

\begin{figure}[h]
    \centering
    \includegraphics[width=15cm]{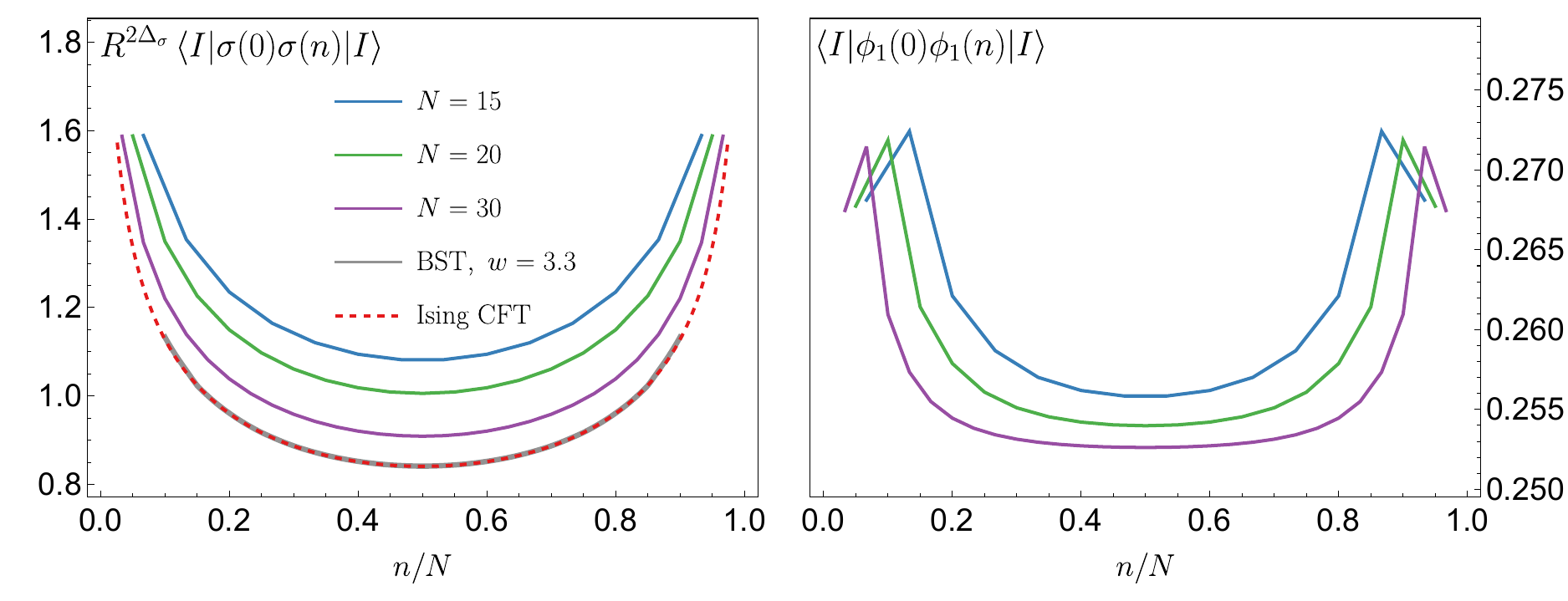}
    \caption{Numerical results for the correlation functions (\ref{sCFTsCFT}) and (\ref{eq_correl_p1p1}) for various system sizes $N$, computed for $J=1$, $f=2$, and pseudo-critical points $h=h_{*}(N)$. We see that correlators show the expected power-law and exponential decays, respectively.}
    \label{fig:correlations}
\end{figure}

Based on symmetry properties, we expect that the massless Ising field $\phi_{2\textbf{r}}$ in the classical action (\ref{LGWCMI}) corresponds to the combination $i(\sigma^{\dagger}_{n} - \sigma_{n})$ of lattice operators. Therefore, we have
\begin{align}
\phi_{2\textbf{r}} \leftrightarrow i(\sigma^{\dagger}_{n} - \sigma_{n}) = a \sigma(n)+\dots\,,
\end{align}
where $\sigma(x)$ is the $\mathbbm{Z}_{2}$-odd field of the Ising CFT on a cylinder, with scaling dimension $\Delta_{\sigma}=1/8$. In turn, we expect that the massive field $\phi_{1\textbf{r}}$ corresponds to the combination $\sigma^{\dagger}_{n} +\sigma_{n}$ on the lattice:
\begin{align}
\phi_{1\textbf{r}} \leftrightarrow \phi_{1}(n)=\sigma^{\dagger}_{n} +\sigma_{n} \,.
\end{align}
The CFT  two-point function of $\sigma(x)$ on a cylinder has the form \cite{Belavin:1984vu}
\begin{align}
\langle I| \sigma(0) \sigma(n) |I\rangle &= R^{-2\Delta_{\sigma}} \left|2\sin{\left( \frac{\pi}{N}n\right)} \right|^{-2\Delta_{\sigma}} \,,  \label{eq_correl_p2p2}
\end{align}
and we can compute it numerically as: 
\begin{align}
R^{2\Delta_{\sigma}}\langle I| \sigma(0) \sigma(n) |I\rangle &  = \lim_{N \to \infty}\frac{\bra{I} i(\sigma^{\dagger}_{0} - \sigma_{0}) i(\sigma^{\dagger}_{n} - \sigma_{n})\ket{I} }{\langle I | i(\sigma^{\dagger} - \sigma)|\sigma \rangle^2}\,.  \label{sCFTsCFT}
\end{align}
By contrast, for the two-point correlation function of the massive operator $\phi_{1}(n)=\sigma^{\dagger}_{n} +\sigma_{n}$, we expect to find an exponential decay
\begin{align}
\langle I| \phi_{1}(0) \phi_{1}(n)|I\rangle \propto c (e^{-n/\xi} + e^{-(N-n)/\xi})\,. 
\label{eq_correl_p1p1}
\end{align}
We show the numerical results for these correlation functions in Figure \ref{fig:correlations}, where one can see that, with increasing $N$, the correlators approach the expected shapes in \eqref{eq_correl_p2p2}  and \eqref{eq_correl_p1p1}. 
Furthermore, in gray, we present the extrapolation to the thermodynamic limit $N\to \infty$ using the BST algorithm with $w = 3.3$.
We  see that this curve overlaps with the theoretical expression in (\ref{eq_correl_p2p2}), showing clear agreement with our expectations.

\section{Concluding Remarks}
In this article, we analyze the Ising and Yang-Lee criticalities in four different quantum models in $1+1$D: the antiferromagnetic quantum Ising chain, the Schwinger model, the spin-1 Blume-Capel model, and the three-state quantum clock model. For all of these models, we derive the corresponding Landau-Ginzburg-Wilson (LGW) theory, obtained using bosonization for the Schwinger model and the Polyakov-Hubbard transformation for the other models. We show that our numerical results are in full agreement with the theoretical expectations obtained from the analysis of the LGW actions. In particular, we show that the Yang-Lee criticality in all of these quantum models is described by the $\mathcal{PT}$-symmetric massless $i\phi^3$ theory. Moreover, for the three-state quantum clock model, we observe massive states in the spectrum in accordance with the LGW action.  In addition,  using the ferromagnetic quantum Ising spin chain at the Yang-Lee critical point, we numerically compute the two-point functions of $\phi$ in the identity state $|I\rangle$ and the ground state $|\phi\rangle$. We show that the numerical results are in complete agreement with the CFT predictions. In the spin-1 Blume-Capel model, we confirm the existence of the YL edge line. It would be interesting to understand a relation between the YL edge line and the Yang-Lee zeros. The Yang-Lee zeros of the Blume-Capel model were studied in \cite{Lee.PRL.1994, Biskup.PRL.2000}, but only in the ferromagnetic phase of the BC model. It would be interesting to better understand their structure in the paramagnetic phase\footnote{We note that the relation between the Yang-Lee zeros and quantum phase transitions in different systems has recently been studied in \cite{Hongchao.PRL.2023, Hongchao.PRB.2025}.}.

It was conjectured in \cite{Gehlen.JMPB.1994} that points on this line are described by the $M(2,7)$ minimal model. It would be interesting to investigate this further.  
It would also be interesting to analyze some of these quantum models in $2+1$D using the fuzzy sphere construction \cite{Zhu:2022gjc, Hu.PRL.2023, Hu.PRB.2025, Fardelli.SP.2025, Lauchli:2025fii, ArguelloCruz:2025zuq} and to observe the Yang-Lee phase transition under similar deformations.

\section*{Acknowledgements}
We are grateful to Shubhayu Chatterjee, Ludo Fraser-Taliente, Igor Klebanov, Vlad Kozii, Margit Lin, and Yuan Xin for many very useful discussions. We also thank
Igor Klebanov for valuable comments on the draft. 
The work of EAC, GT was supported in part by the DOE under Grant No.~DE-SC0010118 and by the 
Simons Foundation under Grant No. 994316.

\appendix

\section{Numerical results for the YL critical point in antiferromagnetic quantum Ising model}
\label{AppAFMYL}

\begin{table}[H]
  \centering
  \begin{tabular}{c}
      \hline\hline
  \begin{tabular}{c|c|c|c|c}
    $N$ & $\Delta_{0}^{k=0}(N)$ & $\Delta_{2}^{k=0}(N)$ & $\Delta_{3}^{k=0}(N)$ & $\Delta_{4}^{k=0}(N)$ \\
    \hline
    $6$       & $-0.375545$ & $1.17143$   & $1.48037$  & $2.13609$ \\
    \hline
    $8$       & $-0.38582$  & $1.30407$   & $1.9738$   & $2.69199$ \\
    \hline
    $10$      & $-0.390724$ & $1.38303$   & $2.47151$  & $3.06374$ \\
    \hline
    $12$      & $-0.393545$ & $1.43585$   & $2.89575$  & $3.25866$ \\
    \hline
    $14$      & $-0.395304$ & $1.47252$   & $3.094$    & $3.43256$ \\
    \hline
    $16$      & $-0.396472$ & $1.49869$   & $3.18385$  & $3.53553$ \\
    \hline
    $18$      & $-0.397271$ & $1.51789$   & $3.247$    & $3.61641$ \\
    \hline
    $20$      & $-0.397845$ & $1.53228$   & $3.29679$  & $3.67905$ \\
    \hline \hline
    $\infty$  & $-0.400304$ & $1.600844$ & $3.571439$ & $3.97963$ \\
    \hline\hline
  \end{tabular}
\end{tabular}
  \caption{Scaling dimensions in $k=0$ sector and at the corresponding pseudo-critical points  of $H_\textrm{AFYL}$ (\ref{AYLHam1}) in Table \ref{tab::crit_pt_locations}, using $J=1$, $h_{x}=2.5$,  $h_{z}=1$.}
  \label{tab::scaling_dim}
\end{table}

\begin{table}[H]
  \centering
  \begin{tabular}{c}
      \hline\hline
  \begin{tabular}{c|c|c}
        $N$ & $v(N)$ & $c_{\textrm{eff}}(N)$ \\ 
        \hline 
        $4$ & $ 2.853238 $ & $ - $ \\
        \hline
        $6$ & $ 3.133181 $ & $ 0.3170004 $ \\
        \hline 
        $8$ & $ 3.161899 $ & $ 0.310329 $ \\
        \hline 
        $10$ & $ 3.182808 $ & $ 0.322839 $ \\
        \hline 
        $12$ & $ 3.196119 $ & $ 0.336297 $ \\
        \hline 
        $14$ & $ 3.205016 $ & $ 0.347738 $ \\
        \hline 
        $16$ & $ 3.211283 $ & $ 0.356510 $ \\
        \hline 
        $18$ & $ 3.215850 $ & $ 0.364088 $ \\
        \hline
        $20$ & $ 3.219315 $ & $ 0.369383 $ \\
        \hline \hline
        $\infty$ & $ 3.237124 $ & $ 0.399646 $ \\
        \hline
    \end{tabular}
  \end{tabular}
  \caption{Speed of light $v$ and  effective central charge $c_{\textrm{eff}}$ for different $N$, computed from \eqref{eq_charge_AFM} and \eqref{eq_v_AFM} at the pseudo-critical points of $H_\textrm{AFYL}$ (\ref{AYLHam1}) in Table \ref{tab::crit_pt_locations}, using $J=1$, $h_{x}=2.5$,  $h_{z}=1$.}
  \label{tab::charge_AFM}
\end{table}

\section{Numerical results for the YL critical point in the Schwinger model}
\label{AppSYL}

\begin{table}[H]
  \centering
  \begin{tabular}{c}
      \hline\hline
  \begin{tabular}{c|c|c|c}
        $N$ & $x=1$ & $x=1.5$ & $x=4$ \\ 
        \hline 
        $4$ & $0.115475 $ & $0.118424$& $0.171068$ \\
        \hline 
        $6$ & $0.109997$ & $0.108235$& $0.121927$ \\
        \hline 
        $8$ & $0.108308$ & $0.105007$& $0.108678$ \\
        \hline 
        $10$ & $0.107672$ & $0.103720$& $0.103154$ \\
        \hline
        $12$ & $0.1073996$ & $ 0.103139$& $0.100424$ \\
        \hline 
        $14$ & $0.107271$ & $0.102852$& $0.098941$ \\
        \hline
        $16$ & $0.107205$ & $0.1027001$ & $0.098083$ \\
        \hline 
        $18$ & $0.107168$ & $0.102615$& $0.097564$ \\
        \hline
        $20$ & $0.107147$ & $0.102564$ & $0.097238$ \\
        \hline 
        $22$ & $0.107134$ & $0.102533$& $0.097026$ \\
        \hline 
        $24$ & $0.107126$ & $0.102513$& $0.096884$ \\
        \hline 
        $26$ & $0.107121$ & $0.1025001$& $0.096788$ \\
        \hline\hline
        $\infty$ & $0.107107$ & $0.102467$ & $0.0965204$\\
        \hline
  \end{tabular}
  \end{tabular}
  \caption{Pseudo-critical points $m_{5*}(N)/e$ at $\theta=\pi$, $h_{\textrm{max}}=5$, and $m/e = 0$, for different $N$ and $x$, using \eqref{eq_FSS_SM} with $10^{-7}$ precision.  Extrapolation to $N =\infty$ is done using the BST algorithm with $w=2$.}
  \label{tab::crit_pt_locations_SM}
\end{table}

\begin{table}[H]
  \centering
  \begin{tabular}{c}
      \hline\hline
  \begin{tabular}{c|c|c|c}
        $N$ & $x=1$ & $x=1.5$ & $x=4$ \\ 
        \hline 
        $6$ & $-0.36927624$ & $-0.34172035$& $-$ \\
        \hline 
        $8$ & $-0.37570276$ & $-0.35405464$& $-0.29366551$ \\
        \hline
        $10$ & $-0.38167691$ & $ -0.36448673$& $-0.31172046$ \\
        \hline
        $12$ & $-0.38613707$ & $ -0.37235968$& $-0.32655435$ \\
        \hline
        $14$ & $-0.38934491$ & $ -0.37817538$& $-0.33855012$ \\
        \hline
        $16$ & $-0.39166001$ & $-0.38248816$ & $-0.34821580$ \\
        \hline
        $18$ & $-0.39335663$ & $ -0.38572733$& $-0.35601790$ \\
        \hline
        $20$ & $-0.39462338$ & $-0.38819809$ & $-0.36234491$ \\
        \hline
        $22$ & $-0.39558673$ & $ -0.39011331$& $-0.36750802$ \\
        \hline
        $24$ & $-0.39633216$ & $ -0.39162074$& $-0.37175160$ \\
        \hline
        $26$ & $-0.39691814$ & $ -0.39282428$& $-0.37526579$ \\
        \hline\hline
        $\infty$ & $-0.40016462$ & $-0.40020174$ & $-0.39963903$\\
        \hline
  \end{tabular}
  \end{tabular}
  \caption{The first scaling dimension $\Delta_{\phi}$, computed at the pseudo-critical points in Table \ref{tab::crit_pt_locations_SM} using $\theta=\pi$, $h_{\textrm{max}}=5$, and $m/e = 0$ for different $N$ and $x$. Extrapolation to $N =\infty$ is done using the BST algorithm with $w=2$.}
  \label{tab::scaling_dims_SM}
\end{table}

\begin{table}[H]
  \centering
  \begin{tabular}{c}
      \hline\hline
    \begin{tabular}[t]{c|c|c}
    $x=1.0$ & $x=1.5$ & $x = 4.0$ \\
    \hline
  \begin{tabular}{c|c|c}
        $N$ & $v(N)$ & $c_{\textrm{eff}}(N)$ \\ 
        \hline 
        $4$ & $ 1.578251 $ & $ - $ \\
        \hline
        $6$ & $ 1.759432 $ & $ - $ \\
        \hline 
        $8$ & $ 1.832138 $ & $ - $ \\
        \hline 
        $10$ & $ 1.868123 $ & $ - $ \\
        \hline 
        $12$ & $ 1.888619 $ & $ 0.311578 $ \\
        \hline 
        $14$ & $ 1.901471 $ & $ 0.318369 $ \\
        \hline 
        $16$ & $ 1.910117 $ & $ 0.327389 $ \\
        \hline 
        $18$ & $ 1.916253 $ & $ 0.336228 $ \\
        \hline
        $20$ & $ 1.920793 $ & $ 0.344113 $ \\
        \hline
        $22$ & $ 1.924264 $ & $ 0.350915 $ \\
        \hline
        $24$ & $ 1.926988 $ & $ 0.356705 $ \\
        \hline
        $26$ & $ 1.929174 $ & $ 0.361617 $ \\
        \hline \hline
        $\infty$ & $ 1.945615 $ & $ 0.398489 $ \\
        \hline
    \end{tabular}
    &
    \begin{tabular}{c|c|c}
        $N$ & $v(N)$ & $c_{\textrm{eff}}(N)$ \\ 
        \hline 
        $4$ & $ 2.390136 $ & $ - $ \\
        \hline
        $6$ & $ 2.666549 $ & $ - $ \\
        \hline 
        $8$ & $ 2.773304 $ & $ - $ \\
        \hline 
        $10$ & $ 2.825886 $ & $ - $ \\
        \hline 
        $12$ & $ 2.855739 $ & $ 0.281057 $ \\
        \hline 
        $14$ & $ 2.874340 $ & $ 0.284073 $ \\
        \hline 
        $16$ & $ 2.886747 $ & $ 0.292877 $ \\
        \hline 
        $18$ & $ 2.895472 $ & $ 0.303298 $ \\
        \hline
        $20$ & $ 2.901872 $ & $ 0.313545 $ \\
        \hline
        $22$ & $ 2.906730 $ & $ 0.322899 $ \\
        \hline
        $24$ & $ 2.910521 $ & $ 0.331191 $ \\
        \hline
        $26$ & $ 2.913549 $ & $ 0.338429 $ \\
        \hline \hline
        $\infty$ & $ 2.934368 $ & $ 0.397529 $ \\
        \hline
    \end{tabular}
    &
    \begin{tabular}{c|c|c}
        $N$ & $v(N)$ & $c_{\textrm{eff}}(N)$ \\ 
        \hline 
        $4$ & $ - $ & $ - $ \\
        \hline
        $6$ & $ - $ & $ - $ \\
        \hline 
        $8$ & $ 7.524468 $ & $ - $ \\
        \hline 
        $10$ & $ 7.654080 $ & $ - $ \\
        \hline 
        $12$ & $ 7.729226 $ & $ - $ \\
        \hline 
        $14$ & $ 7.776843 $ & $ - $ \\
        \hline 
        $16$ & $ 7.808842 $ & $ 0.217483 $ \\
        \hline 
        $18$ & $ 7.831286 $ & $ 0.218439 $ \\
        \hline
        $20$ & $ 7.847565 $ & $ 0.224741 $ \\
        \hline
        $22$ & $ 7.859709 $ & $ 0.233919 $ \\
        \hline
        $24$ & $ 7.868992 $ & $ 0.244390 $ \\
        \hline
        $26$ & $ 7.876242 $ & $ 0.255193 $ \\
        \hline \hline
        $\infty$ & $ 2.9338794 $ & $ 0.397471 $ \\
        \hline
    \end{tabular}
  \end{tabular}
  \end{tabular}
  \caption{    Speed of light $v$ and effective central charge $c_{\textrm{eff}}$, using $\theta=\pi$, $h_{\textrm{max}}=5$, $m/e=0$ for different $N$ and $x$, using \eqref{eq_charge_AFM} and \eqref{eq_v_AFM} at the pseudo-critical points in Table \ref{tab::crit_pt_locations_SM}. Extrapolation to $N =\infty$ is done using the BST algorithm with $w=2$.}
  \label{tab::charge_SM}
\end{table}

\section{Numerical results for the YL critical point in the Blume-Capel model}
\label{AppBCYL}

\begin{table}[H]
  \centering
  \begin{tabular}{c}
      \hline\hline
    \begin{tabular}[t]{c|c}
    $h_{x}=2$ & $h_{x} = 3$ \\
  \begin{tabular}{c|c|c|c}
  \hline
        $N$ & $\alpha=0.25$ & $\alpha=0.5$ & $\alpha=0.6$ \\ 
        \hline 
        $2$ & $ 0.328192 $ & $ 0.484315 $& $ 0.581883 $ \\
        \hline 
        $3$ & $ 0.293781 $ & $ 0.449907 $& $ 0.550561 $ \\
        \hline
        $4$ & $ 0.278729 $ & $ 0.434840 $& $ 0.535881 $ \\
        \hline 
        $5$ & $ 0.271501 $ & $ 0.427663 $& $ 0.528408 $ \\
        \hline 
        $6$ & $ 0.267775 $ & $ 0.424005 $& $ 0.524393 $ \\
        \hline 
        $7$ & $ 0.265740 $ & $ 0.422030 $& $ 0.522137 $ \\
        \hline 
        $8$ & $ 0.264574 $ & $ 0.420908 $& $ 0.520816 $ \\
        \hline 
        $9$ & $ 0.263875 $ & $ 0.420241 $& $ 0.520013 $ \\
        \hline 
        $10$ & $ 0.263441 $ & $ 0.419830 $& $ 0.519509 $ \\
        \hline
        $11$ & $ 0.263162 $ & $ 0.419567 $& $ 0.519181 $ \\
        \hline
        $12$ & $ 0.262977 $ & $ 0.419394 $& $ 0.518963 $ \\
        \hline
        $13$ & $ 0.262852 $ & $ 0.419276 $& $ 0.518814 $ \\
        \hline
        $14$ & $ 0.262765 $ & $ 0.419195 $& $ 0.518709 $ \\
        \hline
        $15$ & $ 0.262703 $ & $ 0.419137 $& $ 0.518635 $ \\
        \hline
        $16$ & $ 0.262658 $ & $ 0.419095 $& $ 0.518581 $ \\
        \hline
        $17$ & $ 0.262625 $ & $ 0.419065 $& $ 0.518541 $ \\
        \hline
        $18$ & $ 0.262600 $ & $ 0.419042 $& $ 0.518511 $ \\
        \hline \hline
        $\infty$ & $ 0.262506 $ & $ 0.418956 $ & $ 0.518397 $\\
        \hline
    \end{tabular}
    &
    \begin{tabular}{c|c|c}
    \hline
        $N$ & $\alpha=0.25$ & $\alpha=0.5$ \\ 
        \hline 
        $2$ & $ - $ & $ - $ \\
        \hline 
        $3$ & $ 0.858398 $ & $ 1.099598 $ \\
        \hline
        $4$ & $ 0.843518 $ & $ 1.083566 $ \\
        \hline 
        $5$ & $ 0.837338 $ & $ 1.076500 $ \\
        \hline 
        $6$ & $ 0.834508 $ & $ 1.073132 $ \\
        \hline 
        $7$ & $ 0.833101 $ & $ 1.071410 $ \\
        \hline 
        $8$ & $ 0.832350 $ & $ 1.070473 $ \\
        \hline 
        $9$ & $ 0.831926 $ & $ 1.069936 $ \\
        \hline 
        $10$ & $ 0.831674 $ & $ 1.069612 $ \\
        \hline
        $11$ & $ 0.831518 $ & $ 1.069410 $ \\
        \hline
        $12$ & $ 0.831417 $ & $ 1.069279 $ \\
        \hline
        $13$ & $ 0.831350 $ & $ 1.069192 $ \\
        \hline
        $14$ & $ 0.831305 $ & $ 1.069132 $ \\
        \hline
        $15$ & $ 0.831273 $ & $ 1.069090 $ \\
        \hline
        $16$ & $ 0.831250 $ & $ 1.069060 $ \\
        \hline
        $17$ & $ 0.831234 $ & $ 1.069038 $ \\
        \hline
        $18$ & $ 0.831221 $ & $ 1.069021 $ \\
        \hline \hline
        $\infty$ & $ 0.831176 $ & $ 1.068961 $ \\
        \hline
    \end{tabular}
  \end{tabular}
  \end{tabular}
  \caption{Pseudo-critical point $h_{z,*}(N)$ for different $N$, $h_{x}$, $\alpha$ and $J=1$, using \eqref{eq_FSS_BC} with $10^{-7}$ precision.}
  \label{tab::crit_pt_locations_BC}
\end{table}

\begin{table}[H]
  \centering
  \begin{tabular}{c}
      \hline\hline
      \begin{tabular}[t]{c}
      $h_{x} = 2$  \\
      \begin{tabular}[t]{cc}
  \begin{tabular}{|c|c|c|c|}
  \hline
        $N$ & $\alpha=0.25$ & $\alpha=0.5$ & $\alpha=0.6$ \\ 
        \hline \hline
        $3$ & $ -0.293812 $ & $ -0.298075 $& $ -0.269338 $ \\
        \hline
        $4$ & $ -0.30556 $ & $ -0.309649 $& $ -0.282935 $ \\
        \hline 
        $5$ & $ -0.320822 $ & $ -0.324802 $& $ -0.303274 $ \\
        \hline 
        $6$ & $ -0.334469 $ & $ -0.338159 $& $ -0.319866 $ \\
        \hline 
        $7$ & $ -0.345695 $ & $ -0.349011 $& $ -0.333299 $ \\
        \hline 
        $8$ & $ -0.354701 $ & $ -0.357634 $& $ -0.344064 $ \\
        \hline 
        $9$ & $ -0.361888 $ & $ -0.364466 $& $ -0.352685 $ \\
        \hline 
        $10$ & $ -0.367643 $ & $ -0.369908 $& $ -0.359623 $ \\
        \hline
        $11$ & $ -0.372283 $ & $ -0.374277 $& $ -0.365247 $ \\
        \hline
        $12$ & $ -0.376056 $ & $ -0.377816 $& $ -0.369844 $ \\
        \hline
        $13$ & $ -0.379152 $ & $ -0.380712 $& $ -0.373635 $ \\
        \hline
        $14$ & $ -0.381715 $ & $ -0.383104 $& $ -0.376789 $ \\
        \hline
        $15$ & $ -0.383855 $ & $ -0.385098 $& $ -0.379433 $ \\
        \hline
        $16$ & $ -0.385657 $ & $ -0.386774 $& $ -0.381669 $ \\
        \hline
        $17$ & $ -0.387185 $ & $ -0.388193 $& $ -0.383572 $ \\
        \hline
        $18$ & $ -0.388492 $ & $ -0.389405 $& $ -0.385204 $ \\
        \hline \hline
        $\infty$ & $ -0.400123 $ & $ -0.399577 $ & $ -0.400648 $\\
        \hline
  \end{tabular} 
    &
  \begin{tabular}{|c|c|c|c|}
  \hline
        $N$ & $\alpha=0.25$ & $\alpha=0.5$ & $\alpha=0.6$ \\ 
        \hline \hline
        $3$ & $ 0.645595 $ & $ 0.569456 $ & $ 0.462623 $\\
        \hline
        $4$ & $ 0.732387 $ & $ 0.655102 $ & $ 0.523917 $ \\
        \hline 
        $5$ & $ 0.825128 $ & $ 0.753247 $ & $ 0.607006 $ \\
        \hline 
        $6$ & $ 0.912516 $ & $ 0.850454 $ & $ 0.692302 $ \\
        \hline 
        $7$ & $ 0.991304 $ & $ 0.941263 $ & $ 0.777746 $ \\
        \hline 
        $8$ & $ 1.060873 $ & $ 1.023121 $ & $ 0.860951 $ \\
        \hline 
        $9$ & $ 1.121653 $ & $ 1.095094 $ & $ 0.939994 $  \\
        \hline 
        $10$ & $ 1.174479 $ & $ 1.157322 $ & $ 1.013387 $  \\
        \hline
        $11$ & $ 1.220301 $ & $ 1.210566 $ & $ 1.080104 $  \\
        \hline
        $12$ & $ 1.260050 $ & $ 1.255908 $ & $ 1.139596 $  \\
        \hline
        $13$ & $ 1.294578 $ & $ 1.294498 $ & $ 1.191828 $ \\
        \hline
        $14$ & $ 1.324636 $ & $ 1.327414 $ & $ 1.237164 $ \\
        \hline
        $15$ & $ 1.350873 $ & $ 1.355597 $ & $ 1.276246 $ \\
        \hline
        $16$ & $ 1.373845 $ & $ 1.379843 $ & $ 1.309848 $ \\
        \hline
        $17$ & $ 1.394023 $ & $ 1.400809 $ & $ 1.338754 $ \\
        \hline
        $18$ & $ 1.411805 $ & $ 1.419034 $ & $ 1.363691 $ \\
        \hline \hline
        $\infty$ & $ 1.600283 $ & $ 1.599754 $ & $ 1.600385 $\\
        \hline
    \end{tabular}
    \end{tabular}

  \end{tabular}
  \end{tabular}

  \caption{Scaling dimensions of the first 2 spinless gaps $\Delta^{k=0}_{i}(N)$ at $h_{z}=h_{z,*}(N)$ for different $N$, $\alpha$, and $h_{x}=2$, $J=1$, using (\ref{eq_scaling_dim_YL}).}
  \label{tab::scaling_dim_BC}
\end{table}

\begin{table}[H]
  \centering
  \begin{tabular}{c}
      \hline\hline
      \begin{tabular}[t]{c}
      $h_{x} = 3$  \\
\begin{tabular}[t]{c c}
    \begin{tabular}{|c|c|c|c|}
  \hline
        $N$ & $\alpha=0.25$ & $\alpha=0.5$ \\ 
        \hline \hline
        $3$ & $ -0.349354 $ & $ -0.324023 $ \\
        \hline
        $4$ & $ -0.355016 $ & $ -0.334174 $ \\
        \hline 
        $5$ & $ -0.363946 $ & $ -0.347277 $ \\
        \hline 
        $6$ & $ -0.371522 $ & $ -0.358050 $ \\
        \hline 
        $7$ & $ -0.377364 $ & $ -0.366331 $ \\
        \hline 
        $8$ & $ -0.381789 $ & $ -0.372632 $ \\
        \hline 
        $9$ & $ -0.385152 $ & $ -0.377455 $ \\
        \hline 
        $10$ & $ -0.387738 $ & $ -0.381193 $ \\
        \hline
        $11$ & $ -0.389750 $ & $ -0.384128 $ \\
        \hline
        $12$ & $ -0.391339 $ & $ -0.386463 $ \\
        \hline
        $13$ & $ -0.392612 $ & $ -0.388343 $ \\
        \hline
        $14$ & $ -0.393639 $ & $ -0.389878 $ \\
        \hline
        $15$ & $ -0.394481 $ & $ -0.391141 $ \\
        \hline
        $16$ & $ -0.395178 $ & $ -0.392192 $ \\
        \hline
        $17$ & $ -0.395759 $ & $ -0.393075 $ \\
        \hline
        $18$ & $ -0.396248 $ & $ -0.393823 $ \\
        \hline \hline
        $\infty$ & $ -0.400216 $ & $ -0.400110 $ \\
        \hline
  \end{tabular} 
  &
  \begin{tabular}{|c|c|c|}
  \hline
        $N$ & $\alpha=0.25$ & $\alpha=0.5$ \\ 
        \hline \hline
        $3$ & $ 0.749389 $ & $ 0.590968 $ \\
        \hline
        $4$ & $ 0.866156 $ & $ 0.687118 $ \\
        \hline 
        $5$ & $ 0.983440 $ & $ 0.799658 $ \\
        \hline 
        $6$ & $ 1.086246 $ & $ 0.911651 $ \\
        \hline 
        $7$ & $ 1.171426 $ & $ 1.015575 $ \\
        \hline 
        $8$ & $ 1.240260 $ & $ 1.107163 $ \\
        \hline 
        $9$ & $ 1.295470 $ & $ 1.184620 $ \\
        \hline 
        $10$ & $ 1.339862 $ & $ 1.248280 $ \\
        \hline
        $11$ & $ 1.375799 $ & $ 1.299834 $  \\
        \hline
        $12$ & $ 1.405157 $ & $ 1.341467 $  \\
        \hline
        $13$ & $ 1.429373 $ & $ 1.375257 $ \\
        \hline
        $14$ & $ 1.449516 $ & $ 1.402938 $ \\
        \hline
        $15$ & $ 1.466430 $ & $ 1.425844 $ \\
        \hline
        $16$ & $ 1.480744 $ & $ 1.444999 $ \\
        \hline
        $17$ & $ 1.492951 $ & $ 1.461178 $ \\
        \hline
        $18$ & $ 1.503433 $ & $ 1.474965 $ \\
        \hline \hline
        $\infty$ & $ 1.600463 $ & $ 1.599982 $ \\
        \hline
    \end{tabular}
  \end{tabular}
   \end{tabular}
  \end{tabular}
  \caption{      Scaling dimensions of the first 2 spinless gaps $\Delta^{k=0}_{i}(N)$ at $h_{z}=h_{z,*}(N)$   for various $N$, $\alpha$ and $h_{x}=3$, $J=1$, using (\ref{eq_scaling_dim_YL}). }
  \label{tab::scaling_dim_BC2}
\end{table}

\begin{table}[H]
  \centering
  \begin{tabular}{c}
      \hline\hline
    \begin{tabular}[t]{c|c|c|c}
    $h_{x}=2$, $\alpha=0.25$ & $h_{x}=2$, $\alpha=0.5$ \\
    \hline
  \begin{tabular}{c|c|c}
        $N$ & $v(N)$ & $c_{\textrm{eff}}(N)$ \\ 
        \hline 
        $3$ & $ 1.2110800 $ & $ - $ \\
        \hline
        $4$ & $ 1.3106532 $ & $ - $ \\
        \hline 
        $5$ & $ 1.3565797 $ & $ - $ \\
        \hline 
        $6$ & $ 1.3811443 $ & $ - $ \\
        \hline 
        $7$ & $ 1.3958265 $ & $ 0.15128969 $ \\
        \hline 
        $8$ & $ 1.4053686 $ & $ 0.15639544 $ \\
        \hline 
        $9$ & $ 1.4119763 $ & $ 0.16936681 $ \\
        \hline 
        $10$ & $ 1.4167800 $ & $ 0.18568469 $ \\
        \hline
        $11$ & $ 1.4204069 $ & $ 0.20286316 $ \\
        \hline
        $12$ & $ 1.4232286 $ & $ 0.21960523 $ \\
        \hline
        $13$ & $ 1.4254776 $ & $ 0.23527836 $ \\
        \hline
        $14$ & $ 1.4273062 $ & $ 0.24964248 $ \\
        \hline
        $15$ & $ 1.4288178 $ & $ 0.26263785 $ \\
        \hline
        $16$ & $ 1.4300849 $ & $ 0.27431626 $ \\
        \hline
        $17$ & $ 1.4311598 $ & $ 0.28479564 $ \\
        \hline
        $18$ & $ 1.4320817 $ & $ 0.29410978 $ \\
        \hline \hline
        $\infty$ & $ 1.4410794 $ & $ 0.3999545 $ \\
    \end{tabular}
    &
    \begin{tabular}{c|c|c}
        $N$ & $v(N)$ & $c_{\textrm{eff}}(N)$ \\ 
        \hline  
        $3$ & $ 1.2431574 $ & $ - $ \\
        \hline
        $4$ & $ 1.3469292 $ & $ - $ \\
        \hline 
        $5$ & $ 1.3938711 $ & $ - $ \\
        \hline 
        $6$ & $ 1.4187837 $ & $ - $ \\
        \hline 
        $7$ & $ 1.4336004 $ & $ 0.10349213 $ \\
        \hline 
        $8$ & $ 1.4431837 $ & $ 0.11746958 $ \\
        \hline 
        $9$ & $ 1.4497928 $ & $ 0.13810300 $ \\
        \hline 
        $10$ & $ 1.4545643 $ & $ 0.16067426 $ \\
        \hline
        $11$ & $ 1.4581535 $ & $ 0.18282880 $ \\
        \hline
        $12$ & $ 1.4609376 $ & $ 0.20348798 $ \\
        \hline
        $13$ & $ 1.4631519 $ & $ 0.2222637 $ \\
        \hline
        $14$ & $ 1.4649457 $ & $ 0.23904516 $ \\
        \hline
        $15$ & $ 1.4664258 $ & $ 0.25398546 $ \\
        \hline
        $16$ & $ 1.4676651 $ & $ 0.26717905 $ \\
        \hline
        $17$ & $ 1.4687153 $ & $ 0.27889341 $ \\
        \hline
        $18$ & $ 1.4696150 $ & $ 0.28925211 $ \\
        \hline \hline
        $\infty$ & $ 1.4782640 $ & $ 0.39849902 $ \\
    \end{tabular}
        \end{tabular}
        \\
        \hline \hline
    \begin{tabular}[t]{c|c}
     $h_{x} = 3$, $\alpha=0.25$ & $h_{x} = 3$, $\alpha=0.5$ \\ \hline
    \begin{tabular}{c|c|c}
        $N$ & $v(N)$ & $c_{\textrm{eff}}(N)$ \\ 
        \hline  
        $3$ & $ 1.5436578 $ & $ - $ \\
        \hline
        $4$ & $ 1.6700197 $ & $ - $ \\
        \hline 
        $5$ & $ 1.7291468 $ & $ - $ \\
        \hline 
        $6$ & $ 1.7616451 $ & $ 0.17875395 $ \\
        \hline 
        $7$ & $ 1.7816099 $ & $ 0.19706057 $ \\
        \hline 
        $8$ & $ 1.7948827 $ & $ 0.21921968 $ \\
        \hline 
        $9$ & $ 1.8042421 $ & $ 0.24075353 $ \\
        \hline 
        $10$ & $ 1.8111293 $ & $ 0.26009701 $ \\
        \hline
        $11$ & $ 1.8163869 $ & $ 0.27692418 $ \\
        \hline
        $12$ & $ 1.8205057 $ & $ 0.29135666 $ \\
        \hline
        $13$ & $ 1.8237948 $ & $ 0.30371045 $ \\
        \hline
        $14$ & $ 1.8264920 $ & $ 0.31421518 $ \\
        \hline
        $15$ & $ 1.8287211 $ & $ 0.32325313 $ \\
        \hline
        $16$ & $ 1.8305929 $ & $ 0.33099988 $ \\
        \hline
        $17$ & $ 1.8321828 $ & $ 0.33769377 $ \\
        \hline
        $18$ & $ 1.8335468 $ & $ 0.34350157 $ \\
        \hline \hline
        $\infty$ & $ 1.8471945 $ & $ 0.3996187 $ \\
        \hline
        \end{tabular}
        &
    \begin{tabular}{c|c|c}
        $N$ & $v(N)$ & $c_{\textrm{eff}}(N)$ \\ 
        \hline  
        $3$ & $ 1.6505627 $ & $ - $ \\
        \hline
        $4$ & $ 1.7746557 $ & $ - $ \\
        \hline 
        $5$ & $ 1.8274514 $ & $ - $ \\
        \hline 
        $6$ & $ 1.8554875 $ & $ 0.09849090 $ \\
        \hline 
        $7$ & $ 1.8723953 $ & $ 0.11957551 $ \\
        \hline 
        $8$ & $ 1.8835159 $ & $ 0.14779712 $ \\
        \hline 
        $9$ & $ 1.8913044 $ & $ 0.17634261 $ \\
        \hline 
        $10$ & $ 1.8970087 $ & $ 0.20260121 $ \\
        \hline
        $11$ & $ 1.9013458 $ & $ 0.22577053 $ \\
        \hline
        $12$ & $ 1.9047398 $ & $ 0.24585163 $ \\
        \hline
        $13$ & $ 1.9074629 $ & $ 0.26311980 $ \\
        \hline
        $14$ & $ 1.9096760 $ & $ 0.27796100 $ \\
        \hline
        $15$ & $ 1.9115161 $ & $ 0.29071567 $ \\
        \hline
        $16$ & $ 1.9130636 $ & $ 0.30167872 $ \\
        \hline
        $17$ & $ 1.9143798 $ & $ 0.31120314 $ \\
        \hline
        $18$ & $ 1.9155108 $ & $ 0.31946779 $ \\
        \hline \hline
        $\infty$ & $ 1.9272942 $ & $ 0.39961648 $ \\
        \hline
    \end{tabular}
  \end{tabular}
  \end{tabular}
  \caption{Speed of light $v$  and effective central charge $c_{\textrm{eff}}$ for different $N$, $\alpha$, and $h_{x}$ and $J=1$, using \eqref{eq_charge_BC} and \eqref{eq_v_AFM} at the pseudo-critical points of Table \ref{tab::crit_pt_locations_BC}.}
  \label{tab::charge_BC}
\end{table}

\section{Numerical results for the YL critical point in the three-state clock model}
\label{AppCMYL}

\begin{table}[H]
  \centering
  \begin{tabular}{c}
      \hline\hline
      \begin{tabular}[t]{c}
  \begin{tabular}{c|c|c|c|c|c|c}
    $N$ &  $8 $ & $9$ & $ 10$ & $11$ & $12$ & $13$ \\
    \hline
    $h_{*}(N)$ & $ 0.71654713 $ & $ 0.71582773 $ & $ 0.71533747 $ & $ 0.71499698 $ & $ 0.71475687 $ & $ 0.71458524 $  \\ 
    \hline\hline
  \end{tabular}
  \\
  \begin{tabular}{c|c|c|c|c|c||c}
    $N$ &  $14$ & $15$ & $16$ & $17$ & $18$ & $\infty$\\
    \hline
    $h_{*}(N)$ & $ 0.71446095 $ & $ 0.71436983 $ & $ 0.71430221 $  & $ 0.71425143 $ & $ 0.71421286 $ & $0.71406120$\\ 
    \hline\hline
  \end{tabular}
  \end{tabular}
  \end{tabular}
  \caption{Yang-Lee pseudo-critical point $h_{*}(N)$ of the Hamiltonian $H_{\textrm{CMYL}}$ in (\ref{eq_H_CM_YL}) for different $N$, using the condition \eqref{eq_FSS_CM} with $J=1$, $f=2$, $\lambda=1$.}
  \label{tab::crit_pt_locations_CM}
\end{table}

\begin{table}[H]
  \centering
  \begin{tabular}{c}
      \hline\hline
  \begin{tabular}{c|c|c}
    $N$  & $\Delta_{0}^{k=0}$ & $\Delta_{0}^{k=2}$ \\ 
    \hline\hline
    $8 $& $ -0.34324937 $ &  $ 1.48067534$\\ 
    \hline
    $9 $& $ -0.34985414 $ &  $ 1.50596149$\\ 
    \hline
    $10 $& $ -0.35574526 $ & $ 1.52362620$\\ 
    \hline
    $11 $& $ -0.36091955 $ &  $1.53641715 $\\ 
    \hline
    $12 $& $ -0.36541986 $ &  $1.54596761 $ \\ 
    \hline
    $13 $& $ -0.36931229 $ & $ 1.55329280$\\ 
    \hline
    $14 $& $ -0.37267031 $ & $ 1.55904582 $\\ 
    \hline
    $16 $& $ -0.37806298 $ & $ 1.56365961 $\\ 
    \hline
    $18 $& $ -0.38204509 $ & $ 1.56742763 $\\ 
    \hline
    $20 $& $ -0.38487196 $ & $ 1.57055417 $\\ 
    \hline\hline
    $\infty $& $ -0.40098786 $ & $ 1.59698429 $\\
    \hline
  \end{tabular}
  \end{tabular}
  \caption{Scaling dimensions of the first excited states in the gapless sector at the corresponding pseudo-critical points.}
  \label{tab::scaling_dim_CM}
\end{table}

\begin{table}[H]
  \centering
  \begin{tabular}{c}
      \hline\hline
  \begin{tabular}{c|c|c}
        $N$ & $v(N)$ & $c_{\textrm{eff}}(N)$ \\ 
        \hline 
        $8$ & $ 3.6483823 $ & $ - $ \\
        \hline 
        $9$ & $ 3.6612549 $ & $ - $ \\
        \hline 
        $10$ & $ 3.6710777 $ & $ - $ \\
        \hline
        $11$ & $ 3.6788948 $ & $ - $ \\
        \hline
        $12$ & $ 3.6852771 $ & $ - $ \\
        \hline
        $13$ & $ 3.6905708 $ & $ 0.26750511 $ \\
        \hline
        $14$ & $ 3.6950078 $ & $ 0.26951454 $ \\
        \hline
        $15$ & $ 3.6987554 $ & $ 0.27359545 $ \\
        \hline
        $16$ & $ 3.7019420 $ & $ 0.27889058 $ \\
        \hline
        $17$ & $ 3.7046679 $ & $ 0.28482511 $ \\
        \hline
        $18$ & $ 3.7070138 $ & $ 0.29100050 $ \\
        \hline
        $19$ & $ 3.7090925 $ & $ 0.29558494 $ \\
        \hline \hline
        $\infty$ & $ 3.7285092 $ & $ 0.40863643 $ \\
        \hline
    \end{tabular}
  \end{tabular}
  \caption{Speed of light $v$ and effective central charge $c_{\textrm{eff}}$ for different $N$ at $J=1$, $f=2$ and $\lambda=1$, using \eqref{eq_v_AFM} and \eqref{eq_charge_BC} at the pseudo-critical points in Table \ref{tab::crit_pt_locations_CM}.}
  \label{tab::charge_CM}
\end{table}

\section{Numerical results for the Ising critical point in the three-state clock model}
\label{AppCMI}

\begin{table}[H]
  \centering
  \begin{tabular}{c}
      \hline\hline
  \begin{tabular}{c|c|c|c|c|c|c}
    $N$ &  $8 $ & $9$ & $ 10$ & $11$ & $12$ & $13$  \\
    \hline
    $h^{*}(N)$ & $ 1.5726535 $ & $ 1.5730529 $ & $ 1.5732883 $ & $ 1.5734338 $ & $ 1.5735276 $ & $ 1.5735899 $  \\ 
    \hline
  \end{tabular}
  \\
  \\
  \begin{tabular}{c|c|c|c|c||c}
  \hline
    $N$ & $14$ & $16$ & $18$ & $20$& $\infty$ \\
    \hline
    $h_{*}(N)$& $ 1.5736328 $ & $ 1.5736848 $ & $ 1.5737128 $ & $ 1.57372898 $ & $ 1.5737608 $\\ 
    \hline
  \end{tabular}
  \end{tabular}
  \caption{Ising pseudo-critical point $h_{*}(N)$ of the Hamiltonian $H_{\textrm{CMI}}$ in (\ref{eq_H_CM_Ising}) for different $N$, using the condition \eqref{eq_FSS_CM}, with $10^{-5}$ precision and parameters $J=1$, $f=2$.}
  \label{tab::crit_pt_locations_Ising}
\end{table}

\begin{table}[H]
  \centering
  \begin{tabular}{c}
      \hline\hline
  \begin{tabular}{c|c|c|c|c|c|c}
    $N$  & $\Delta_{1}^{k=0}$ & $\Delta_{2}^{k=0}$ & $\Delta_{1}^{k=1}$ & $\Delta_{3}^{k=0}$ & $\Delta_{0}^{k=2}$ & $\Delta_{1}^{k=2}$ \\ 
    \hline\hline
    $8 $& $ 0.13348060 $ & $ 0.93582406 $ & $ 1.50992726 $ & $ 1.63854428 $ & $ 1.68461706 $ & $ 1.79359816 $\\ 
    \hline
    $9 $& $ 0.13154906 $ & $ 0.94865093 $ & $ 1.59787855 $ & $ 1.71255708 $& $ 1.83669093 $ & $ 1.87654768 $\\ 
    \hline
    $10 $& $ 0.13021805 $ & $ 0.95826896 $ & $ 1.67126687 $ & $ 1.77689889 $ & $ 1.93530904 $ & $ 1.95799000 $\\ 
    \hline
    $11 $& $ 0.12926044 $ & $ 0.96554982 $ & $ 1.73069068 $ & $ 1.83118343 $ & $ 1.97589789 $ & $ 1.98956009 $\\ 
    \hline
    $12 $& $ 0.12854746 $ & $ 0.97114293 $ & $ 1.77789358$ & $ 1.87612027 $ & $ 1.99450424 $ & $ 2.00444729 $\\ 
    \hline
    $13 $& $ 0.12800171 $ & $ 0.97550898 $ & $ 1.81506784$ & $ 1.91294053 $ & $ 1.99631829 $ & $ 2.02520168 $\\ 
    \hline
    $14 $& $ 0.12757422 $ & $ 0.97897092 $ & $ 1.84434179 $ & $ 1.94300708 $ & $ 1.99727171 $ & $ 2.04079375 $\\ 
    \hline
    $15 $& $ 0.12723283 $ & $ 0.98175641 $ & $ 1.86752697 $ & $ 1.96759470 $ & $ 1.99786525 $ & $ 2.05284546 $\\ 
    \hline
    $16 $& $ 0.12695583 $ & $ 0.98402809 $ & $ 1.88605812 $ & $ 1.98779626 $ & $ 1.99827216 $ & $ 2.06238418 $\\ 
    \hline
    $17 $& $ 0.12672774 $ & $ 0.98590312 $ & $ 1.90102822 $ & $ 2.00450256 $ & $ 1.99856801 $ & $ 2.07008328 $\\ 
    \hline
    $18 $& $ 0.12653774 $ & $ 0.98746805 $ & $1.91325754 $ & $ 2.01842291 $ & $ 1.99879251 $ & $ 2.07640087  $\\ 
    \hline
    $20 $& $ 0.12624150 $ & $ 0.98990879 $ & $ 1.93178615$ & $ 2.04000702 $ & $ 1.99910773 $ & $ 2.08608350 $\\ 
    \hline
    $30 $& $ 0.12554944 $ & $ 0.99559773 $ & $ 1.97198825 $ & $ 2.08902361 $ & $ 1.99970539 $ & $ 2.10806707 $\\ 
    \hline\hline
    $\infty $& $ 0.12500519 $ & $ 1.0000254 $ & $ 2.00008016 $ & $ 2.12510859 $ & $ 2.000049 $ & $ 2.12493391 $\\ 
    \hline
  \end{tabular}
  \end{tabular}
  \caption{Scaling dimensions of the first excited states in the gapless sector of the Hamiltonian $H_{\textrm{CMI}}$ in (\ref{eq_H_CM_Ising}) at the corresponding pseudo-critical points of Table \ref{tab::crit_pt_locations_Ising} and $J=1$, $f=2$.}
  \label{tab::scaling_dim_Ising}
\end{table}

\begin{table}[H]
  \centering
  \begin{tabular}{c}
      \hline\hline
  \begin{tabular}{c|c|c}
  $N$ & $v(N)$ & $c(N)$ \\ 
        \hline 
        $9$ & $ 5.0331777 $ & $ - $ \\
        \hline 
        $10$ & $ 5.0820419 $ & $ 0.50357234 $ \\
        \hline
        $11$ & $ 5.1179115 $ & $ 0.50269290 $ \\
        \hline
        $12$ & $ 5.1450343 $ & $ 0.50211373 $ \\
        \hline
        $13$ & $ 5.1660521 $ & $ 0.50174022 $ \\
        \hline
        $14$ & $ 5.1826785 $ & $ 0.50143236 $ \\
        \hline
        $15$ & $ 5.1960631 $ & $ 0.50118348 $ \\
        \hline
        $16$ & $ 5.2070001 $ & $ 0.50109110 $ \\
        \hline
        $17$ & $ 5.2160557 $ & $ 0.50088145 $ \\
        \hline \hline
        $\infty$ & $ 5.2858696 $ & $ 0.49974482 $ \\
        \hline
    \end{tabular}
  \end{tabular}
  \caption{Speed of light $v$ and the central charge $c$  for different $N$, using \eqref{eq_v_Ising} and \eqref{eq_charge_BC} at the pseudo-critical points of Table \ref{tab::crit_pt_locations_Ising} for $J=1$, $f=2$.}
  \label{tab::charge_CM_Ising}
\end{table}

\bibliography{ylqm.bib}
\end{document}